\documentclass[11pt,a4paper]{article}
\pdfoutput=1
\usepackage{jcappub}
\allowdisplaybreaks
\usepackage{xcolor}
\usepackage{amsmath}
\usepackage{graphicx}
\usepackage{esint}
\usepackage{slashed}
\usepackage[normalem]{ulem}
\usepackage[english]{babel}

\graphicspath{{./figures/}}

\begin{document}
{\hfill INT-PUB-21-024}

\title{Light scalars in neutron star mergers}

\author[a]{P. S. Bhupal Dev,}
\author[b]{Jean-Fran\c cois Fortin,}
\author[c]{Steven P. Harris,}
\author[d]{Kuver Sinha,}
\author[e]{Yongchao Zhang}

\affiliation[a]{Department of Physics and McDonnell Center for the Space Sciences, Washington University, St. Louis, MO 63130, USA}

\affiliation[b]{D\'epartement de Physique, de G\'enie Physique et d'Optique, Universit\'e Laval, Qu\'ebec, QC G1V 0A6, Canada}

\affiliation[c]{Institute for Nuclear Theory, University of Washington, Seattle, WA 98195, USA}

\affiliation[d]{Department of Physics and Astronomy, University of Oklahoma, Norman, OK 73019, USA}

\affiliation[e]{School of Physics, Southeast University, Nanjing 211189, China}

\emailAdd{bdev@wustl.edu}
\emailAdd{jean-francois.fortin@phy.ulaval.ca}
\emailAdd{harrissp@uw.edu}
\emailAdd{kuver.sinha@ou.edu}
\emailAdd{zhangyongchao@seu.edu.cn}

\abstract{
Due to their unique set of multimessenger signals, neutron star mergers have emerged as a novel environment for studies of new physics beyond the Standard Model (SM).  As a case study, we consider the simplest extension of the SM scalar sector involving a light CP-even scalar singlet $S$ mixing with the SM Higgs boson.  These $S$ particles can be produced abundantly in neutron star mergers via the nucleon bremsstrahlung process.  We show that the $S$ particles may either be trapped in or stream freely out of the merger remnant, depending on the $S$ mass, its mixing with the SM Higgs boson, and the temperature and baryon density in the merger.  In the free-streaming region, the scalar $S$ will provide an extra channel to cool down the merger remnant, with cooling timescales as small as ${\cal O}$(ms). On the other hand, in the trapped region, the Bose gas of $S$ particles could contribute a larger thermal conductivity than the trapped neutrinos in some parts of the  parameter space, thus leading to faster thermal equilibration than expected. Therefore, future observations of the early postmerger phase of a neutron star merger could effectively probe a unique range of the $S$ parameter space, largely complementary to the existing and future laboratory and supernova limits. In view of these results, we hope the merger simulation community will be motivated to implement the effects of light CP-even scalars into their simulations in both the free-streaming and trapped regimes.
}

\maketitle

\section{Introduction}

Dense astrophysical environments have proved extraordinarily useful in constraining light beyond the Standard Model (BSM) particles (such as axions and dark photons)~\cite{Kolb:1990vq,Raffelt:1990yz,Raffelt:1996wa}.  If the BSM particles are sufficiently light and couple to nucleons, leptons, or photons,  they can be readily produced in the hot and dense nuclear matter in the cores of stars, supernovae and neutron stars. If the BSM particles interact very weakly with the SM particles, they will free-stream out of the astrophysical bodies, thus providing an additional energy loss mechanism that would affect their evolution. Limits can then be derived on the couplings of these BSM particles to nucleons, leptons and photons from the requirement that stellar lifetimes and energy-loss rates should not conflict with observations. There exists a suite of astrophysical limits from the Sun, from evolved low-mass stars---such as red giants and horizontal-branch stars in globular clusters, or white dwarfs---as well as from neutron stars and supernova 1987A~\cite{Batista:2021ctk}.

Since the 2017 detection of gravitational waves from the inspiral phase of the neutron star merger GW170817~\cite{TheLIGOScientific:2017qsa} along with an electromagnetic counterpart~\cite{LIGOScientific:2017ync, LIGOScientific:2017zic}, it has become clear that neutron star mergers provide an interestingly new astrophysical environment to study light BSM physics~\cite{Fortin:2021cog}, apart from being a powerful tool for exploring other fundamental physics topics~\cite{Burns:2019byj, Evans:2021gyd}, such as constraining the nuclear equation of state (EoS)~\cite{De:2018uhw, LIGOScientific:2018cki}, understanding the $r$-process~\cite{Kasen:2017sxr}, measuring the Hubble constant~\cite{LIGOScientific:2017adf}, studying dark matter capture~\cite{Bezares:2019jcb}, placing bounds on the violation of Lorentz invariance~\cite{LIGOScientific:2017zic} and on gravitational parity violation~\cite{Alexander:2017jmt}, and testing alternative theories of gravity~\cite{Creminelli:2017sry, Sakstein:2017xjx,Ezquiaga:2017ekz, Boran:2017rdn, Baker:2017hug, Sennett:2019bpc}. Recently, the possible role of axions~\cite{Hook:2017psm,Huang:2018pbu,Dietrich:2019shr,Harris:2020qim,Fortin:2021cog,Fiorillo:2021gsw} and dark gauge bosons~\cite{Croon:2017zcu,Dror:2019uea, Diamond:2021ekg} in the merger environment has been investigated.  In this work we will consider the impact of a light CP-even scalar on mergers.  We will not attempt to set limits on the properties of scalars from mergers, but instead make a case for implementing the effects of scalar particles in numerical simulations of mergers, which can be used in combination with future observations to constrain or detect the presence of scalars in mergers.

During the merger, the two constituent neutron stars collide and form a hot, dense remnant containing nuclear matter where BSM particles can be copiously produced, similar to the environment that exists in a supernova core, though perhaps reaching higher densities and temperatures~\cite{Bernuzzi:2020tgt} (see the discussion in Section~\ref{sec:nuclear_matter_in_mergers}).  Neutron star mergers also complement supernovae in terms of their multimessenger signals.  Supernovae produce an electromagnetic signal which provides information about the surface of last scattering for photons, as well as a neutrino signal which is sensitive to the temperature of the supernovae neutrinosphere.  Core-collapse supernovae with aspherical mass-energy dynamics are also believed to generate gravitational waves, though a signal has not yet been detected~\cite{LIGOScientific:2016jvu} and is expected to be beyond the current detection capabilities unless the supernova occurs close by in our galaxy~\cite{Abdikamalov:2020jzn}.  Neutron star mergers, on the other hand, produce a stronger gravitational wave signal from both the inspiral (measured by LIGO and VIRGO~\cite{TheLIGOScientific:2017qsa,LIGOScientific:2020aai}) and the (unmeasured, thus far) postmerger phase~\cite{Baiotti:2019sew}.  In addition, they produce electromagnetic signals in the form of a gamma-ray burst~\cite{LIGOScientific:2017ync} and the kilonova~\cite{Metzger:2019zeh}.  Neutron star mergers produce neutrinos as well, but since the number of neutrinos per event is less than for a supernova, and the merger rate is much less than the supernova rate, the expected number of detections of merger-originated neutrinos is, even with a network of gravitational wave detectors and a megaton neutrino detector operating in tandem, a few per century~\cite{Kyutoku:2017wnb, Schilbach:2018bsg, Lin:2019piz}.  

A program of using neutron star mergers to discover or constrain BSM physics must identify the impact a BSM particle will have on either the gravitational wave or the electromagnetic signal coming from the merger. The case of QCD axions in the context of mergers was investigated by some of us in Ref.~\cite{Harris:2020qim} and it was found that the possibility of axion trapping in the merger remnant is ruled out by the SN1987A constraint~\cite{Turner:1987by,Raffelt:1987yt,Mayle:1987as,Burrows:1988ah,Mayle:1989yx,Burrows:1990pk}. This is also expected to be the case for very light axion-like particles (ALPs)~\cite{Giannotti:2005tn}, with no allowed region left between the supernova and laboratory constraints~\cite{Chang:2018rso}.  
They can still be produced copiously in the merger remnant and then escape, cooling the hottest regions of the remnant in millisecond timescales.  However, simulations including cooling due to radiation of axions showed little effect on the gravitational signal or the amount of ejected material~\cite{Dietrich:2019shr}. The effect of ultralight axions on neutron star inspiral phase has been studied in Refs.~\cite{Hook:2017psm,Huang:2018pbu}.

In this work, we investigate the impact of light CP-even scalars on the cooling and thermal transport properties of a neutron star merger. Such CP-even scalar particles contrast in several ways with CP-odd axions (and ALPs in general):  
\begin{enumerate}
    \item [i.] The pseudo-scalar nature of the axion leads to qualitatively different interactions with SM particles.  In particular, in the nucleon bremsstrahlung process, the axions can only be emitted from either initial or final state nucleons, but not from the mediator pions, because the axion-pion coupling is forbidden at leading order. On the other hand, a CP-even scalar can be emitted from any of the nucleons, as well as from the pion mediator, thus making its production rate very different from the axion case.  

\item [ii.] Light axions can oscillate into SM photons when propagating through background electromagnetic fields.  This property leads to modified axion and SM photon luminosities at the observation point.  On the other hand, the CP-even scalars in our scenario dominantly decay into $e^+e^-$ pairs (above the 1 MeV threshold); their coupling to photons and any associated scalar-photon oscillation effect (see e.g.\ Ref.~\cite{Jaiswal:2021exp}) will be loop-suppressed.

\item [iii.] There exist strong astrophysical~\cite{Raffelt:2006cw, Fortin:2018ehg, Fortin:2021sst} and laboratory~\cite{Irastorza:2018dyq, Choi:2020rgn} bounds on axion couplings, which preclude the possibility of QCD axion (or light ALP) trapping in neutron star mergers. On the contrary, as we will show in this work, current constraints on the CP-even scalar, mainly from their decay into SM particles, still allow for the interesting possibility of trapped scalars in the merger. 
\end{enumerate}

Depending on the thermodynamic conditions in the merger, we find that the CP-even scalar can have a wide range of mean free path (MFP), for instance---for a particular choice of mixing angle $\sin{\theta}$---the MFP can range from hundreds of kilometers to below one meter (see Figure~\ref{fig:S_mfp_nB_T}). The dependence of the MFP on the scalar mass $m_S$ and its mixing angle $\sin\theta$ with the SM Higgs is exemplified in Figure~\ref{fig:S_mfp_ms_sintheta} for some benchmark baryon densities and temperatures. Even after taking into account all the laboratory and supernova limits on the scalar, there is still some parameter space left where the scalar can either be trapped or free-stream out of the merger. In the free-streaming region, a given scalar will very likely escape the merger remnant and cool it down. The emissivity is illustrated in Figure~\ref{fig:S_emissivity_cooling_1d} and the corresponding cooling timescales are shown in Figure~\ref{fig:cooling}. It is remarkable that some regions of the parameter space with the cooling timescale of ${\cal O} (1\, {\rm ms})$ to ${\cal O} (10\, {\rm ms})$ can be directly probed at future laboratory experiments, such as DUNE (cf.~the lower left panel of Figure~\ref{fig:cooling}). For sufficiently small MFP, the scalar $S$ will form a Bose gas and contribute to thermal conductivity of the neutron star merger remnants. As shown in Figures~\ref{fig:thermal_eq_1d},~\ref{fig:thermal_eq2}, and~\ref{fig:thermal_eq}, the thermal equilibration time due to the scalar $S$ can reach ${\cal O} ({\rm ms})$ or even smaller in some regions of the parameter space that are allowed by current laboratory constraints. The thermal conductivities due to the trapped scalars and trapped neutrinos are compared in Figure~\ref{fig:kappa_ratio}. It turns out that in a large region of parameters, in particular when the temperature is hotter than a few tens of MeV, the thermal equilibration will be dominated by the scalars, which could potentially lead to some detectable effects in future neutron star merger observations.

The rest of the paper is organized as follows: In Section~\ref{sec:scalar_model}, we briefly review the CP-even scalar scenario and its couplings to the SM from a model-independent perspective. We give an overview of the conditions inside neutron star mergers in Section~\ref{sec:nuclear_matter_in_mergers} and  calculate the MFP of the scalar in these conditions in Section~\ref{sec:mfp}.  If free-streaming, the scalar will radiatively cool the hot regions of the merger.  In Section~\ref{sec:scalar_cooling} we calculate the timescale of this cooling.  If trapped, the scalar could contribute to the thermal equilibration of the interior of the merger remnant, which we discuss and calculate the associated timescale in Section~\ref{sec:thermal_conduction}.  Our conclusions and outlook are given in Section~\ref{sec:conclusion}. Details on various aspects of the MFP, emissivity, and thermal conductivity calculations are presented in several appendices. Throughout this paper, we work in natural units, where $\hbar=c=k_B=1$.

\section{The CP-even scalar}
\label{sec:scalar_model}
After the discovery of a Higgs-like particle at the LHC~\cite{ATLAS:2012yve, CMS:2012qbp}, the SM is now complete.  However, there are compelling theoretical arguments and experimental observations suggesting the existence of BSM physics, although the energy scale at which it could manifest is still unknown. In a bottom-up phenomenological approach, i.e.\ without referring to any specific ultraviolet (UV) completion, the effects of BSM physics could be captured simply by extending the SM scalar, fermion and/or gauge sectors. The minimal extension of the SM scalar sector consists of adding a real SM-gauge-singlet scalar field (denoted hereafter by $S$) that mixes with the SM Higgs boson~\cite{Silveira:1985rk, Davoudiasl:2004be, Schabinger:2005ei, OConnell:2006rsp, Barger:2007im, Robens:2015gla}. This $S$ particle does not have to be a fundamental degree of freedom and might actually be the remnant of a more complicated scalar sector at higher scales. Depending on the model parameters, singlet scalars can be useful for improving the vacuum stability of the SM~\cite{Gonderinger:2009jp, Gonderinger:2012rd, Lebedev:2012zw, Elias-Miro:2012eoi, Khan:2014kba, Falkowski:2015iwa, Ghorbani:2021rgs}, first order electroweak phase transition and electroweak baryogenesis~\cite{Espinosa:1993bs, Choi:1993cv, Ham:2004cf,Profumo:2007wc,Espinosa:2011ax, Barger:2011vm,  Profumo:2014opa,Curtin:2014jma, Kotwal:2016tex,Chen:2017qcz}, addressing the hierarchy problem in relaxion models~\cite{Graham:2015cka, Flacke:2016szy, Frugiuele:2018coc, Banerjee:2020kww, Brax:2021rwk}, addressing the cosmological constant problem through radiative breaking of classical scale invariance~\cite{Foot:2011et, Heikinheimo:2013fta, Wang:2015cda}, or even as a mediator which links the dark and visible sectors~\cite{Pospelov:2007mp, Baek:2011aa, Baek:2012uj, Baek:2012se, Schmidt-Hoberg:2013hba, Krnjaic:2015mbs, Beniwal:2015sdl}.  

When the mass of the new scalar becomes much larger than the electroweak scale, the theory can be mapped onto an effective field theory~\cite{Contino:2013kra, Chiang:2015ura, Dawson:2017vgm, Jiang:2018pbd}. If the scalar mass is around the electroweak scale, collider constraints on extended Higgs sector apply~\cite{LHCHiggsCrossSectionWorkingGroup:2011wcg, Dawson:2013bba, Proceedings:2018het}. However, for a light scalar well below the electroweak scale, the constraints from low-energy high-intensity experiments, such as those from meson decays and beam-dump experiments~\cite{Krnjaic:2015mbs,Winkler:2018qyg, Egana-Ugrinovic:2019wzj,Dev:2019hho, Foroughi-Abari:2020gju, Lanfranchi:2020crw}, as well as the astrophysical~\cite{Krnjaic:2015mbs, Dev:2020eam, Dev:2020jkh} and cosmological~\cite{Krnjaic:2015mbs, Fradette:2017sdd} limits offer a powerful probe. Such a light singlet scalar may couple to the SM particles via the following ways:
\begin{itemize}
    \item Through mixing with the SM Higgs $h$. In this case, all the couplings to the SM particles are proportional to the corresponding SM couplings, rescaled by the mixing angle $\sin\theta$ of $S$ with the SM Higgs $h$~\cite{Silveira:1985rk, Davoudiasl:2004be, Schabinger:2005ei, OConnell:2006rsp, Barger:2007im, Robens:2015gla}.
    \item With couplings exclusively or predominantly to some specific SM flavors, such as the hadronic and leptonic scalars~\cite{Dev:2017ftk,Batell:2017kty, deGouvea:2019qaz,Dev:2021axj}.
    \item With couplings to some SM particles through some heavy BSM particle loops, such as the $SU(2)_R$-breaking scalar in the left-right symmetric model (LRSM) based on the gauge group $SU(2)_L \times SU(2)_R \times U(1)_{B-L}$~\cite{Pati:1974yy, Mohapatra:1974gc, Senjanovic:1975rk}, which couples to photons via the heavy $W_R$ and charged scalars in the LRSM~\cite{Nemevsek:2012cd, Maiezza:2016ybz, BhupalDev:2016nfr, Dev:2017dui}.
\end{itemize}
For simplicity we will consider only the first case above, where all the phenomenologies of $S$ are determined by only two parameters, i.e.\ its mass $m_S$ and the mixing angle $\sin\theta$ of $S$ with $h$. 

\begin{figure}[t!]
  \centering
  \includegraphics[width=0.4\textwidth]{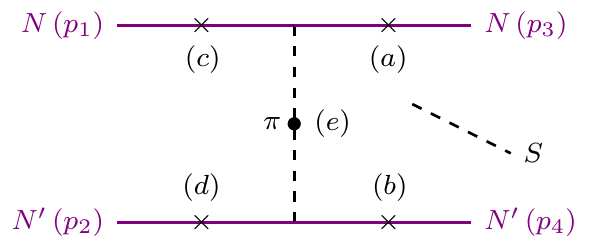}
  \includegraphics[width=0.4\textwidth]{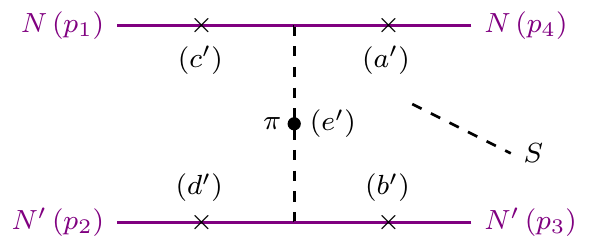}
  \caption{Feynman diagrams for the production of light scalar $S$ in the nucleon bremsstrahlung process $N + N' \to N + N' +S$ (with $N,N' = p,\, n$) in  the neutron star merger. The scalar $S$ can be attached to any of the nucleon lines $(a)$, $(b)$, $(c)$, $(d)$, $(a')$, $(b')$, $(c')$, $(d')$, as denoted by the crosses ($\times$), or to the pion mediator $(e)$, $(e')$, as denoted by the blobs ($\bullet$). The left diagram is for the $t$-channel process, whereas the right diagram is for the $u$-channel, with the final-state nucleon 4-momenta interchanged.}
  \label{fig:diagram}
\end{figure}

In neutron stars, the dominant production channel for the light $S$ is the nucleon bremsstrahlung process (just like in the supernova case~\cite{Dev:2020eam})
\begin{eqnarray}
\label{eqn:brem}
N + N' \to N + N' + S \,,
\end{eqnarray}
where the scalar $S$ can couple to any of the four external nucleon legs ($N, N'$ can be either a neutron or a proton) or to the intermediate pion mediator, as shown in Figure~\ref{fig:diagram}. The $S$ coupling strengths are given by the Lagrangian
\begin{eqnarray}
{\cal L} \supset \sin\theta \left[ y_{hNN} S \bar{N} N + A_{\pi} (S\pi^0 \pi^0 + S\pi^+ \pi^-)  \right] \,,
\end{eqnarray}
with $y_{hNN} \simeq 10^{-3}$ being the coupling of $h$ with nucleons~\cite{Shifman:1978zn, Cheng:1988im}. The coupling of $S$ with pions can be calculated from the chiral perturbation theory~\cite{Voloshin:1985tc, Donoghue:1990xh}:
\begin{eqnarray}
A_\pi = \frac{m_S^2}{v_{\rm EW}} \left( \frac29 + \frac{11}{9} \frac{m_\pi^2}{m_S^2} \right) \,,
\end{eqnarray}
with $v_{\rm EW} = (\sqrt2 G_F)^{-1/2}$ being the electroweak scale ($G_F$ is the Fermi constant). 

After being produced, the light scalar $S$ can decay into the SM particles via its mixing with the SM Higgs $h$. For a sub-GeV scale mass, the scalar $S$ can decay into charged lepton pairs $\ell^+ \ell^- = e^+ e^-$, $\mu^+ \mu^-$ and pion pairs $\pi^0 \pi^0$, $\pi^+ \pi^-$ at the tree level, and into the photon pairs $\gamma\gamma$ at one-loop level. The corresponding MFPs can be found in Appendix~\ref{appendix:decays}.

Let us also summarize the existing (and future) laboratory, astrophysical and cosmological limits on the light scalar $S$, which will be used later for comparison with our results:
\begin{itemize}
    \item The scalar $S$ can obtain flavor-changing neutral current (FCNC) couplings to the SM quarks at one-loop level via mixing with the SM Higgs, and contribute to rare FCNC decays of mesons, such as $K \to \pi +X$, $B \to K + X$ and $B \to \pi + X$ with $X = e^+ e^-,\, \mu^+ \mu^-,\, \gamma\gamma$ or missing energy if $X$ decays outside the detectors.  There are stringent limits from $K^+ \to \pi^+ e^+ e^-,\; \pi^+ \mu^+ \mu^-$ in NA48/2~\cite{Batley:2009aa,Batley:2011zz}, $K^+ \to \pi^+ \gamma\gamma,\; \pi^+ \nu \bar{\nu}$ at NA62~\cite{Ceccucci:2014oza,NA62:2020fhy} and E949~\cite{Artamonov:2009sz}, $K_L \to \pi^0 \chi\chi$ (with $\chi\chi = e^+ e^-,\; \mu^+ \mu^-,\; \gamma\gamma$) in KTeV~\cite{AlaviHarati:2003mr,AlaviHarati:2000hs,Alexopoulos:2004sx, Abouzaid:2008xm}, $K_L \to \pi^0 \nu \bar{\nu},\; \pi^0 X$ (with $X$ a long-lived particle) at KOTO~\cite{Ahn:2018mvc}, $B \to K +X$ with $X = e^+ e^-,\, \mu^+ \mu^-,\, \nu \bar\nu$ at LHCb~\cite{Aaij:2012vr},  BaBar~\cite{Aubert:2003cm,Lees:2013kla}, Belle~\cite{Wei:2009zv}. The MicroBooNE collaboration has recently performed the searches of light monoenergetic scalars from kaon decay at rest $K^+ \to \pi^+ + S$ with $S \to e^+ e^-$~\cite{MicroBooNE:2021ewq}. The calculation details of these FCNC decays can be found e.g.\ in Refs.~\cite{Dev:2017dui,Dev:2019hho, Egana-Ugrinovic:2019wzj}.  The most stringent limits from NA62, E949, KOTO, MicroBooNE and LHCb are shown in Figure~\ref{fig:S_mfp_ms_sintheta} respectively as the shaded blue, light green, orange, dark green and pink regions. The remaining limits are relatively weaker and are not shown in these figures.
    
    \item  The light scalar $S$ can also be produced in the high-intensity beam-dump experiments. The current most stringent limits are from kaon decays $K \to \pi + S$ in the CHARM experiment~\cite{CHARM:1985anb}, which is presented as the magenta shaded regions in Figure~\ref{fig:S_mfp_ms_sintheta}. At the LSND experiment, the scalar $S$ is predominately produced by the proton bremsstrahlung process, and the current LSND electron and muon data~\cite{LSND:2001aii, LSND:2003mon} have excluded the brown shaded regions in Figure~\ref{fig:S_mfp_ms_sintheta}~\cite{Foroughi-Abari:2020gju}. The data from the fixed target experiment PS191 have also be reinterpreted to set limits on the light scalar from kaon decays $K \to \pi + S$, and the resultant limits are shown as the red shaded regions in Figure~\ref{fig:S_mfp_ms_sintheta}.
    
    \item On the astrophysical side, the scalar $S$ can be produced abundantly in the supernova cores with temperature $T \sim 30$ MeV via the nucleon bremsstrahlung process (\ref{eqn:brem}). The observation of neutrino luminosity ${\cal L}_\nu$ of SN1987A can be used to set limits on the mixing angle $\sin\theta$ of $S$ with the SM Higgs~\cite{Ishizuka:1989ts, Diener:2013xpa, Krnjaic:2015mbs, Lee:2018lcj, Arndt:2002yg, Dev:2020eam}. The most recent supernova limits can be found in Ref.~\cite{Dev:2020eam}. Setting conservatively ${\cal L}_\nu = 3\times 10^{53}$ erg/sec and $5\times 10^{53}$ erg/sec, the corresponding supernova limits on $m_S$ and $\sin\theta$ are shown in Figure~\ref{fig:S_mfp_ms_sintheta} respectively as the lighter and darker black shaded regions, which exclude the mixing angle $\sin\theta$ from roughly $5.9\times10^{-7}$ to $7.0\times10^{-6}$ with scalar mass up to 148 MeV. With a less conservative luminosity limit, say $3 \times 10^{52}$ erg/sec, the SN1987A observations could exclude larger regions of $m_S$ and $\sin\theta$. 
    Inside the Sun, white dwarfs, red giants and horizontal-branch stars where the temperature is ${\cal O}({\rm keV})$, the production of $S$ is expected to be dominated by the process $e + N_i \to e + N_i +S$ with $N_i$ here denoting different nuclei in the stars~\cite{Dev:2020jkh}. However, the limits on $S$ from these keV-temperature stars can reach only up to ${\cal O}(100\, {\rm keV})$, and are thus not shown in Figure~\ref{fig:S_mfp_ms_sintheta}. 
    
    \item In the early universe, if the light $S$ keeps in thermal equilibrium with the SM particles and has a lifetime $\tau_S \gtrsim 1$ sec, it will contribute an extra degree of freedom and spoil the success of big bang nucleosynthesis (BBN)~\cite{Kainulainen:2015sva, Fradette:2017sdd}. However, it turns out that there is no parameter space of $m_S$ and $\sin\theta$ that satisfies both the conditions above, so the generic BBN limit is not applicable in our setup~\cite{Dev:2020eam}. Moreover, the actual BBN constraint will depend on the particular thermal history of the scalar $S$ in the early universe and can be evaded, without affecting the late-time phenomenology we are interested in here. Therefore,  we do not show the BBN limits in Figure~\ref{fig:S_mfp_ms_sintheta}. In addition, a sufficiently light scalar $S$ in thermal equilibrium contributes also to the light degrees of freedom $N_{\rm eff}$ at the CMB epoch, and the mixing angle $\sin\theta$ is constrained by the precision Planck data~\cite{Planck:2018vyg}. However, this limit on the mixing from $\Delta N_{\rm eff}$ is very weak, of order ${\cal O}(0.01)$~\cite{Dev:2019hho}, therefore it is not shown in Figure~\ref{fig:S_mfp_ms_sintheta}.
    
    \item  The mixing of light $S$ with the SM Higgs $h$ will contribute to the invisible decay of the SM Higgs via $h \to SS$; as a result, the mixing angle $\sin\theta$ is constrained by the precision Higgs data at the LHC. The current limit of ${\rm BR} (h \to {\rm inv.}) < 0.11$~\cite{ATLAS:2020kdi} leads to the limit of $\sin\theta < 0.079$~\cite{Dev:2020jkh}. As this limit is comparatively weaker than the other limits discussed above, it is not shown in Figure~\ref{fig:S_mfp_ms_sintheta}.
\end{itemize}

It is expected that the laboratory limits on $S$ will be significantly improved in future high-intensity experiments. For instance, the Fermilab SBN detectors, which is a combination of using the SBND and ICARUS experiments with the BNB and NuMI beams respectively, can probe a mixing angle down to $1.7\times10^{-7}$~\cite{Batell:2019nwo}. The near detector at DUNE can further improve the prospects by over one order of magnitude~\cite{Berryman:2019dme}, assuming at least 10 signal events. Similarly, the FASER 2 experiment at the LHC can detect light long-lived particles produced in the high-energy $pp$ collisions. Benefiting from the high colliding energy, the FASER 2 experiment can probe a relatively heavier $S$ with mass up to few GeV~\cite{Feng:2017vli, FASER:2018eoc}. The corresponding prospects of $m_S$ and $\sin\theta$ at SNB, DUNE and FASER 2 are denoted respectively by the purple, green and light blue lines in Figure~\ref{fig:S_mfp_ms_sintheta}. The light scalar $S$ can also be searched for in some other future high-intensity experiments, such as NA62~\cite{Dobrich:2018ezn}, LHCb~\cite{Gligorov:2017nwh}, SHiP~\cite{Alekhin:2015byh}, CODEX-b~\cite{Gligorov:2017nwh} and
MATHUSLA~\cite{Curtin:2018mvb, Chou:2016lxi}. For the sake of clarity, these prospects are not shown in Figure~\ref{fig:S_mfp_ms_sintheta}. For a summary of these constraints, see e.g.\ \cite{Anchordoqui:2021ghd}. 

\section{Nuclear matter in neutron star mergers}\label{sec:nuclear_matter_in_mergers}
The matter inside cold neutron stars is a degenerate Fermi liquid of neutrons and protons along with a nearly ideal, degenerate Fermi gas of electrons and muons.  We ignore the possibility of exotic phases of matter, such as those containing quarks, hyperons, quarkyonic matter, or Bose condensates~\cite{McLerran:2018hbz,Annala:2019puf,Alford:2019oge, Schaffner-Bielich:2020psc}.  In the interior of the star, the density can reach several times the nuclear saturation density $n_0\equiv 0.16 \text{ fm}^{-3}$.  Throughout the neutron star inspiral, there is little change in the thermodynamic conditions of the stars.  When the two stars collide, numerical simulations~\cite{Rezzolla_Zanotti_book,Baiotti:2016qnr, Duez:2018jaf} indicate that significant shock heating occurs and the temperature of the nuclear matter at the collision interface rises to several tens of MeV, while the nuclear matter in the densest part of the neutron star cores only heats modestly, to perhaps 5--10 MeV~\cite{Hanauske:2019qgs, Perego:2019adq, Most:2021zvc, Most:2021ktk}.  Therefore, in contrast to supernovae (nondegenerate) or cold neutron stars (degenerate), the nuclear matter inside the neutron star merger remnant spans the entire range of degeneracies~\cite{Harris:2020qim,Du:2018vyp}.

In this work, we will focus on the regions of the neutron star merger remnant where the temperature exceeds 5--10 MeV.  In these conditions, the neutrino MFP is well under one kilometer~\cite{Roberts:2016mwj, Alford:2018lhf} and therefore neutrinos are trapped inside the remnant through their charged and neutral current interactions, and attain a Fermi-Dirac distribution.  We will consider the neutron star merger matter to consist of neutrons $n$, protons $p$, electrons $e^-$, muons $\mu^-$, electron neutrinos $\nu_e$, and muon neutrinos $\nu_{\mu}$ (and the corresponding antileptons).  The content of this matter at a given temperature $T$ is determined by choosing a baryon density $n_B$, an electron lepton fraction $Y_{L_e}\equiv (n_e+n_{\nu_e})/n_B$, and a muon lepton fraction $Y_{L_\mu}\equiv (n_{\mu}+n_{\nu_{\mu}})/n_B$.  In this work we choose $Y_{L_e} = Y_{L_\mu}=0.1$~\cite{Perego:2019adq}, though the value of $Y_{L_\mu}$ throughout the neutrino-trapped regions of a neutron star merger remnant is, as far as we know, unstudied.  The choice of conserved lepton fractions has a small effect on the production rate and the MFP of the $S$ particle, typically up to a factor of 2 to 3.  This is about the same level of variation caused by uncertainty in the nuclear EoS (see Appendix F of Ref.~\cite{Fortin:2021sst} where this was studied for axion production in degenerate nuclear matter).

We model the nuclear matter with the IUF EoS~\cite{Fattoyev:2010mx}, which is a relativistic mean field (RMF) theory~\cite{Glendenning:1997wn,Schmitt:2010pn, Dutra:2014qga}.  In this type of model, the strong interaction between nucleons is treated as meson exchange, where the meson fields are frozen to their vacuum expectation values, which vary with density and temperature.  In this mean field approximation, the nucleons have energy dispersion relationships
\begin{align}
    E_n &= U_n + E^*_n = U_n + \sqrt{{\bf p}_n^2+m_*^2} \, ,\\
    E_p &= U_p + E^*_p = U_p + \sqrt{{\bf p}_p^2+m_*^2} \, ,
\end{align}
where $U_n$ and $U_p$ are the nuclear mean fields experienced by the neutron and proton, respectively.  The mean fields $U_i$ depend on the baryon density and temperature, but not the nucleon momentum.  In the IUF RMF theory that we use here, the Dirac effective mass $m_*$ of the neutron and proton are equal.  At $n_B=n_0$, $m_* \approx 560$ MeV and as density rises, $m_*$ drops dramatically, falling to $m_*\approx 185$ MeV  at $n_B=5n_0$.  As the neutron Fermi momentum rises from 330 to 540 MeV in this density range, neutrons cannot be considered nonrelativistic for densities $n_B \gtrsim 3n_0$~\cite{Alford:2021ogv}.  The leptons are treated as free Fermi gases.  We will discuss the calculation of rate integrals in the RMF theory in the following section.

\section{Scalar mean free path}\label{sec:mfp}
The MFP of the light CP-even scalar $S$ in hot, dense nuclear matter is primarily determined by the rate of absorption in inverse nucleon bremsstrahlung processes $N+N'+S\rightarrow N+N'$ (with $N,N'=n,p$), and to some extent, by the total decay rate of $S$ into SM particles, such as $S\rightarrow \gamma\gamma, \ e^+e^-, \ \mu^+\mu^-, \ \pi^0\pi^0, \ \pi^+\pi^-$.  In this work, we consider scalars with masses $m_S < 500 \text{ MeV}$, as scalars in this mass range are light enough to be produced in significant quantities in neutron star mergers.  However, scalars with masses $m_S > 2m_{\pi^0}$ ($m_S > 2m_{\pi^{\pm}}$) can decay into neutral pions (charged pions), yet our nuclear EoS does not account for thermal populations of pions~\cite{Fore:2019wib} nor a pion condensate~\cite{Shapiro:1983du, Pethick:2015jma,Glendenning:1997wn}.  Inclusion of a thermal population of pions has been shown to slightly increase the proton fraction and soften the EoS, especially at high temperatures where the pion population becomes significant \cite{Fore:2019wib}.  These effects will not significantly change the MFP of the scalar due to the nucleon bremsstrahlung processes.  However, as we discuss later in this section, the added possibility of the $S$ decaying to pions can, in certain conditions, noticeably shorten the MFP of the scalar particle.  Additionally, the matrix element for $S$ production from nucleon bremsstrahlung is only valid for scalar masses that are small compared to the nucleon effective mass (see the description of the calculation in Appendix~\ref{appendix:relativistic_corrections}).  Therefore, our results for the production rate and MFP of the scalar particle are most robust for $m_S \lesssim 2m_{\pi}$.  Since the MFP of the scalar $S$ is predominantly dictated by absorption via the inverse nucleon bremsstrahlung processes, we discuss that process here and leave the contributions of the various decay processes to Appendix~\ref{appendix:decays}.

The inverse MFP of the CP-even scalar $S$ due to the absorption process $S+N+N'\rightarrow N+N'$ is given by
\begin{align}
    \lambda_{NN'}^{-1} = \int & \frac{\mathop{d^3 {\bf p}_1}}{(2\pi)^3}\frac{\mathop{d^3{\bf p}_2}}{(2\pi)^3}\frac{\mathop{d^3 {\bf p}_3}}{(2\pi)^3}\frac{\mathop{d^3 {\bf p}_4}}{(2\pi)^3}\frac{S_{NN'}\sum_{\text{spins}}\vert\mathcal{M}_{NN'}\vert^2}{2^5E_1^*E_2^*E_3^*E_4^*E_S}\label{eq:MFP_integral}\\
    &\times (2\pi)^4\delta^4(p_S+p_1+p_2-p_3-p_4)f_1f_2(1-f_3)(1-f_4),\nonumber
\end{align}
where $p_i=(E_i,{\bf p}_i)$ are the four-momenta,  $f_i$ represents a Fermi-Dirac factor and  $E_i^*=\sqrt{{\bf p}_i^2+m_*^2}$ the energy of the $i$th nucleon in the reaction (without the nuclear mean field contribution), $E_S = \sqrt{{\bf p}_S^2+m_S^2}$ is 
the energy of $S$, and $S_{NN'}$ is a symmetry factor for identical particles. To calculate the matrix elements ${\cal M}_{NN'}$ for the nucleon bremsstrahlung process, the nucleon-nucleon scattering via the strong interaction was modeled with the one-pion-exchange (OPE) approximation.  For the reactions $S+n+n\rightarrow n+n$, $S+p+p\rightarrow p+p$ and $S+n+p\rightarrow n+p$, the matrix elements are respectively given by~\cite{Dev:2020eam}
\begin{align}
    S_{nn/pp}\sum_{\text{spins}}\vert \mathcal{M}_{nn/pp}\vert^2 &= \frac{64\pi^2 \alpha_{\pi}^2f^4\sin^2{\theta}}{m_*^2} \nonumber \\
    &\times \left(\frac{1}{4}\frac{y_{hnn}^2m_S^4}{E_S^4}I_A^{nn/pp} + \frac{1}{2}\frac{m_*^2}{81v_{\rm EW}^2}I_B^{nn/pp} +\frac{2y_{hnn}m_S^2m_*}{9E_S^2v_{\rm EW}}I_C^{nn/pp}      \right),  \label{eq:nn_matrix}  \\ 
S_{np}\sum_{\text{spins}}\vert \mathcal{M}_{np}\vert^2 &= \frac{256\pi^2 \alpha_{\pi}^2f^4\sin^2{\theta}}{m_*^2} \nonumber \\
&\times 
\left(\frac{1}{4}\frac{y_{hnn}^2m_S^4}{E_S^4}I_A^{np} + \frac{1}{2}\frac{m_*^2}{81v_{\rm EW}^2}I_B^{np} +\frac{2y_{hnn}m_S^2m_*}{9E_S^2v_{\rm EW}}I_C^{np}      \right) \,. \label{eq:np_matrix}
\end{align}
The symmetry factors for identical particles are $S_{nn/pp} = 1/4$ and $S_{np}=1$.  The nucleon-pion coupling $f\approx 1$ and $\alpha_{\pi} \equiv (2m_N/m_{\pi})^2/(4\pi)\approx 15$ (the nucleon and pion masses in $\alpha_{\pi}$ are the values in vacuum).  The prefactors of $1/4$ and $1/2$ in the $I_A$ and $I_B$ terms respectively were absent in the earlier calculation~\cite{Dev:2020eam}, and come from relativistic corrections to the nucleons, as discussed in Appendix~\ref{appendix:relativistic_corrections}.  The dimensionless functions $I_A$, $I_B$, and $I_C$ can be written in terms of the scalar mass, the pion mass, and the nucleon momentum transfers $\mathbf{k}=\mathbf{p}_2-\mathbf{p}_4$ and $\mathbf{l}=\mathbf{p}_2-\mathbf{p}_3$, and are listed in Appendix~\ref{appendix:nondimensional}.  In the phase space integral Eq.~(\ref{eq:MFP_integral}), the factors of energy in the denominator (or $E^*$ for the nucleons) relate to the normalization of the fermion spinors~\cite{Roberts:2016mwj}. The energies in the four-momentum conserving $\delta$-function are the true particle energies $E$, not $E^*$.  The twelve-dimensional phase space integral in Eq.~(\ref{eq:MFP_integral}) can be reduced to a five-dimensional integral, the details of which are given in Appendix~\ref{appendix:MFP}.

\begin{figure}
    \centering
    \includegraphics[width=0.5\textwidth]{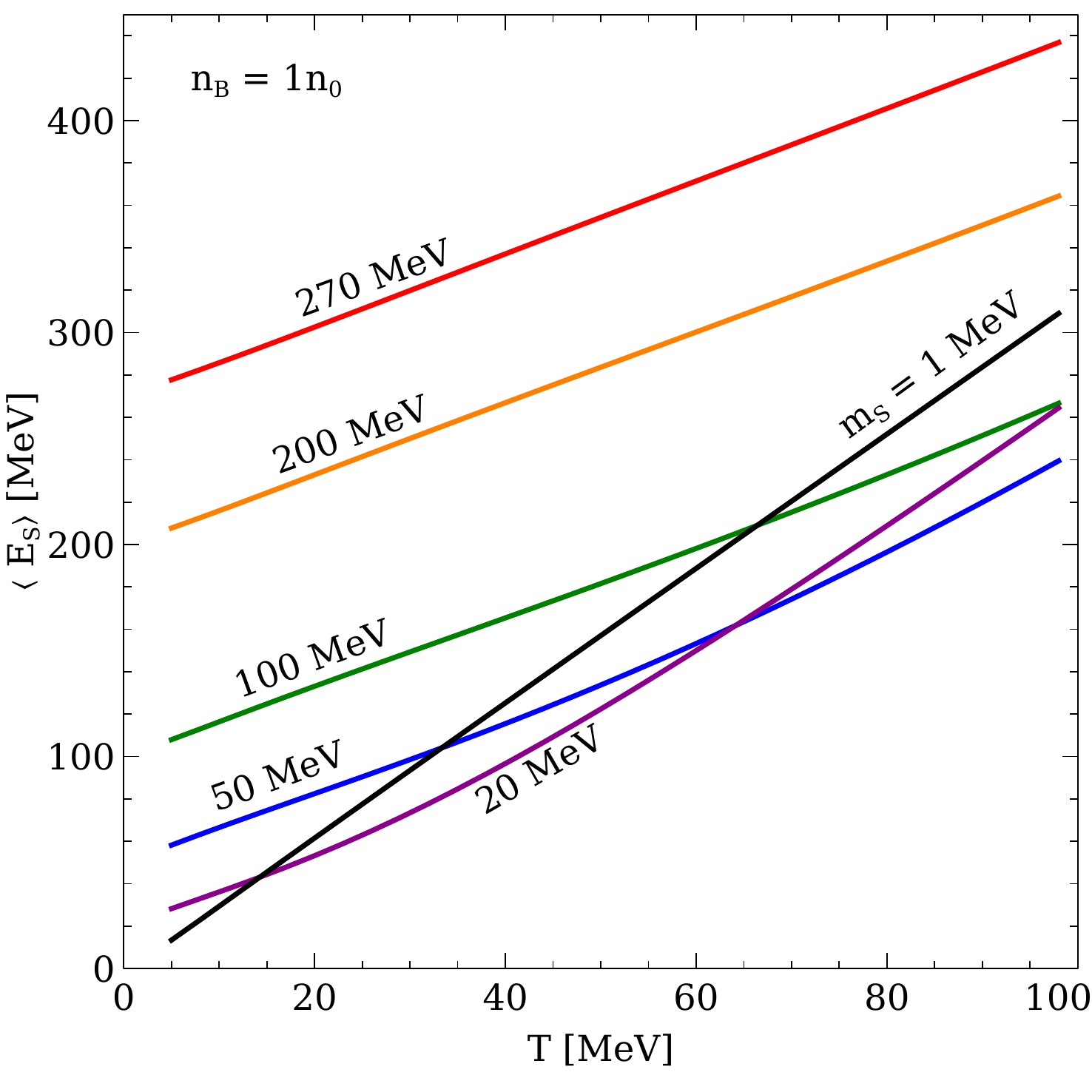}
    \caption{Average energy $\langle E_S \rangle$ of $S$ particles produced at $n_B=n_0$ as a function of temperature $T$ of the fluid element.  This result is independent of $\sin{\theta}$.  The nuclear matter is assumed to be strongly degenerate, making this result independent of density. In the limit of $m_S\to 0$, the average energy $\langle E_S \rangle \approx 3T$.
    }
    \label{fig:ES}
\end{figure}

The total MFP of $S$ is calculated by adding together (in inverse~\cite{ziman2001electrons}) the MFP of each absorption and decay channel, i.e.\
\begin{equation}
    \lambda_S^{-1} = \lambda_{nn}^{-1}+\lambda_{pp}^{-1}+\lambda_{np}^{-1}+\lambda_{\gamma\gamma}^{-1}+\lambda_{e^+e^-}^{-1}+\lambda_{\mu^+\mu^-}^{-1}+\lambda_{\pi^0\pi^0}^{-1}+\lambda_{\pi^+\pi^-}^{-1}.
\end{equation}
The MFP is a function of the baryon density $n_B$, the temperature $T$, the mixing angle $\sin{\theta}$, and the mass $m_S$ and energy $E_S$ of the scalar particle.  To illustrate the ``typical'' MFP of the $S$ particle at a given temperature and density, we choose to evaluate the MFP at the average energy $\langle E_S\rangle$ of the scalar particle emitted in those thermodynamic conditions, 
\begin{equation}
    \langle E_S\rangle = Q_S/\Gamma_S, \label{eq:ES_avg}
\end{equation}
where $\Gamma_S$ is the production rate of scalars per unit volume, and $Q_S$ is the rate of energy production in scalars per unit volume (the emissivity, see Eq.~(\ref{eqn:QS}) below).  The production rate per unit volume $\Gamma_S$ is given by the same phase space integral as $Q_S$ (Eq.~\ref{eq:emissivity_integral}) but without the factor of $E_S$. The average energy of the produced $S$ particle is plotted at nuclear saturation density $n_0$ in Figure~\ref{fig:ES}.  Low-mass scalars are emitted with typical energies $\langle E_S\rangle \approx 3T$, independent of density (assuming the nuclear matter is still strongly degenerate).  This low-mass limit seems to match the result obtained for axions produced by nucleon bremsstrahlung~\cite{Ishizuka:1989ts}.  As the scalar mass increases, the rest mass becomes the dominant contribution to the energy of the scalar, but the energy still increases with temperature on top of the rest mass contribution, although with a slope less than in the low-mass limit.  

\begin{figure}\centering
\includegraphics[width=0.49\textwidth]{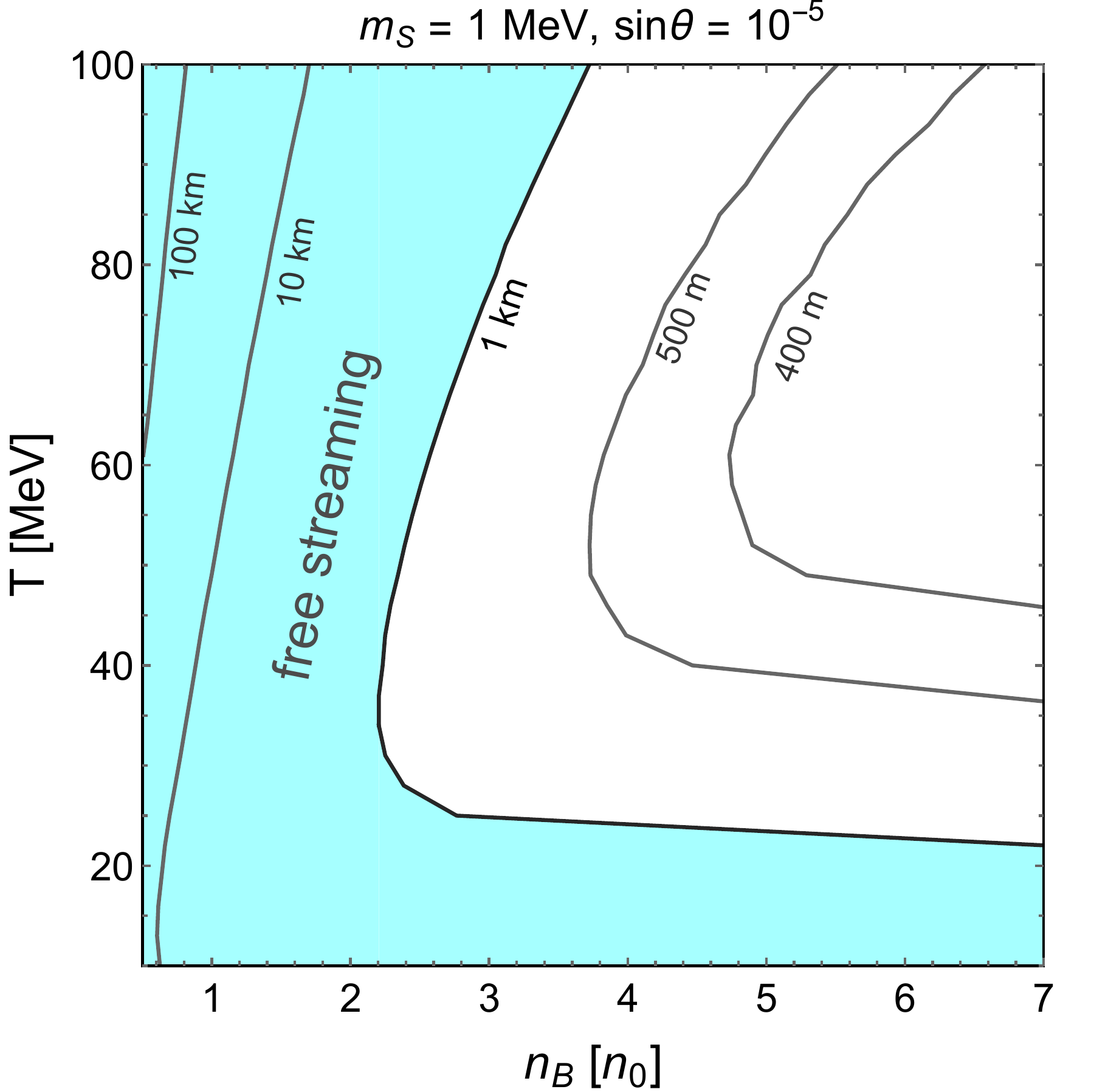}
\includegraphics[width=0.49\textwidth]{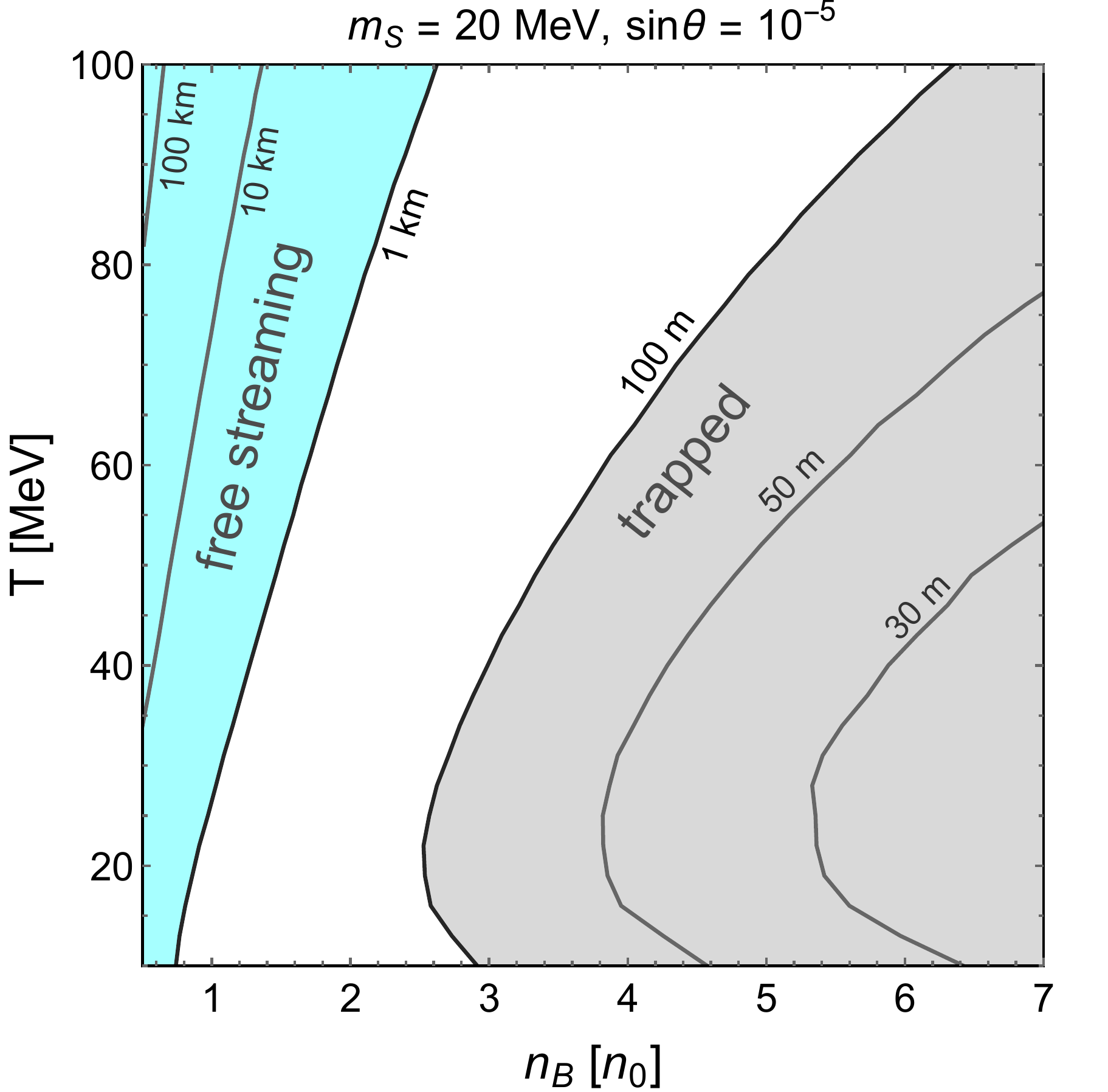}
\includegraphics[width=0.49\textwidth]{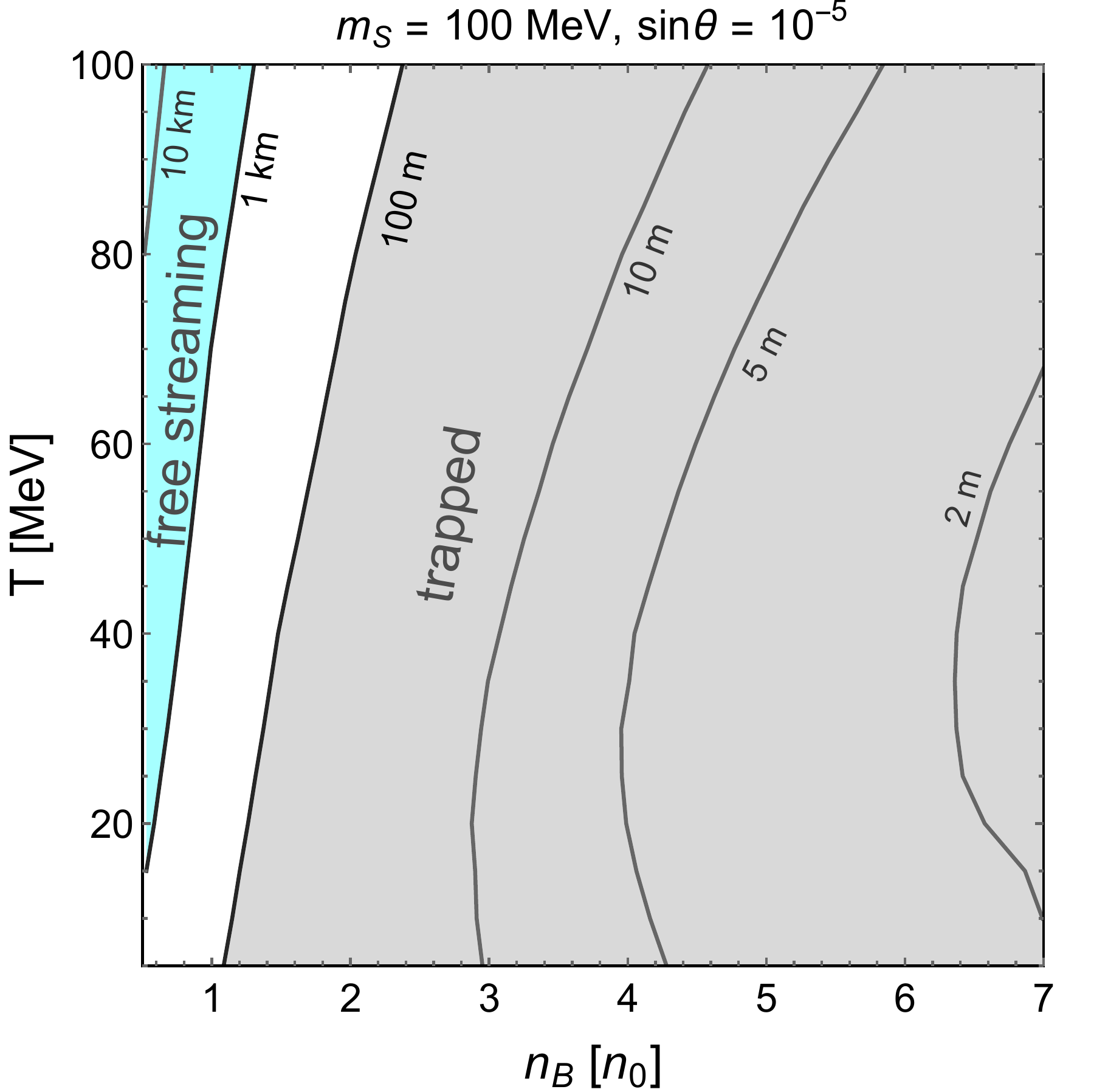}
\includegraphics[width=0.49\textwidth]{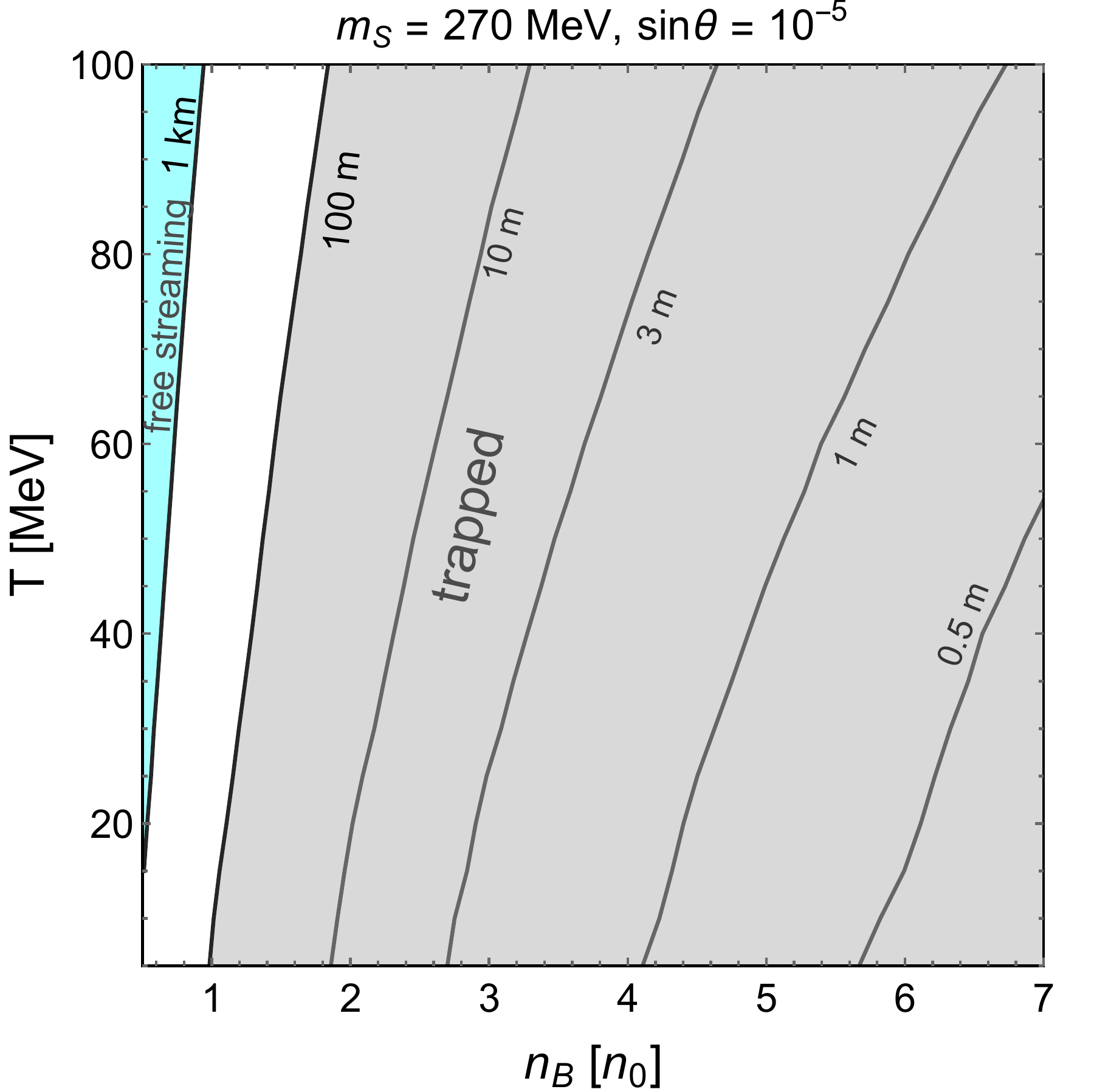}
\caption{MFP of the scalar $S$ in the $n_B -T$ plane with the mixing angle fixed at $\sin\theta = 10^{-5}$, and the scalar mass $m_S = 1$ MeV (upper left), 20 MeV (upper right), 100 MeV (lower left) and 270 MeV (lower right).  For the trapped (shaded grey) and free streaming (shaded cyan) regions, the MFP is assumed to be smaller than 100 m and larger than 1 km  respectively, although this demarcation is somewhat arbitrary.}
\label{fig:S_mfp_nB_T}
\end{figure}

In Figure~\ref{fig:S_mfp_nB_T} we present contour plots of the MFP $\lambda_S$ of the scalar in the $n_B$ (in units of $n_0$)--$T$ plane, for four different values of the scalar masses $m_S = 1$ MeV (upper left panel), 20 MeV (upper right panel), 100 MeV (lower left panel) and 270 MeV (lower right panel).  The energy of the scalar is chosen according to Eq.~(\ref{eq:ES_avg}).  We have chosen $\sin{\theta} = 10^{-5}$ in each panel, but since $\lambda^{-1} \propto \sin^2{\theta}$, it is easy to get the MFP for other values of $\sin{\theta}$.  In each panel, we label regions where the MFP is longer than 1 km as ``free-streaming", i.e.\ the $S$ particles created in a particular region of the merger escapes that region---and perhaps the merger remnant as a whole---without interacting with the nuclear matter. Regions where the MFP is shorter than 100 m are labeled as ``trapped'', where the $S$ particles form a thermal Bose-Einstein distribution inside a region of the merger remnant.  The situation in between these two regimes is difficult to treat---scalar particles can be absorbed or decay, but not frequently enough where they would establish a Bose-Einstein distribution. Please note that the demarcation chosen here between the trapped and free-streaming regions is just for concreteness, and our main results will not be significantly affected by a slightly different choice.

For light scalars, like the $m_S = 1 \text{ MeV}$ case in the upper left panel of Figure~\ref{fig:S_mfp_nB_T}, the MFP contours indicate that as the conditions in the merger remnant get hotter and denser, the MFP grows shorter.  At high densities and low temperatures, where the nuclear matter is strongly degenerate, the MFP becomes strongly dependent on temperature and weakly dependent on density, which is expected in degenerate systems because the number of available states for nucleons to scatter into near the Fermi surface grows as $T$ for each degenerate nucleon in the reaction~\cite{Page:2013hxa,Shapiro:1983du}.  At low densities and high temperatures, where the nuclear matter is nondegenerate, the MFP depends somewhat strongly on density and has a weaker temperature dependence.  This behavior of the MFP in nondegenerate matter can be seen in Eq.~(4.3) in Ref.~\cite{Dev:2020eam}.  In fact, the features of this plot match closely the MFP of light QCD axions (see Figure~2 in Ref.~\cite{Harris:2020qim}).  As the scalar mass increases, the MFP shrinks and becomes largely dependent only on the baryon density, even in the degenerate regime. 

\begin{figure}
\centering
\includegraphics[width=0.47\textwidth]{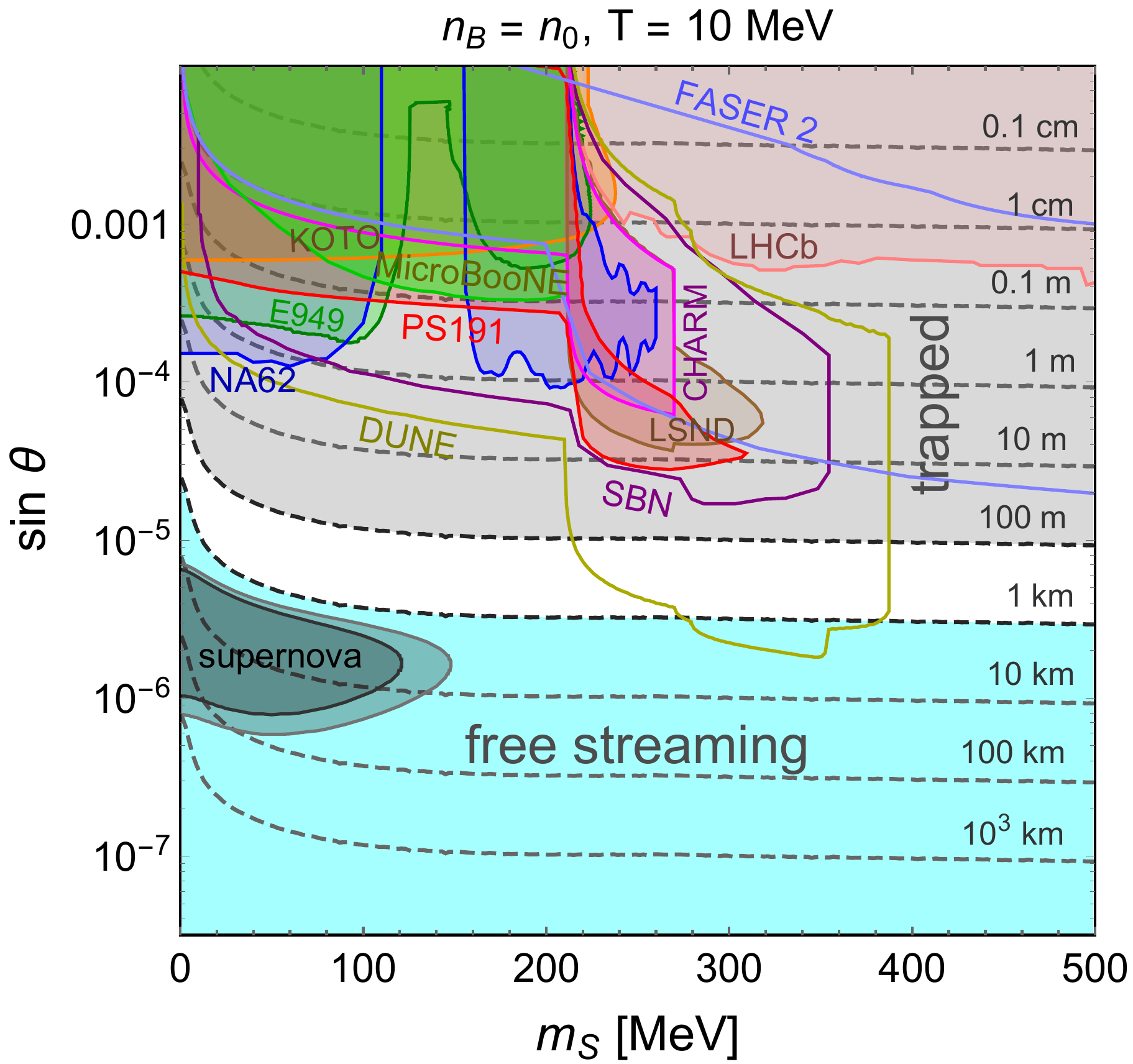}
\includegraphics[width=0.47\textwidth]{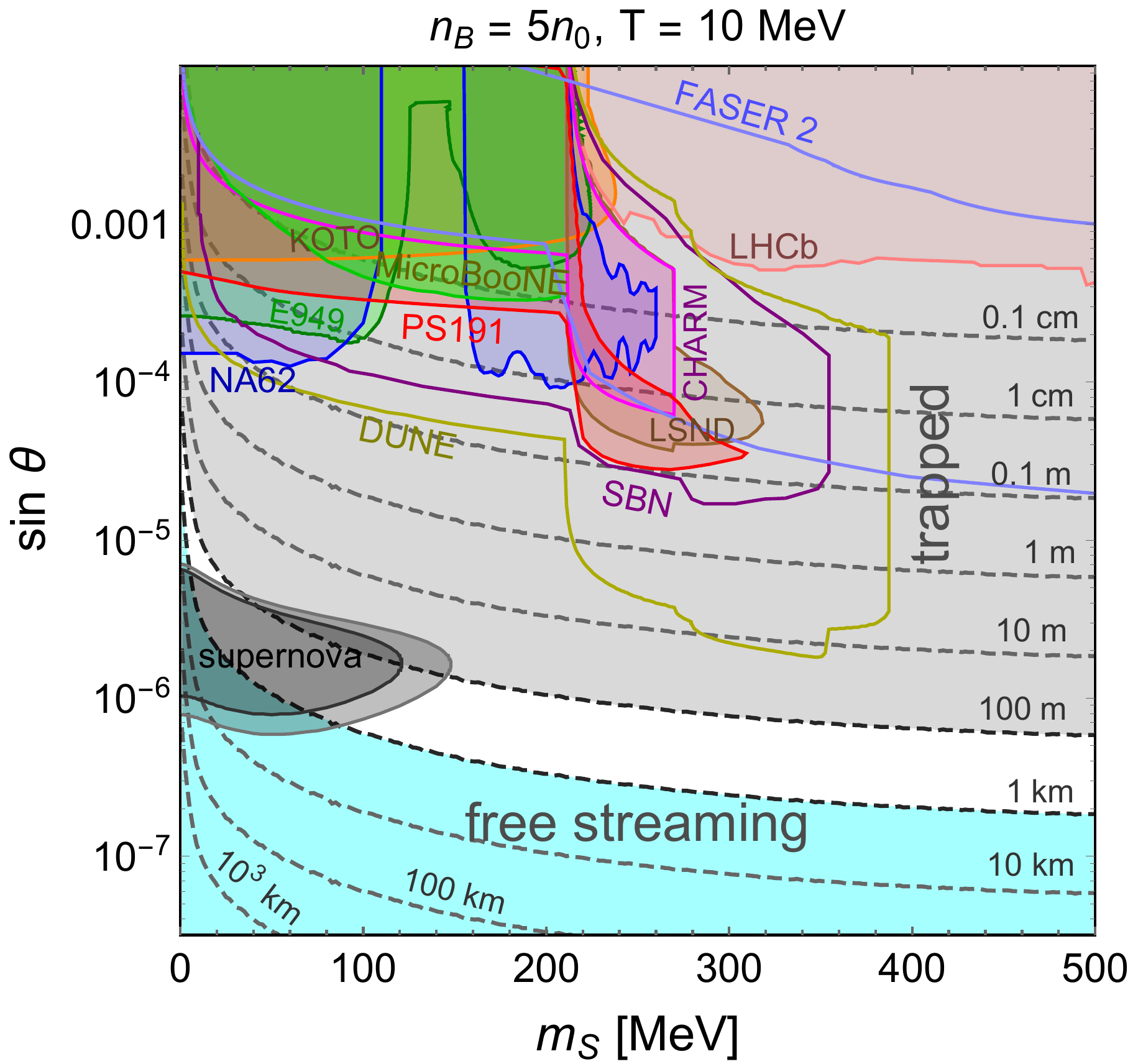}
\includegraphics[width=0.47\textwidth]{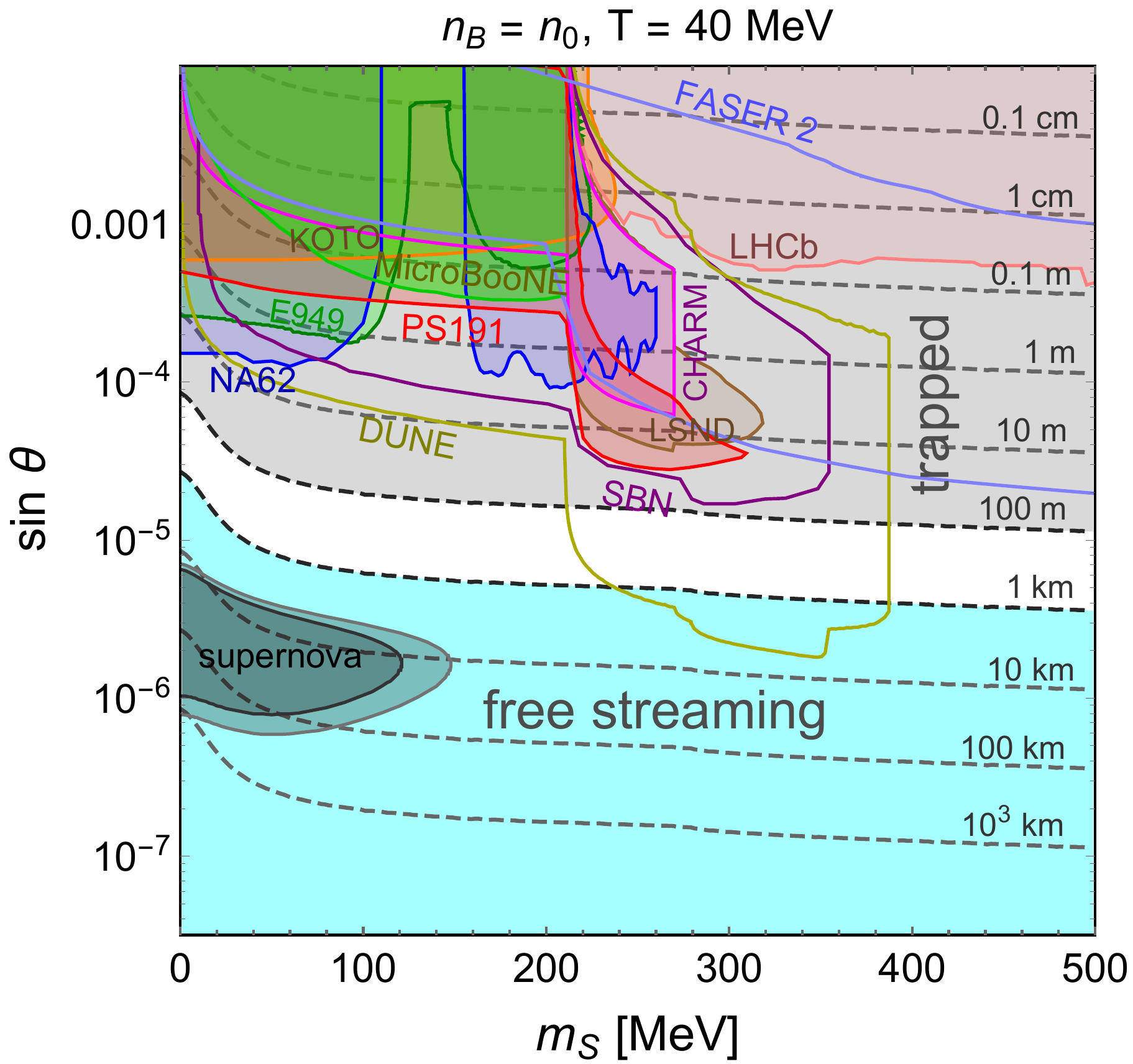} 
\includegraphics[width=0.47\textwidth]{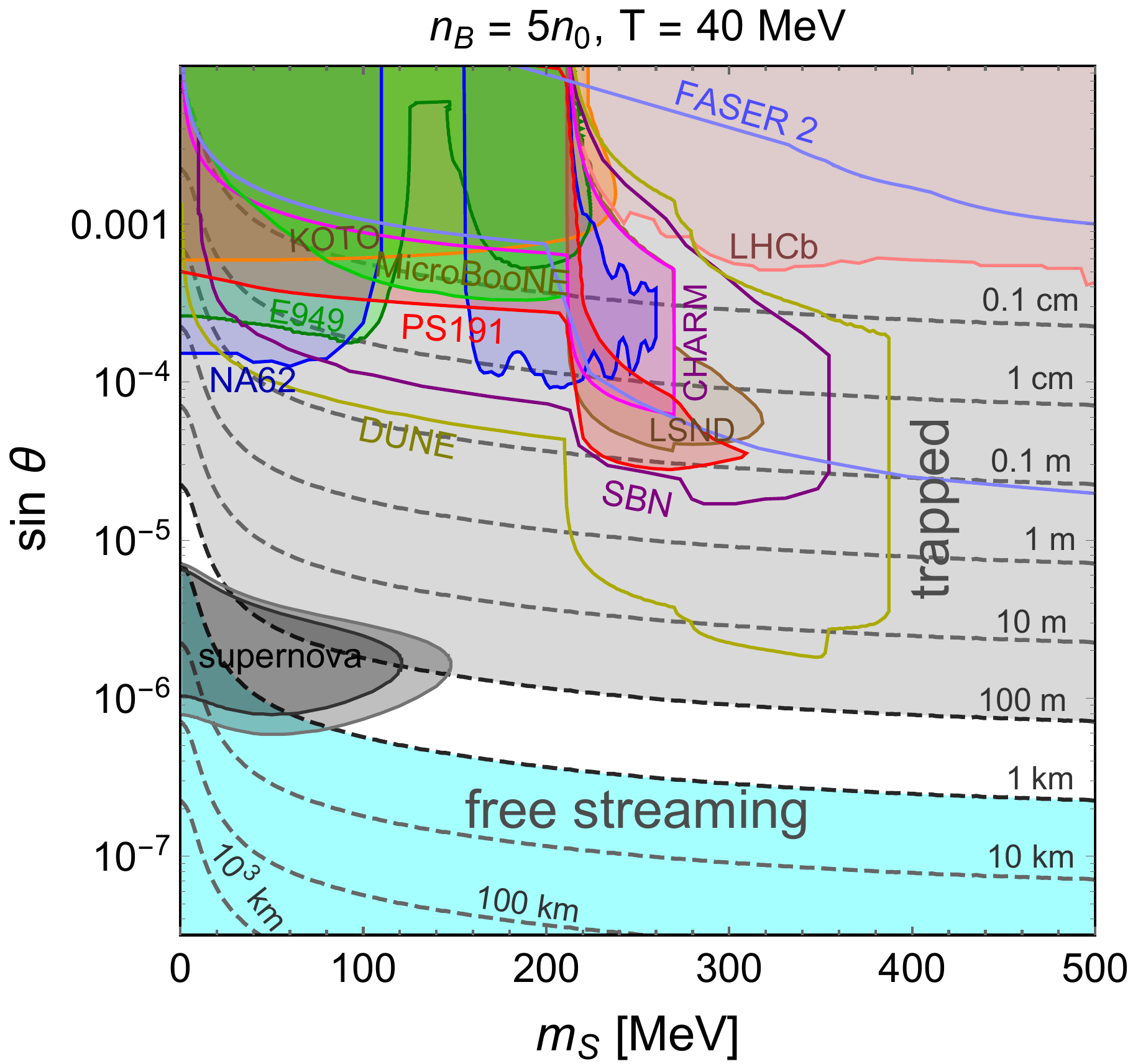}
\includegraphics[width=0.47\textwidth]{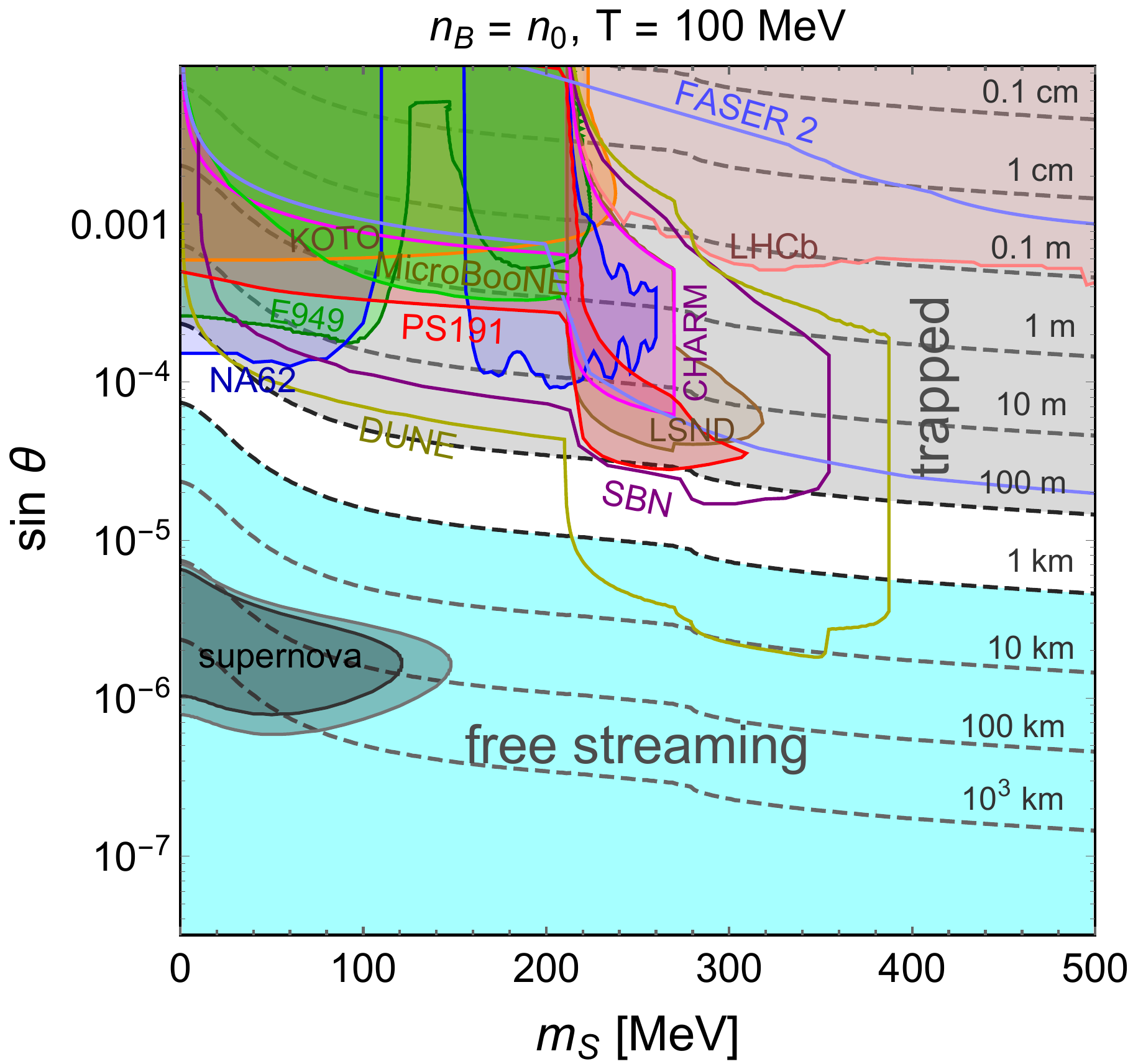}
\includegraphics[width=0.47\textwidth]{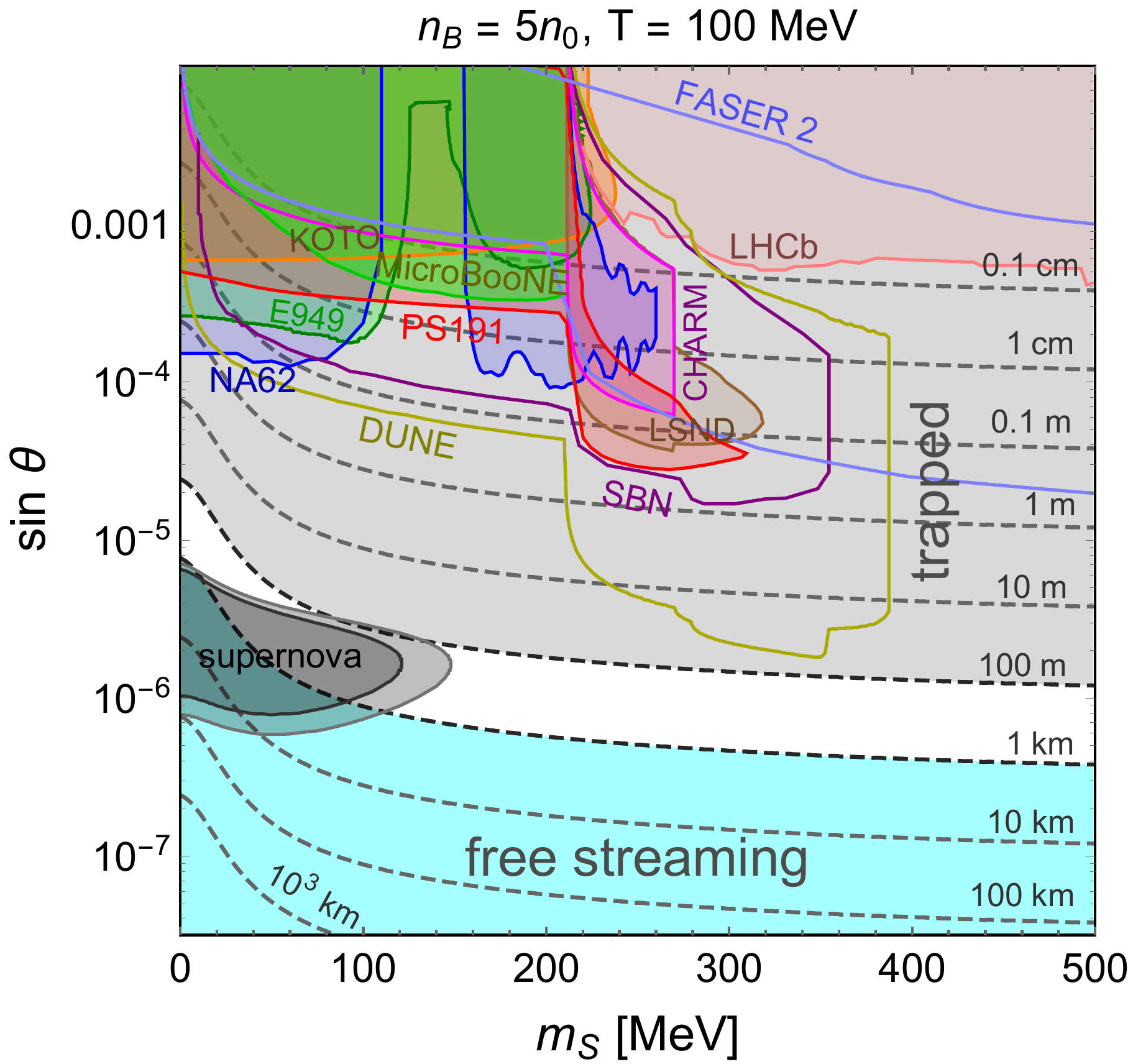}
\caption{MFP of the scalar $S$ in the $m_S-\sin{\theta}$ plane, with baryon density $n_B = n_0$ (left) or $5n_0$ (right), and temperature $T = 10$ MeV (upper), 40 MeV (middle) or 100 MeV (lower).  For the trapped (shaded grey) and free streaming (shaded cyan) regions, the MFP is smaller than 100 m and larger than 1 km, respectively.  The current laboratory and astrophysical constraints are overlaid as the other shaded regions, and the future prospects at SBN, DUNE and FASER 2 are shown as the solid purple, green and light blue lines (see more details in Section~\ref{sec:scalar_model}).}
\label{fig:S_mfp_ms_sintheta}
\end{figure}

In Figure~\ref{fig:S_mfp_ms_sintheta} we plot the MFP of $S$ in the $m_S-\sin{\theta}$ plane, evaluated at the average $S$ energy given in Eq.~(\ref{eq:ES_avg}).  We show the results at several different densities and temperatures possibly encountered in neutron star mergers: the three left panels are for the baryon density $n_B = n_0$, while the three right panels are at $n_B = 5n_0$. The upper, middle and lower panels are with the temperatures $T = 10$ MeV, 40 MeV and 100 MeV, respectively.  The MFP of the scalar is dominated by absorption from inverse nucleon bremsstrahlung processes $S+N+N'\rightarrow N+N'$.  All contributions to the MFP go as $\lambda^{-1}\propto \sin^2{\theta}$, making the mixing angle the dominant factor in the $m_S-\sin{\theta}$ plane.  In all panels, as the mixing angle $\sin{\theta}$ grows larger, the MFP of the scalar shrinks.  Likewise, though much less dramatically, as the scalar mass grows, the MFP shrinks.  This behavior was seen in nondegenerate nuclear matter (see Figure~4 in Ref.~\cite{Dev:2020eam}).  The contribution to the MFP from decay processes is negligible, with the exception of the decay to pions, which is only possible when the scalar mass exceeds twice the pion mass, roughly 270 MeV.  This decay becomes significant at high temperatures, because at these temperatures the thermal pion population is large and the decay of $S$ is subject to Bose enhancement~\cite{Burrows:1990pk,Kolb:1990vq,Kachelriess:2017cfe}.  A small but sudden decrease in the MFP can be seen in the bottom left plot ($n_B=1n_0$ and $T=100 \text{ MeV}$) in Figure~\ref{fig:S_mfp_ms_sintheta} at twice the pion mass.  At lower temperatures, the thermal population of pions is not large enough to stimulate sufficient decay to pions to affect the MFP.  At higher densities, the $\pi^-$ would likely form a condensate as $\mu_{\pi^-} = \mu_n-\mu_p$ rises above $m_{\pi^-}$, complicating the calculation of the scalar decay rate to pions, so we neglect this decay channel at high densities.

Current constraints on the scalar mass $m_S$ and the mixing angle $\sin\theta$ from the laboratory and supernova observations discussed in Section~\ref{sec:scalar_model} are displayed on top of the MFP contours in Figure~\ref{fig:S_mfp_ms_sintheta} as the shaded regions. The representative prospects at future experiments SBN, DUNE and FASER 2 are also indicated by the solid purple, green and light blue lines, respectively. It is clear that there is unconstrained parameter space in both the trapped and free-streaming regions of a neutron star merger, where mergers can in principle provide complementary information.  In particular, we find that mergers have the potential to probe regions of the parameter space of this model that no other experiment can, including the future experiments DUNE, SBN, and FASER 2.  This motivates a more detailed study of the role of scalars in the regions of the merger where they free-stream and the regions where they are trapped.  In the rest of the paper, we pursue this analysis, following the study in the case of axions~\cite{Harris:2020qim}.

\section{Scalar contribution to cooling}\label{sec:scalar_cooling}
If the scalar is produced in a region of the merger and the MFP of the scalar is longer than the size of that region, the scalar has a high likelihood of escaping, cooling down that part of the merger.  If the scalar continues to encounter regions where it has a long MFP, it is likely that the scalar will free-stream through the merger entirely.  The changing conditions the scalar encounters throughout its trajectory are encapsulated in the optical depth~\cite{Janka:2017vlw} 
\begin{align}
    \tau(r,E_S) = \int_r^{\infty}\mathop{dr'}\lambda_S^{-1}(r',E_S) \, ,
\end{align}
but in this work we will consider trapping or free streaming on the level of individual fluid elements in the merger, and in this section we will consider the question of how quickly a fluid element cools when it emits scalar particles.  The decrease in temperature of a fluid element emitting scalars is related to the emissivity $Q_S$ through 
\begin{eqnarray}
\label{eqn:dTdt}
\frac{d T}{dt} \ = \
- \frac{Q_S(n_B,T)}{c_V (n_B,T)} \,,
\end{eqnarray}
where 
\begin{eqnarray}
\label{eqn:QS}
Q_S = Q_{nn}+Q_{np}+Q_{pp}
\end{eqnarray}
incorporates all the contributions from the $pp$, $nn$ and $np$ bremsstrahlung processes, with ($NN' = pp,\, nn,\, np$) 
\begin{eqnarray}
\label{eq:emissivity_integral}
    Q_{NN'} &=& \int \frac{\mathop{d^3{\bf p}_1}}{(2\pi)^3}\frac{\mathop{d^3{\bf p}_2}}{(2\pi)^3}\frac{\mathop{d^3{\bf p}_3}}{(2\pi)^3}\frac{\mathop{d^3{\bf p}}_4}{(2\pi)^3}\frac{\mathop{d^3{\bf p}_S}}{(2\pi)^3}\frac{S_{NN'}\sum_{\text{spins}}\vert\mathcal{M}_{NN'}\vert^2}{2^5E_1^*E_2^*E_3^*E_4^*E_S} \nonumber  \\
    &&\times (2\pi)^4\delta^4(p_1+p_2-p_3-p_4-p_S)f_1f_2(1-f_3)(1-f_4)E_S \,,
\end{eqnarray}
with the matrix elements ${\cal M}_{NN'}$ same as in Eqs.~\eqref{eq:nn_matrix} and \eqref{eq:np_matrix}. 
The specific heat $c_V$ per unit volume of degenerate nuclear matter in Eq.~(\ref{eqn:dTdt}) is dominated by the neutrons, as they have the largest Fermi sphere, and is given by~\cite{1994ARep...38..247L}
\begin{eqnarray}
c_V (n_B,T) \approx \frac13 m_n^L p_{Fn} T,\label{eq:specific_heat}
\end{eqnarray}
where $m_n^L=\sqrt{p_{Fn}^2+m_*^2}$ is the Landau effective mass of the neutron~\cite{Li:2018lpy}, and $p_{Fn}$ is the Fermi momentum of the neutron.  The specific heat does not deviate much from this expression even at the highest temperatures encountered in mergers (see Ref.~\cite{Harris:2020qim} for an expression valid at arbitrary degeneracy).  From Eq.~(\ref{eqn:dTdt}), one can integrate over the temperature evolution and estimate the timescale $\tau_{1/2}$ for the fluid element temperature to drop by a factor of two via 
\begin{eqnarray}
\tau_{1/2} = \int_0^{\tau_{1/2}} dt =  \int_{T_0/2}^{T_0} dT \frac{c_V (n_B,T)}{Q_S(n_B,T)} \,.
\end{eqnarray}
In this integral over temperature, we make the (reasonable) approximation that the neutron Fermi momentum and Dirac effective mass do not change significantly with temperature.

\begin{figure}
  \centering
  \includegraphics[width=0.49\textwidth]{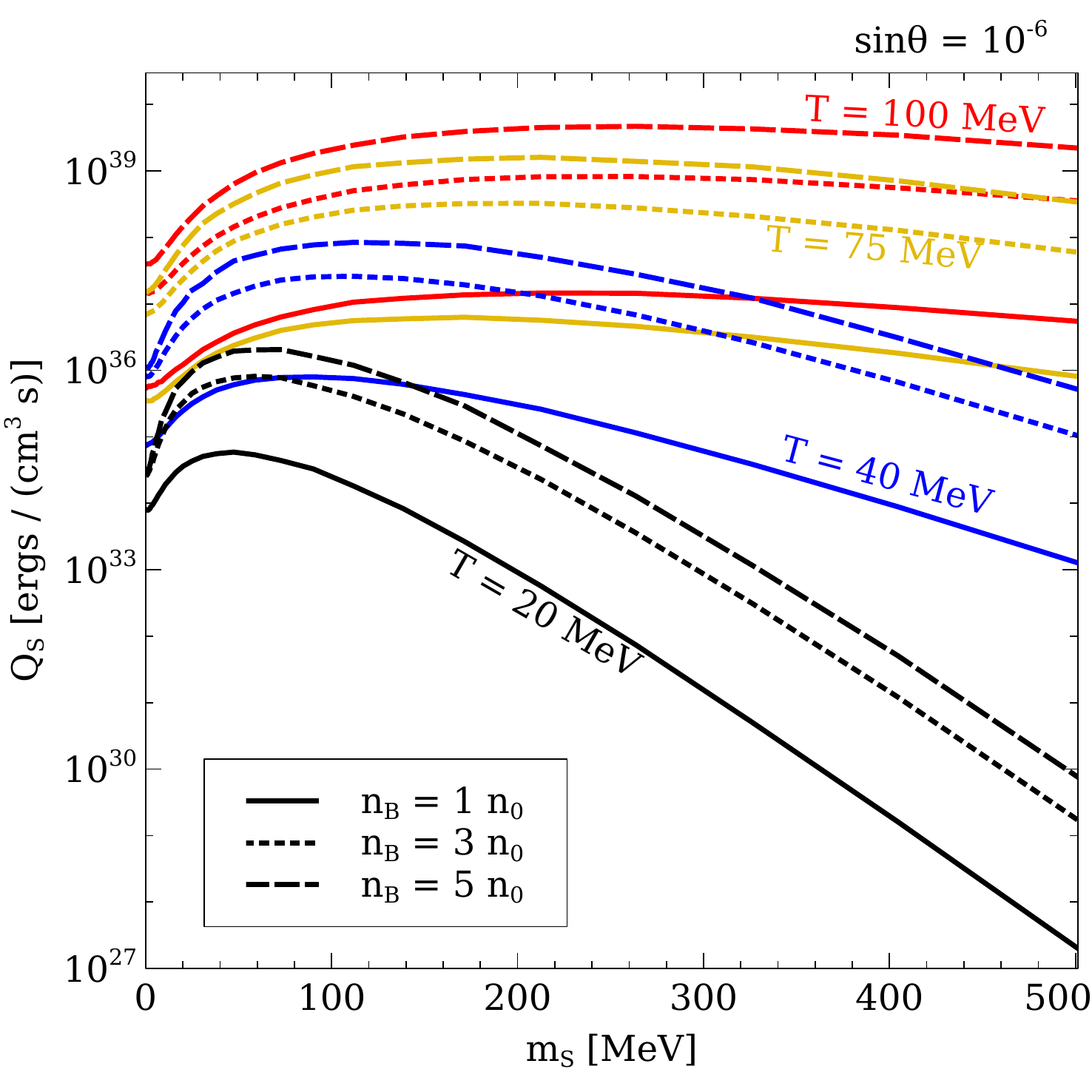}
  \includegraphics[width=0.49\textwidth]{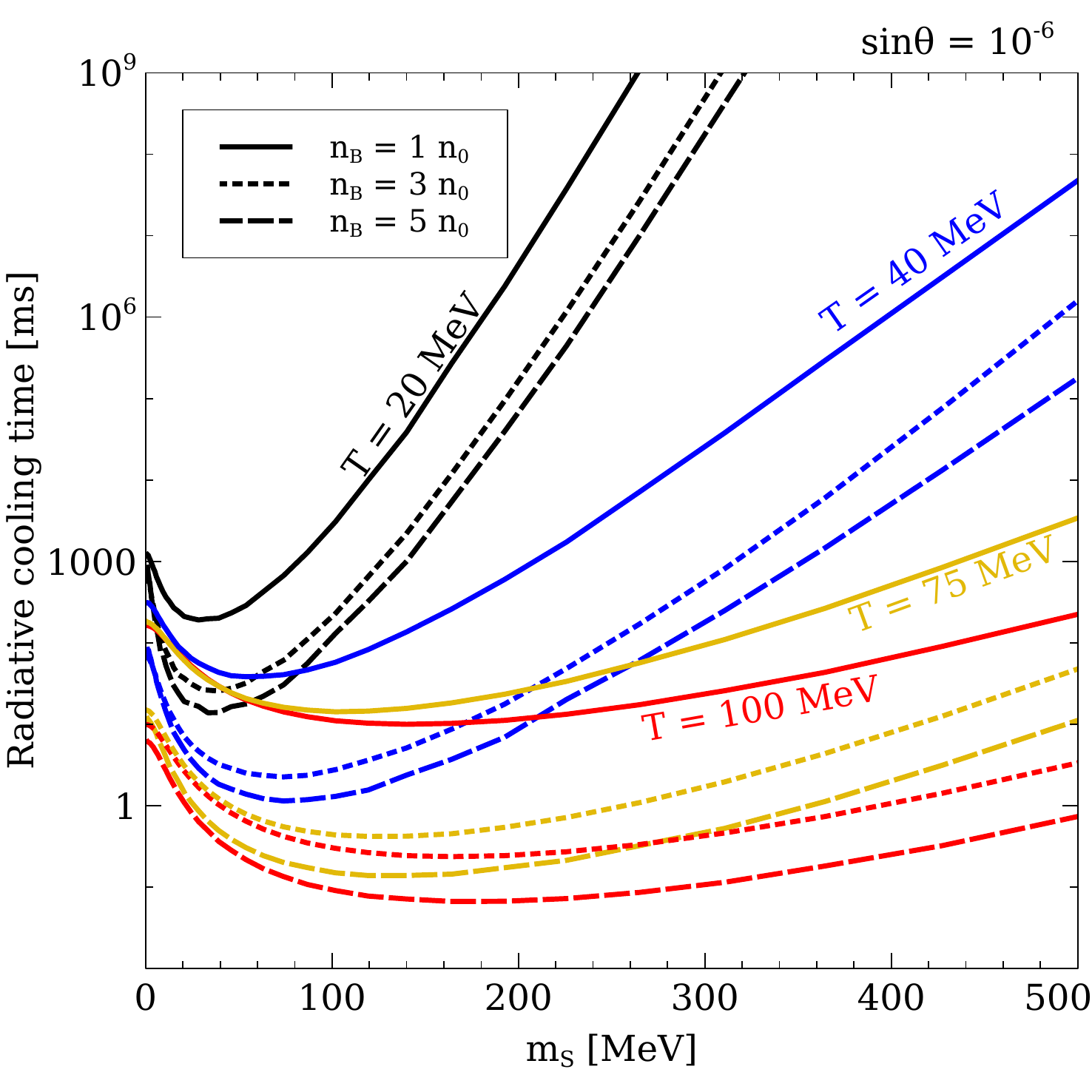}
  \caption{Emissivity of the scalar $S$ (left) and resultant radiative cooling time of fluid elements (right) in the neutron star merger remnant, as a function of the scalar mass $m_S$. In both panels, 
  the solid, short-dashed and long-dashed curves are respectively for $n_B = n_0$, $3n_0$ and $5n_0$, whereas the black, blue, yellow and red curves represent respectively $T =20$ MeV, 40 MeV, 75 MeV and 100 MeV. The mixing is fixed at  $\sin{\theta}=10^{-6}$ to ensure that the scalars free-stream in all thermodynamic conditions for all scalar masses shown in the plot (cf.~Figure~\ref{fig:S_mfp_ms_sintheta}).}
  \label{fig:S_emissivity_cooling_1d}
\end{figure}

In the left panel of Figure~\ref{fig:S_emissivity_cooling_1d} we plot the emissivity $Q_S$ of the CP-even scalar as a function of its mass $m_S$. The black, blue, yellow and red curves are respectively for the temperatures $T = 20$ MeV, 40 MeV, 75 MeV and 100 MeV, and the solid, short-dashed and long-dashed curves correspond respectively to the baryon densities of $n_B = n_0$, $3n_0$ and $5n_0$.  We see that for all thermodynamic conditions shown in this panel, the emissivity rises with the scalar mass, reaches a maximum, and then falls off.  This trend is seen in the nondegenerate nuclear matter limit as well (cf.~Figure~2 of Ref.~\cite{Dev:2020eam}).  This mass dependence is more complicated than the monotonic behavior seen in the emissivity of massive axions~\cite{Giannotti:2005tn}.  The reason stems from the CP-even nature of the scalar, thus the matrix element for $N + N' \to N + N' + S$ has a nontrivial dependence on the scalar mass $m_S$~\cite{Dev:2020eam}.  For $m_S\gtrsim 10$ MeV, the production of $S$ will be dominated by the $a^{(')}$, $b^{(')}$, $c^{(')}$ and $d^{(')}$ diagrams in Figure~\ref{fig:diagram} (see Figure~2 of Ref.~\cite{Dev:2020eam}); as a result of the partial cancellation of these diagrams, the emissivity $Q_S$ will have the mass-dependence in the form of $(m_S/E_S)^4$ for $10 \text{ MeV} \lesssim m_S \lesssim 3T$ (cf. Ref.~\cite{Dev:2020eam} and Appendix~\ref{appendix:relativistic_corrections}). This is clear for all the lines in the left panel of Figure~\ref{fig:S_emissivity_cooling_1d}. The emissivity increases strongly with temperature; however, as the mass increases and becomes large compared to the temperature, the production rate of the scalar becomes Boltzmann suppressed and falls off as $e^{-m_S/T}$.  This behavior is evident for the $T=20\text{ MeV}$ and $T=40\text{ MeV}$ curves, as shown in the left panel of Figure~\ref{fig:S_emissivity_cooling_1d}.  For the hottest temperatures encountered in mergers, say $75$ MeV or even $100\text{ MeV}$ in the left panel of Figure~\ref{fig:S_emissivity_cooling_1d}, the production rate seems to drop more slowly as the mass $m_S$ increases.  This is because at these high temperatures, the Boltzmann suppression is less dramatic for the range of $m_S$ we consider.  However, as the mass of the scalar increases beyond 500 MeV, the emissivity will fall off as $e^{-m_S/T}$.  For all the temperatures in Figure~\ref{fig:S_emissivity_cooling_1d}, the largest emissivity can be found when the scalar mass $m_S \sim 3 T$, which is a direct result of the CP-even nature of the scalar and the resultant $m_S$-dependence of the matrix element of the nucleon bremsstrahlung processes, and is very different from the axion case.  In the right panel of Figure~\ref{fig:S_emissivity_cooling_1d}, we plot the corresponding radiative cooling time $\tau_{1/2}$ as a function of the scalar mass $m_S$, for the benchmark temperatures and baryon densities in the left panel of Figure~\ref{fig:S_emissivity_cooling_1d}.  The nonmonotonicity of the emissivity as a function of $m_S$ is imprinted (in inverse) on the cooling time.  

\begin{figure}\centering
\includegraphics[width=0.49\textwidth]{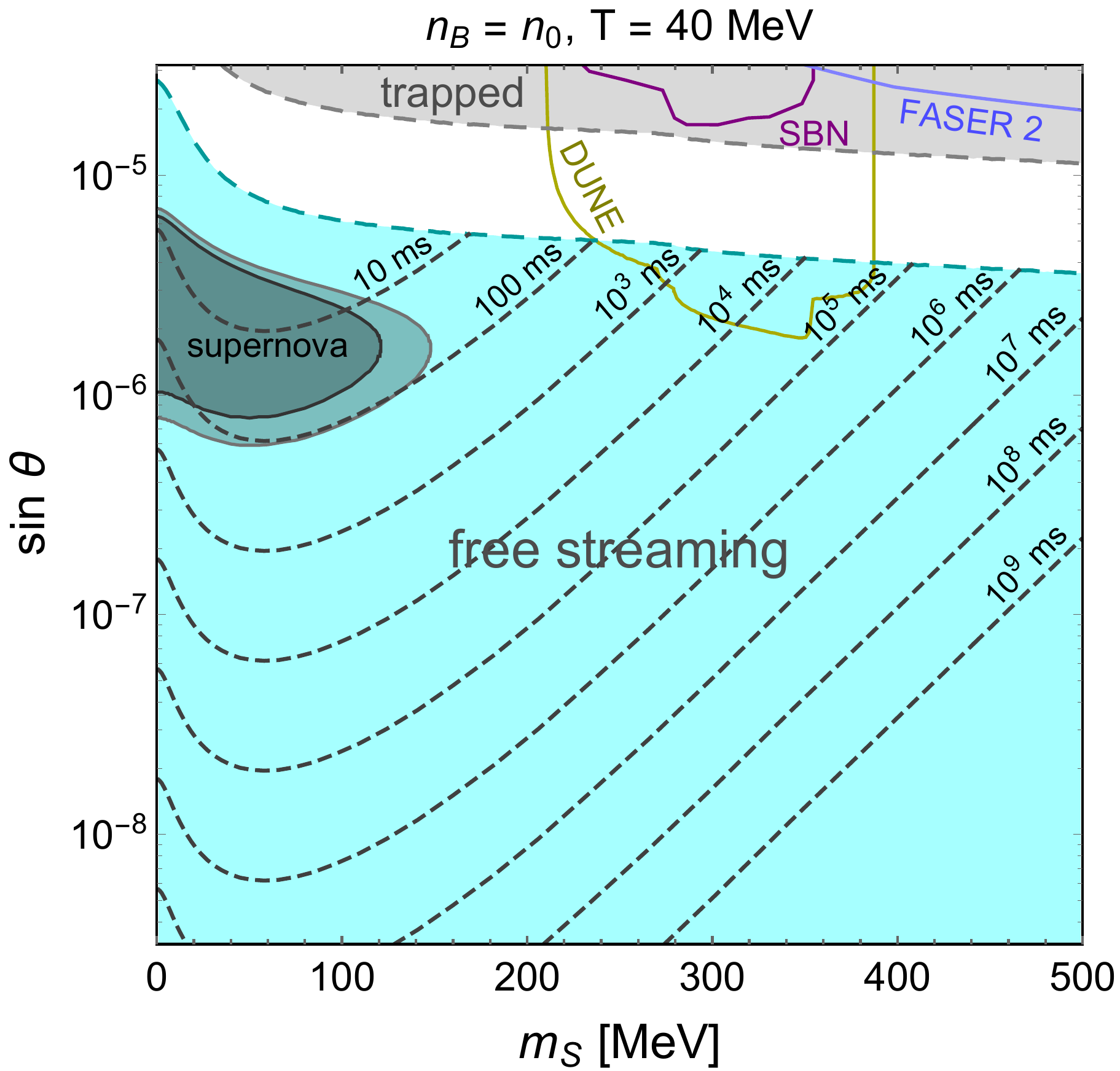}
\includegraphics[width=0.49\textwidth]{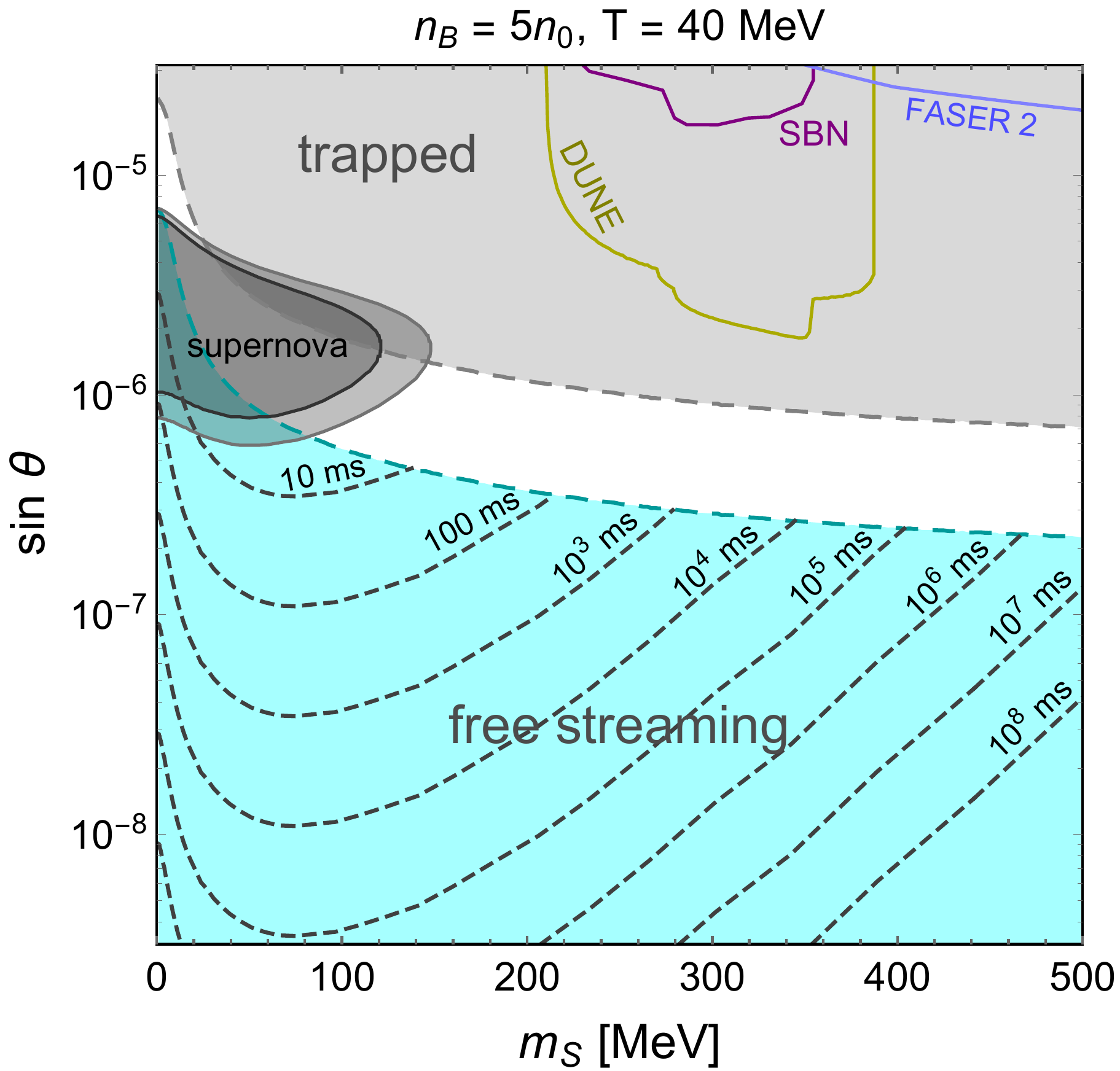} 
\includegraphics[width=0.49\textwidth]{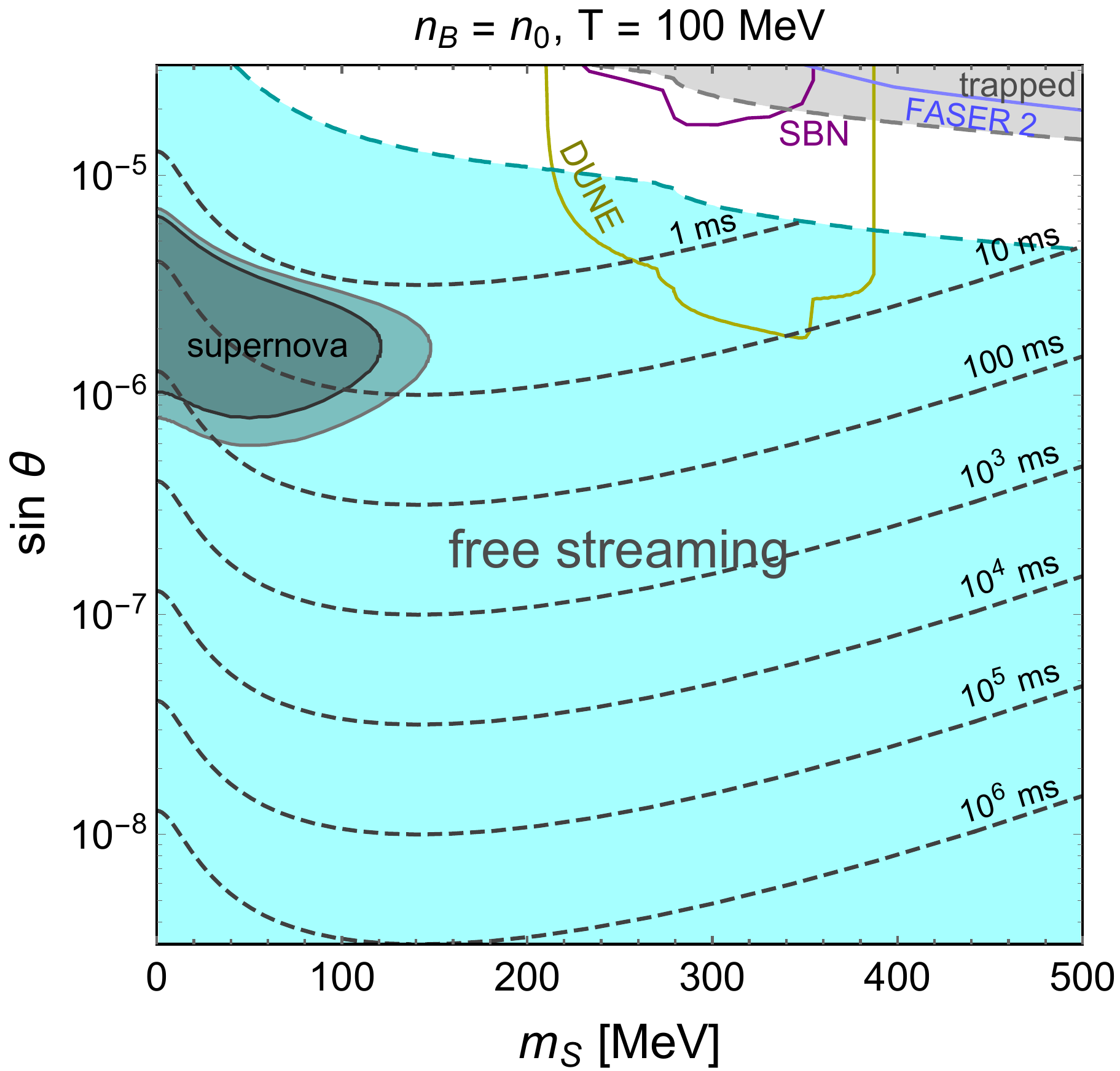}
\includegraphics[width=0.49\textwidth]{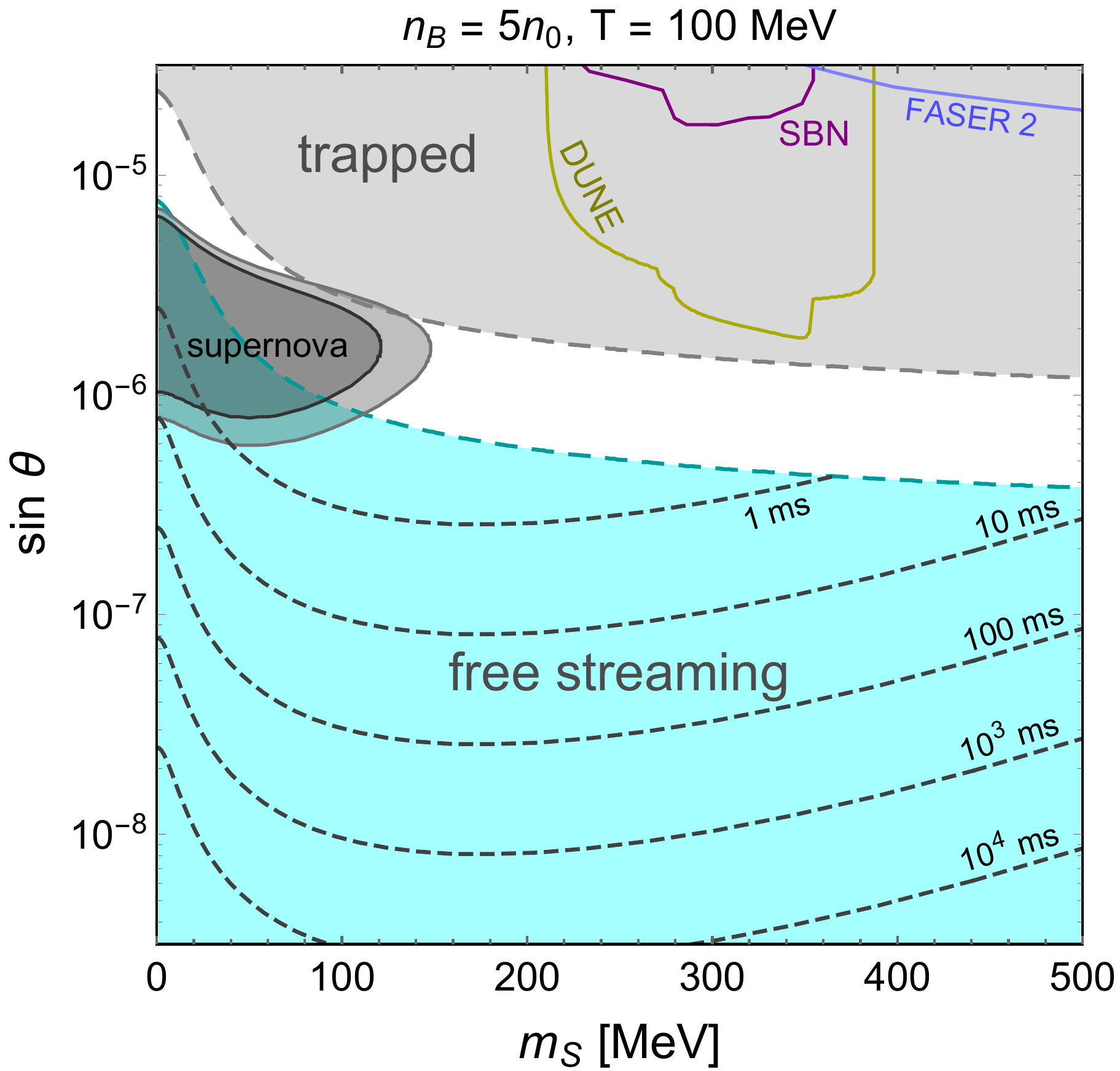}
\caption{Timescale for fluid elements to cool due to radiating scalar particles, in the $m_S-\sin{\theta}$ plane.  Each panel corresponds to a different density and temperature possibly encountered in a neutron star merger.  The radiative cooling times are only valid for parameters (shown in cyan) where the scalars free-stream through the matter in the merger.  The parameter space for trapped scalars is also shown, in grey.  Constraints from SN1987a and expected constraints from future collider experiments are overlaid.}
\label{fig:cooling}
\end{figure}

To get a holistic view of the behavior of the radiative cooling time, we plot the contours of cooling time $\tau_{1/2}$ in the $m_S-\sin{\theta}$ plane in Figure~\ref{fig:cooling}. The left and right panels are respectively for the baryon densities $n_B = n_0$ and $5n_0$.  The temperatures chosen reflect the hottest sections of the merger: in the two upper panels we take $T = 40$ MeV, which can essentially be reached in all numerical simulations for tens of milliseconds (or until the remnant begins to collapse)~\cite{Most:2021zvc,Most:2021ktk,Hanauske:2019qgs,Figura:2021bcn,Prakash:2021wpz,Hammond:2021vtv}; in the two lower panels the temperature is $T =100$ MeV, which can be present in some regions of matter in a few simulations for a few milliseconds, and also as the remnant begins to collapse to a black hole~\cite{Perego:2019adq,Bernuzzi:2020tgt}.  At these high temperatures, neutrinos are trapped (and exit the region on the much slower diffusion timescale~\cite{Janka:2017vlw}) so it would be left to exotic particle species with long MFPs to radiatively cool the hot fluid elements.  Radiative cooling is only a well-defined concept if the scalars free-stream through the remnant.  In each panel of Figure~\ref{fig:cooling} we color in cyan  the parameter space where the scalars are free-streaming, and the trapped regions are shown in grey.  The cooling time is proportional to $\sin^2{\theta}$ (coming from the expression for the scalar emissivity in Eq.~(\ref{eq:emissivity_integral})), so the cooling timescale dramatically shortens as the mixing $\sin{\theta}$ increases, up until the MFP shortens to the point where the scalars cannot free-stream anymore.  

Merger remnants can survive for tens, hundreds, or thousands of milliseconds~\cite{Lucca:2019ohp,Murguia-Berthier:2020tfs} (or much longer if the remnant mass is below the limit for a nonrotating star~\cite{Lattimer:2012nd,Riley:2021pdl}), so radiative cooling that occurs on the timescales shown in Figure~\ref{fig:cooling} is potentially relevant to the postmerger evolution.  We see from the two upper panels of Figure~\ref{fig:cooling} that areas of the merger at $T=40\text{ MeV}$ can cool to 20 MeV on timescales shorter than 1 second only if the scalar is relatively light, say $m_S\lesssim 250 \text{ MeV}$. The mixing angle would need to be roughly $10^{-7}\lesssim \sin{\theta} \lesssim 10^{-5}$, although a fraction of the relevant parameter space is ruled out by the SN1987A constraint~\cite{Dev:2020eam}.  If the remnant is short-lived, say lasting for 10 ms, only scalars with $m_S \lesssim 100 \text{ MeV}$ can significantly cool the 40 MeV regions of the merger before it collapses, and only in a very limited range of $\sin\theta$.  As expected, cooling of a fluid element from $T=100\text{ MeV}$ down to $50 \text{ MeV}$ would occur much more rapidly for a given choice of scalar mass and coupling, as shown in  the two lower panels of Figure~\ref{fig:cooling}.  Rapid cooling of these extremely hot fluid elements can occur due to radiating scalars of the full range of masses we discuss in this work. As shown in the lower left panel of Figure~\ref{fig:cooling}, a sizable region of parameter space with ${\cal O} (1 \, {\rm ms}) \lesssim \tau_{1/2} \lesssim {\cal O} (10 \, {\rm ms})$ can also be probed at the DUNE experiment. 

Cooling of fluid elements due to axion emission was implemented in a neutron star merger simulation~\cite{Dietrich:2019shr}. Even though axion radiation can quickly cool the merger remnant (see semianalytic calculations in Ref.~\cite{Harris:2020qim} and the simulation results in Ref.~\cite{Dietrich:2019shr}), it does not lead to significant changes in the dynamics of the merger remnant.  In particular, while the cooling leads to a small sphericalization of the merger remnant, the gravitational wave signal, remnant lifetime, and ejecta mass were close to unchanged (compared to the case without radiative cooling from axion emission), even if the rate of axion emission was chosen to be unrealistically high~\cite{Dietrich:2019shr,Fortin:2021cog}.  Therefore, we do not expect radiative cooling to have a direct impact on the dynamics of the merger, unless it involves physics that was not included in the merger simulation in Ref.~\cite{Dietrich:2019shr}.  For example, cooling due to the emission of scalars could trigger a phase transition in nuclear matter or change the temperature-dependent transport properties, which could then lead to an observable signature of the presence of scalars.  It is also possible that emitted scalars can decay once they have escaped the merger remnant.  Similar to the discussion in the case of dark photon emission~\cite{Diamond:2021ekg}, scalars that escape the merger remnant can decay into photons directly, or into leptons that scatter and create photons.  These decays may result in distinct $\gamma$-ray signals, the properties of which deserve further investigation, but beyond the scope of the current work.

\section{Scalar contribution to thermal conduction}
\label{sec:thermal_conduction}

While experimental data forbid the QCD axion and light ALPs from being trapped in neutron star merger remnants~\cite{Harris:2020qim}, the CP-even scalar discussed in this paper has, in certain conditions, a short enough MFP to reach thermal equilibrium and form a Bose-Einstein distribution.  For a given scalar mass and mixing angle (or, equivalently, coupling strengths), the temperatures and densities where scalars are trapped are mapped out in Figure~\ref{fig:S_mfp_nB_T}, and for fixed densities and temperatures, the values of $m_S$ and $\sin{\theta}$ for which scalars are trapped are mapped out in Figure~\ref{fig:S_mfp_ms_sintheta}. It is clear that current laboratory and astrophysical constraints do not rule out the possibility of trapped scalars.  

In regions of the merger remnant where the scalar MFP is sufficiently short, the scalars will form a free Bose gas with zero chemical potential set by nucleon bremsstrahlung reactions. Throughout this work, chemical potentials are defined relativistically, that is, they include the rest mass of the particle. 

In this section, we will calculate the impact of a trapped scalar gas on transport properties in the merger remnant, and will need a variety of thermodynamic properties of a quantum gas of scalar particles, which we detail below.

The number density of a (noninteracting) massive Bose gas with zero chemical potential is
\begin{equation}
    n_S = \int \dfrac{\mathop{d^3{\bf k}}}{(2\pi)^3}\frac{1}{e^{E_S/T}-1} = \frac{1}{2\pi^2}\int_{m_S}^{\infty}\mathop{dE_S}\frac{E_S\sqrt{E_S^2-m_S^2}}{e^{E_S/T}-1} \, ,
    \label{eq:scalar_n}
\end{equation}
and the energy density is
\begin{equation}
    \varepsilon = \int \dfrac{\mathop{d^3{\bf k}}}{(2\pi)^3}\frac{E_S}{e^{E_S/T}-1} = \frac{1}{2\pi^2}\int_{m_S}^{\infty}\mathop{dE_S}\frac{E_S^2\sqrt{E_S^2-m_S^2}}{e^{E_S/T}-1} \, .
    \label{eq:scalar_eps}
\end{equation}
The heat capacity at constant volume (or at constant baryon number density, as baryon number is conserved) is given by
\begin{equation}
    c_{V,S} = \dfrac{\partial\varepsilon}{\partial T}= \frac{1}{8\pi^2T^2}\int_{m_S}^{\infty}\mathop{dE_S}\dfrac{E_S^3\sqrt{E_S^2-m_S^2}}{\sinh^2{[E_S/(2T)]}} \, .
    \label{eq:scalar_cv}
\end{equation}
The typical velocity of a scalar particle with mass $m_S$ and energy $E_S$ is
\begin{equation}
    v_S = \frac{\partial E_S(p)}{\partial p} =\sqrt{1-(m_S/E_S)^2} \, .
    \label{eq:scalar_velocity}
\end{equation}

Trapped species of particles in the merger remnant can participate in various transport processes, including thermal conductivity, shear viscosity, and electrical conductivity~\cite{Alford:2017rxf}.  Typically particle species with long MFPs (but still shorter than the system size) will dominate transport, though other factors are also important for each transport process~\cite{Alford:2017rxf,Schmitt:2017efp}.  In this section, we will consider the impact of the trapped CP-even scalar on the thermal conductivity $\kappa$ of the matter in the merger remnant.  In the standard consideration of neutron star mergers, where only neutrons, protons, electrons, and neutrinos are involved, the thermal conductivity is dominated by either electrons (in regions of the star where the neutrinos are free-streaming) or the neutrinos (if they are trapped)~\cite{Alford:2017rxf}.  It is useful to quantify the effect of thermal conductivity by calculating the timescale for two adjacent fluid elements at slightly different temperatures to reach thermal equilibrium with each other.  A hot fluid element of size $z$ (or volume $z^3$)  has excess thermal energy $E_{\text{th}} = c_Vz^3\Delta T$ compared to its neighbors.  This heat diffuses through the boundary of the fluid element (with surface area $6z^2$) at rate $W_{\text{th}} = \kappa (\Delta T/z) 6z^2$.  This thermal equilibration timescale is given by the ratio of these quantities
\begin{equation}
    \tau_{\kappa} = \frac{E_{\text{th}}}{W_{\text{th}}} = \frac{c_Vz^2}{6\kappa} \, .
    \label{eq:eqtime}
\end{equation}
The specific heat of nuclear matter $c_V$ is given by Eq.~(\ref{eq:specific_heat}).  The thermal conductivity is the sum of the thermal conductivity generated by each particle species in the remnant 
\begin{equation}
    \kappa_i = \frac13 c_{V,i}v_i\lambda_i \, ,
    \label{eq:kappa_i}
\end{equation}
where $c_{V,i}$, $v_i$ and $\lambda_i$ are respectively the specific heat, speed and MFP of the $i$th particle species.  Particle species with the optimal combination of large density and long MFP (but still shorter than the system size) dominate the thermal conductivity.  In this paper, we focus on the neutrino-trapped regime with temperature $T>5$ to 10 MeV, so out of the SM particles, neutrinos dominate the thermal conductivity. The details of the neutrino thermal conductivity calculation can be found in Appendix~\ref{app:neutrino_kappa}.

\begin{figure}
\centering
\includegraphics[width=0.5\textwidth]{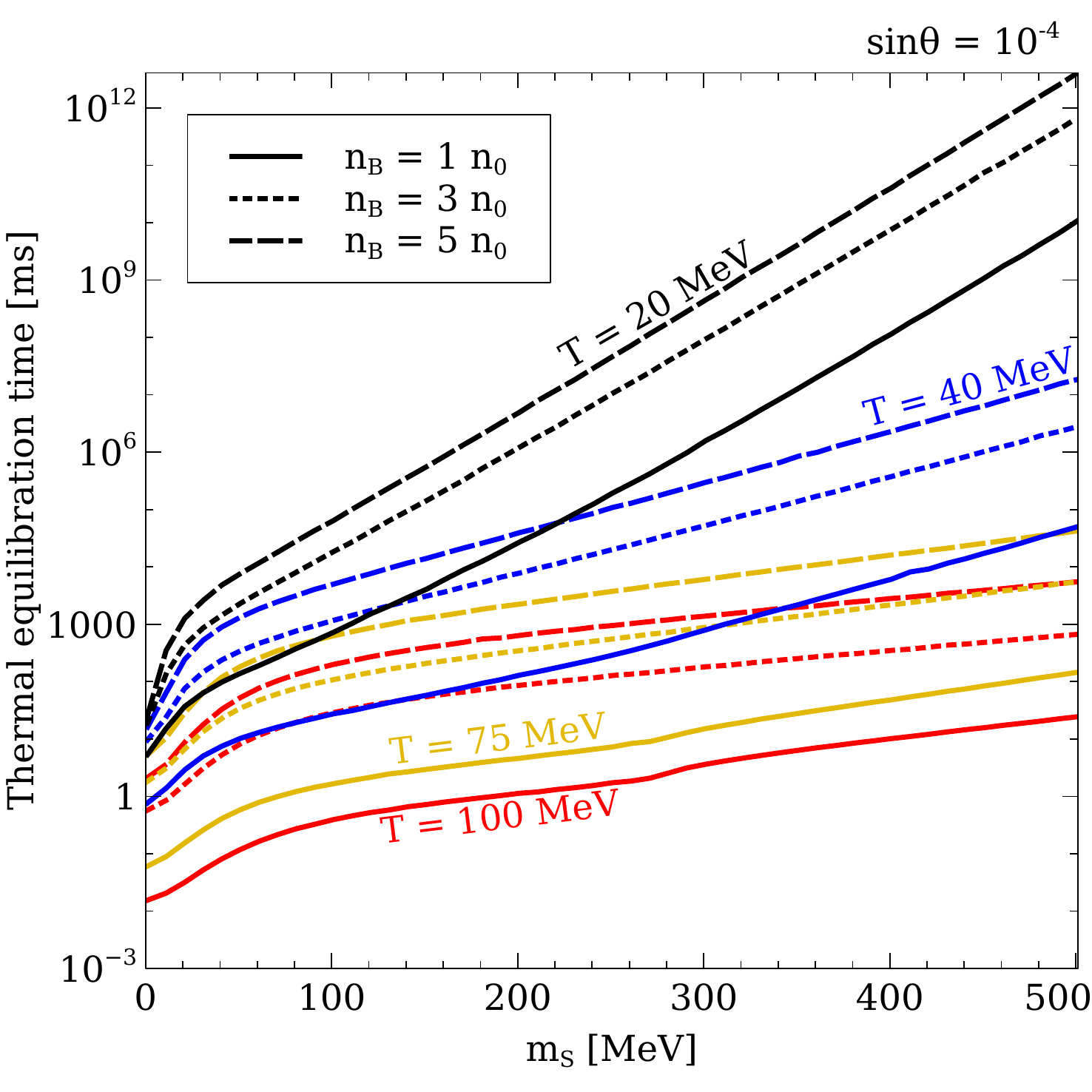}
\caption{Thermal equilibration timescale due to the scalar $S$ over a distance of 1 km 
as a function of the scalar mass $m_S$.  The solid, short-dashed and long-dashed curves are respectively for $n_B = n_0$, $3n_0$ and $5n_0$, the black, blue, yellow and red curves represent respectively $T =20$ MeV, 40 MeV, 75 MeV and 100 MeV.  
The mixing $\sin{\theta}=10^{-4}$ is chosen so that the scalar particles are trapped for all displayed thermodynamic conditions and scalar masses. }
\label{fig:thermal_eq_1d}
\end{figure}

Under certain conditions, trapped CP-even scalars would give rise to a larger thermal conductivity than the trapped neutrinos.  The thermal conductivity of scalars can be calculated using Eq.~(\ref{eq:kappa_i}), with input from  Eqs.~(\ref{eq:scalar_cv}) and   (\ref{eq:scalar_velocity}), as well as the scalar MFP calculated in Section~\ref{sec:mfp}.   
To obtain an estimate for the thermal equilibration timescale due to diffusion of a gas of CP-even scalars in a neutron star merger remnant, we consider a temperature gradient occurring over a distance scale of $z = 1 \text{ km}$.  For example, numerical simulations indicate the presence of a ring-like hot region in the merger remnant, occurring 4 to 7 km out from the remnant's center~\cite{Kastaun:2016yaf,Hanauske:2016gia,Perego:2019adq,Hanauske:2019qgs, Camelio:2020mdi, Raithel:2021hye}.  This ring has a thickness of 1 to 3 km.  We will show results only for $z=1\text{ km}$, but since $\tau_{\kappa}\propto z^2$, the thermal equilibration times for other distance scales can be easily obtained.  In particular, it is clear that temperature gradients on shorter distance scales, for example from turbulence~\cite{Radice:2020ids}, will equilibrate more quickly.

In Figure~\ref{fig:thermal_eq_1d} we plot the timescale for thermal equilibration of matter in the merger due to diffusion of trapped CP-even scalars (cf.~Eq.~\eqref{eq:eqtime}), over a distance scale of 1 km, as a function of the scalar mass $m_S$. As in Figure~\ref{fig:S_emissivity_cooling_1d}, we have chosen the benchmark temperatures $T = 20$ MeV (black), 40 MeV (blue), 75 MeV (yellow) and 100 MeV (red), and the densities $n_B = n_0$ (solid), $3n_0$ (short-dashed) and $5n_0$ (long-dashed). The mixing angle of $\sin\theta = 10^{-4}$ has been chosen such that the scalar $S$ is trapped in the neutron star merger remnant in the whole mass range we are considering, cf.~Figure~\ref{fig:S_mfp_ms_sintheta}. It should be noted, however, that for $\sin\theta =10^{-4}$ the scalar mass range of $220 \, {\rm MeV} \lesssim m_S \lesssim 310$ MeV has been excluded by the limits from PS191, CHARM, and LSND (see Figure~\ref{fig:S_mfp_ms_sintheta}). It is clear that higher temperature regions in the merger remnant have a higher density of scalars (cf.~Eq.~(\ref{eq:scalar_n})) and thus thermally equilibrate faster.  The density of scalars drops as their mass increases, causing the thermal equilibration time to grow longer as the scalar mass $m_S$ increases.  

\begin{figure}\centering
\includegraphics[width=0.47\textwidth]{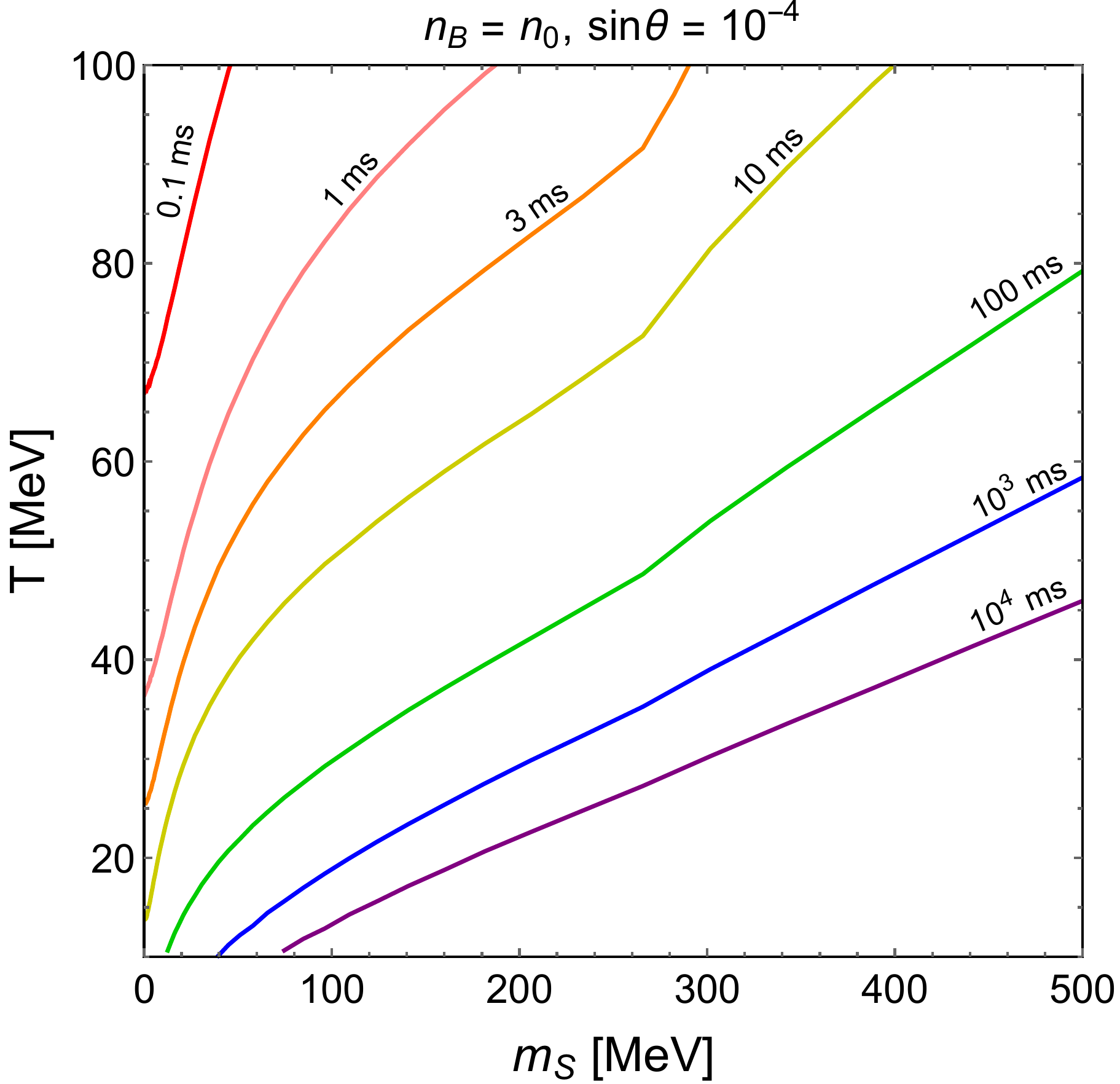} 
\includegraphics[width=0.47\textwidth]{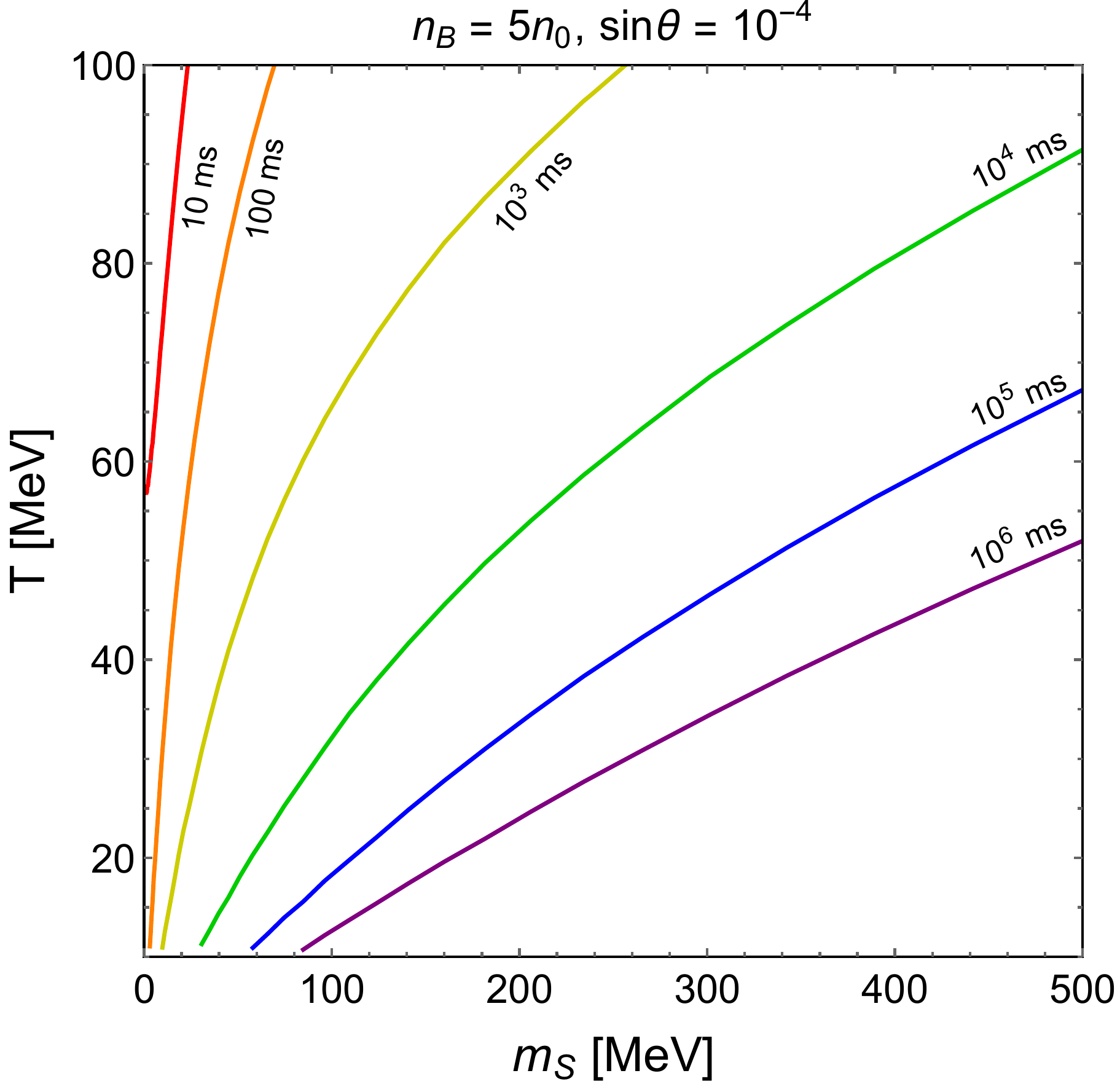}
\caption{Thermal equilibration timescale due to the scalar $S$ over a distance of 1 km 
in the $m_S-T$ plane, for the baryon density of $n_B =n_0$ (left) and $5n_0$ (right).  
Here $\sin{\theta}=10^{-4}$ is chosen to ensure that the scalar is trapped for all displayed thermodynamic conditions and scalar masses.}
\label{fig:thermal_eq2}
\end{figure}

In Figure~\ref{fig:thermal_eq2}, we show contour plots of the thermal equilibration time in the $m_S-T$ plane, for two choices of density $n_B = n_0$ (left panel) and $5n_0$ (right panel), again with $\sin\theta = 10^{-4}$ to ensure the scalars are trapped for all conditions shown in the plots. 
This presentation of the thermal equilibration time makes clear that hotter regions of the merger reach thermal equilibrium more quickly, for the reasons mentioned above, especially when the scalars are light.  The kink in the contours in the left panel at $m_S=2m_{\pi^{\pm}}\simeq 270$ MeV is due to the appearance of the decay channel $S\rightarrow \pi^+\pi^-$, which shortens the MFP of the scalar, slowing its ability to attenuate temperature gradients.

\begin{figure}\centering
\includegraphics[width=0.47\textwidth]{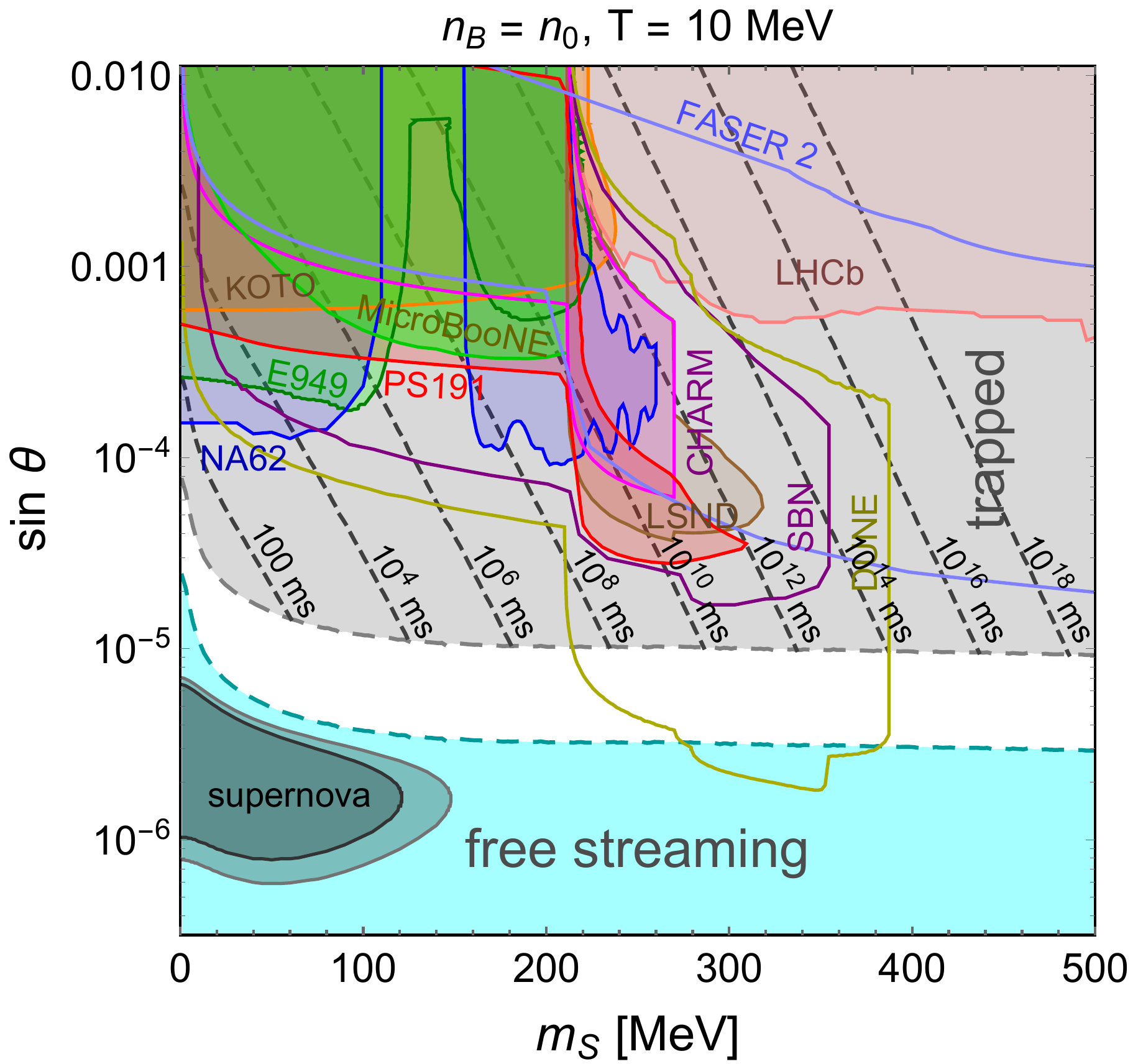}
\includegraphics[width=0.47\textwidth]{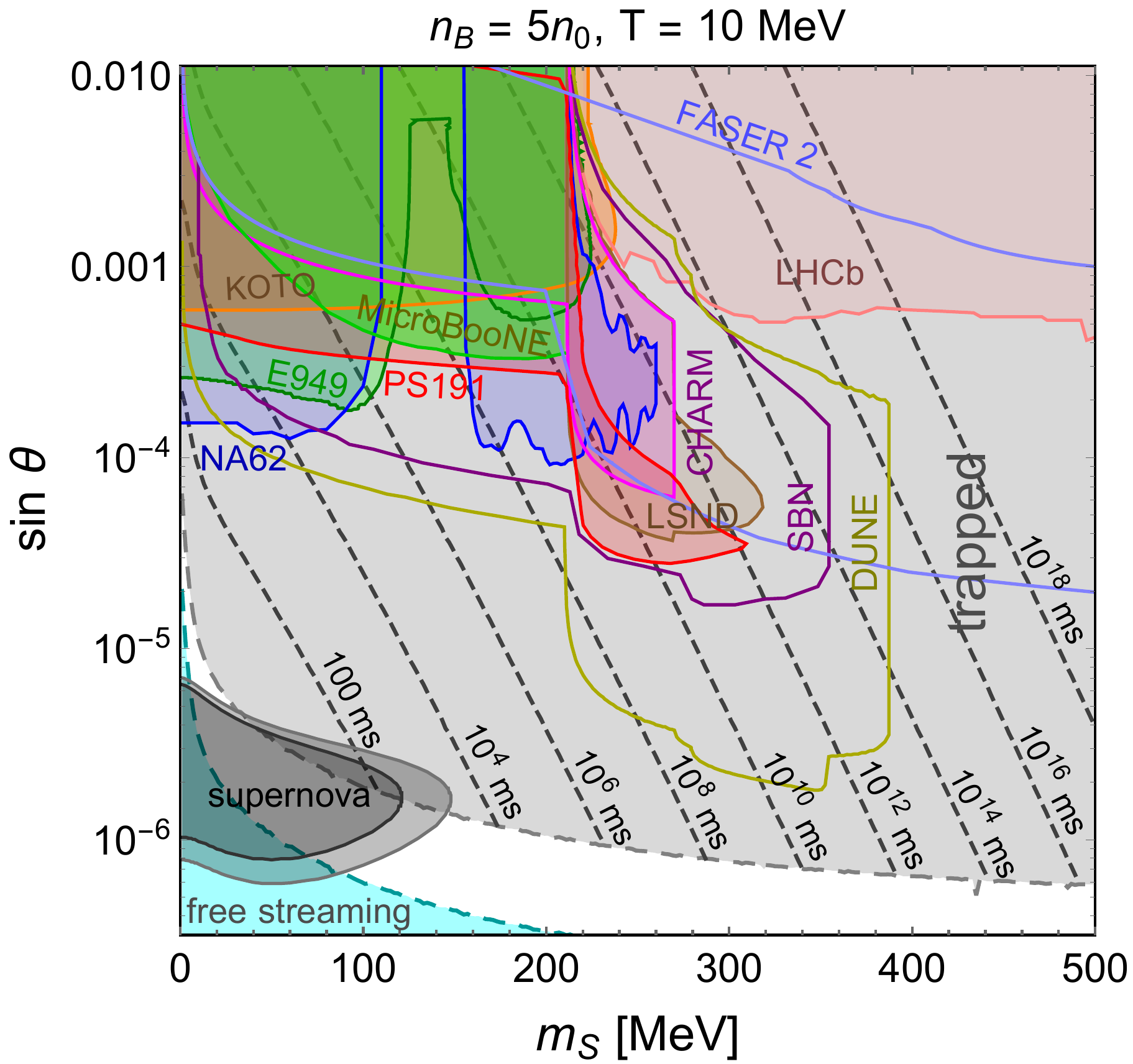}
\includegraphics[width=0.47\textwidth]{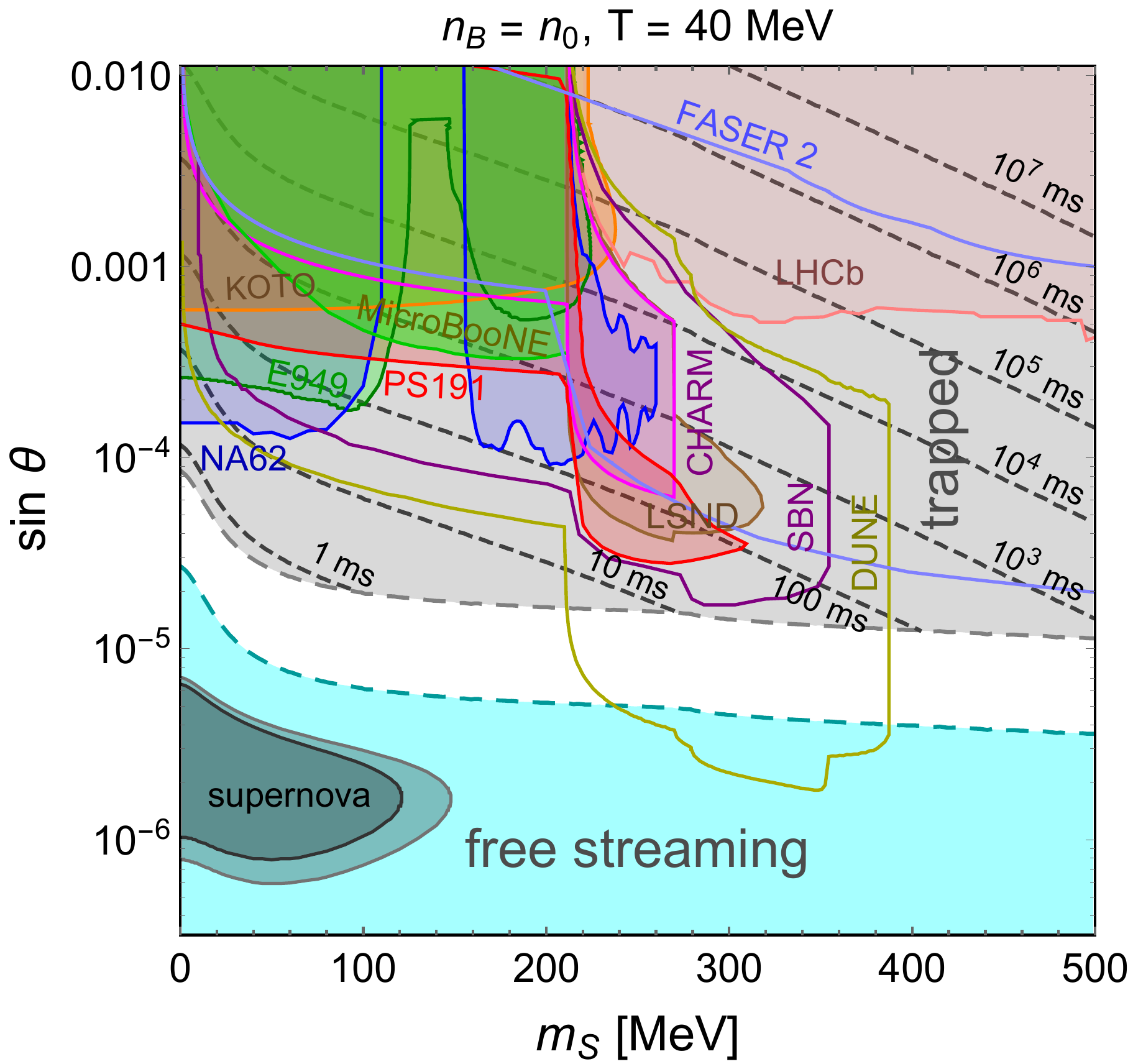}
\includegraphics[width=0.47\textwidth]{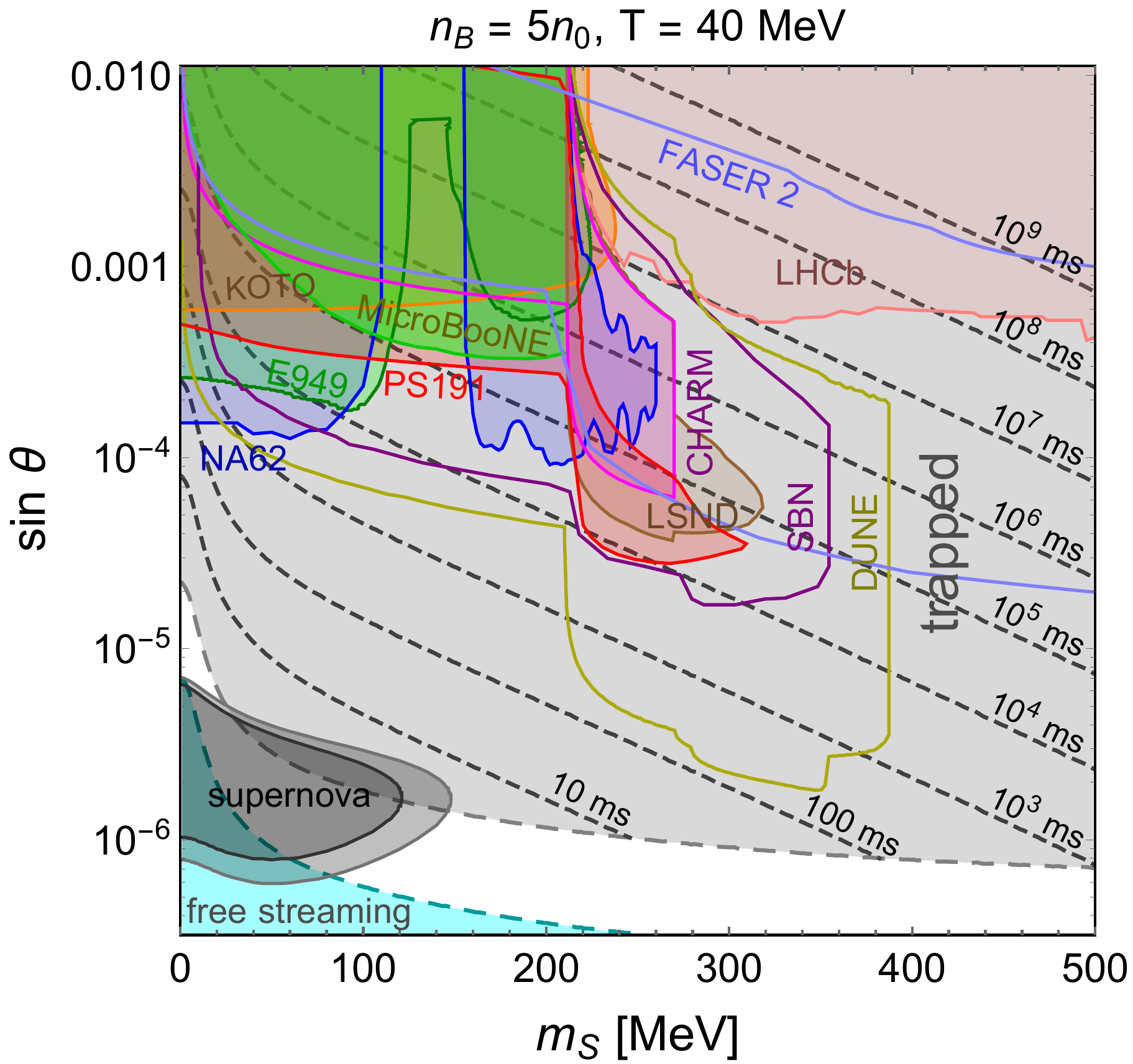}
\includegraphics[width=0.47\textwidth]{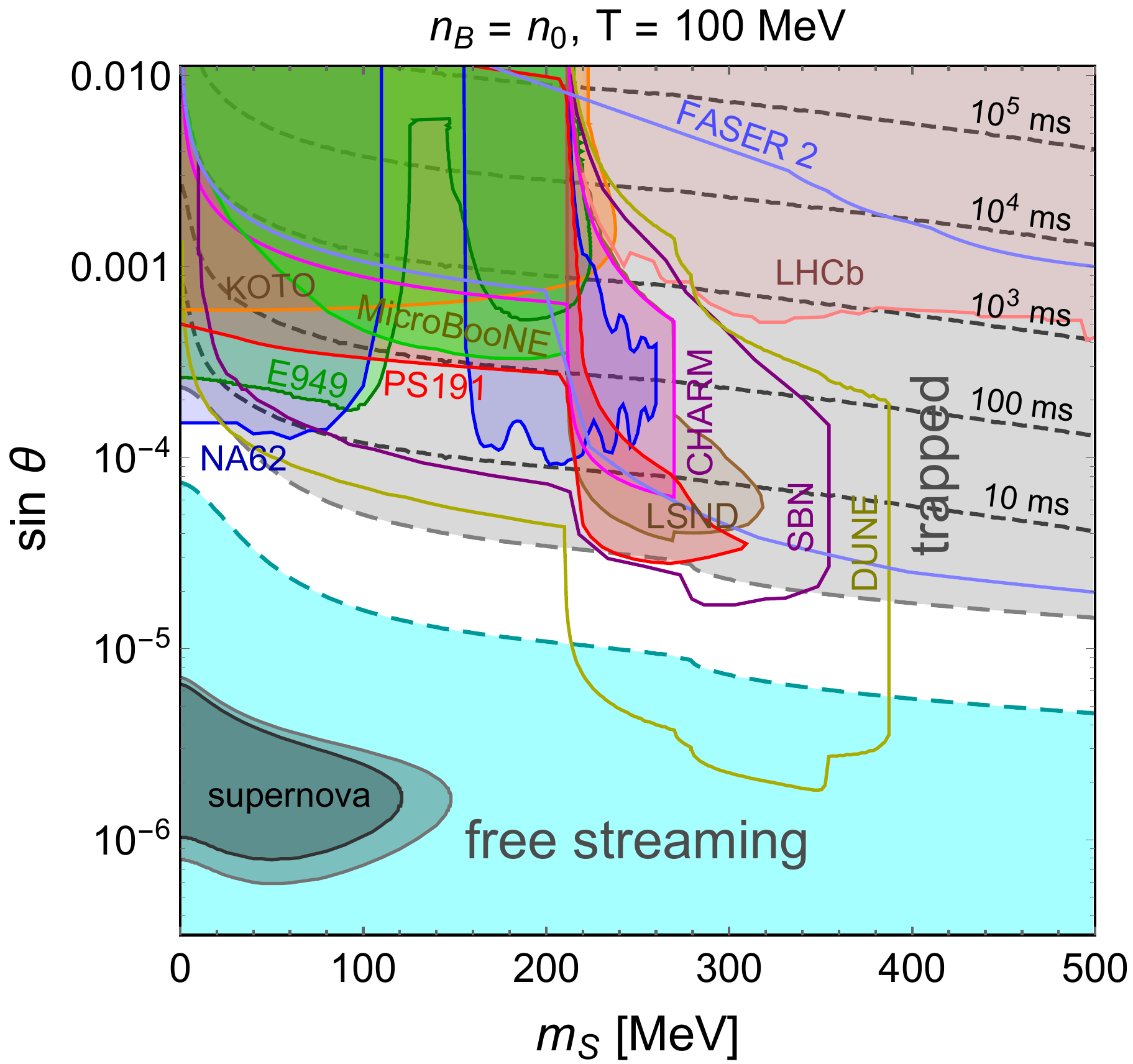}
\includegraphics[width=0.47\textwidth]{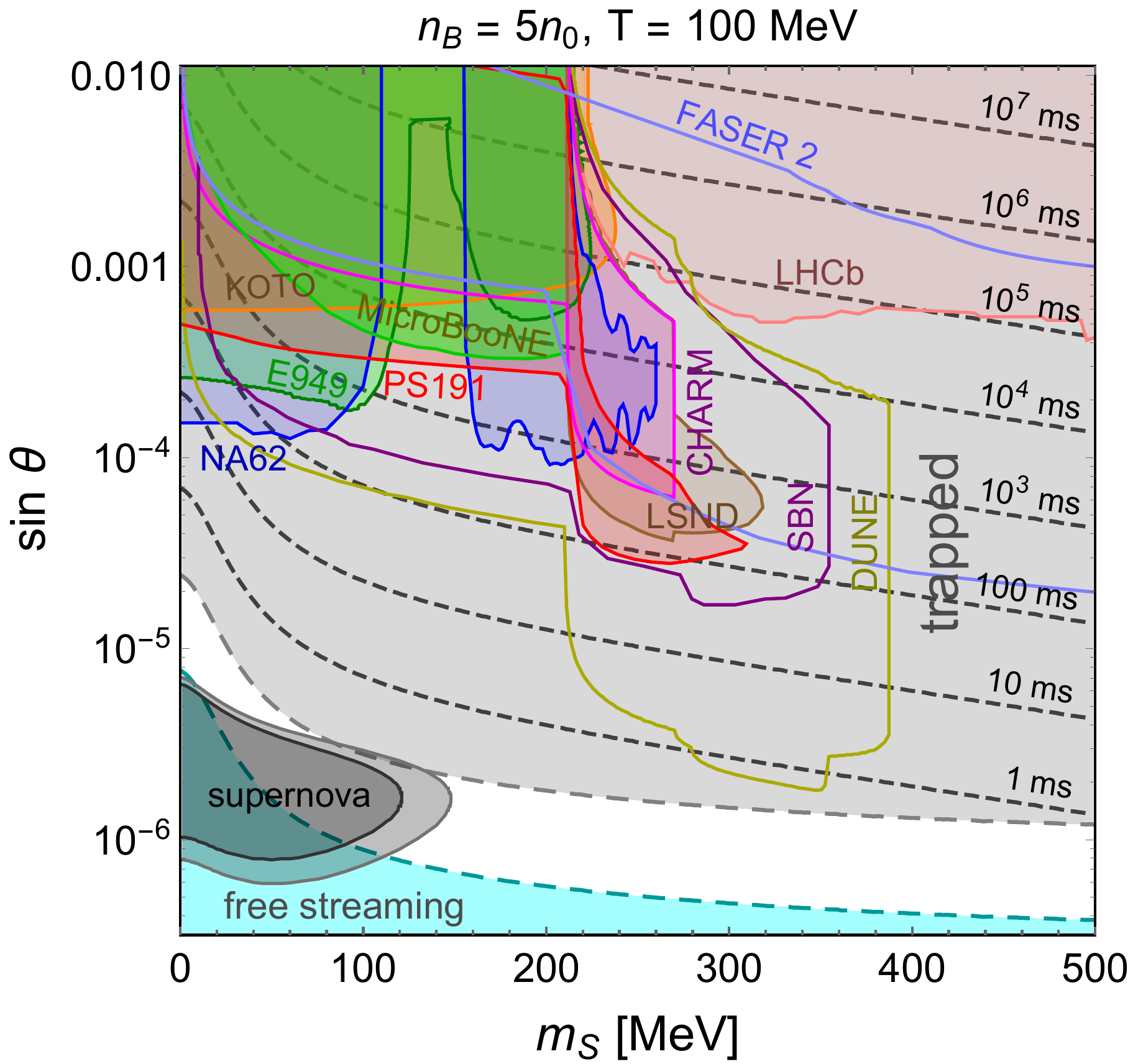}
\caption{Thermal equilibration timescale due to the scalar $S$ over a distance of 1 km in the $m_S-\sin\theta$ plane, for the baryon density of $n_B =n_0$ (left) and $5n_0$ (right), and temperatures $T = 10$ MeV (upper), 40 MeV (middle) and 100 MeV (bottom).
The thermal equilibration times are only valid for parameters where the scalars are trapped in the matter in the merger (shown in grey).  The other labels are the same as in Figure~\ref{fig:S_mfp_ms_sintheta}.
}
\label{fig:thermal_eq}
\end{figure}

The thermal equilibration time due to a Bose gas of scalars in the $m_S-\sin{\theta}$ plane is shown in  Figure~\ref{fig:thermal_eq}. As in Figure~\ref{fig:S_mfp_ms_sintheta}, the left and right panels are respectively for the baryon density $n_B = n_0$ and $5n_0$, and the upper, middle and lower panels correspond to the temperatures of $T = 10$ MeV, 40 MeV and 100 MeV, respectively. The labels for all the current laboratory constraints and future prospects, the supernova limits and the trapped and free streaming regions are the same as in Figure~\ref{fig:S_mfp_ms_sintheta}. 
The thermal equilibration time depends on the mixing angle via $\tau_{\kappa} \propto \sin^{2}{\theta}$, meaning that as the coupling between nucleons and scalars grows weaker, the MFP of the scalar grows longer, shortening the timescale needed to even out temperature gradients in the remnant (provided that the scalar MFP remains much shorter than the system size).   

\begin{figure}\centering
\includegraphics[width=0.47\textwidth]{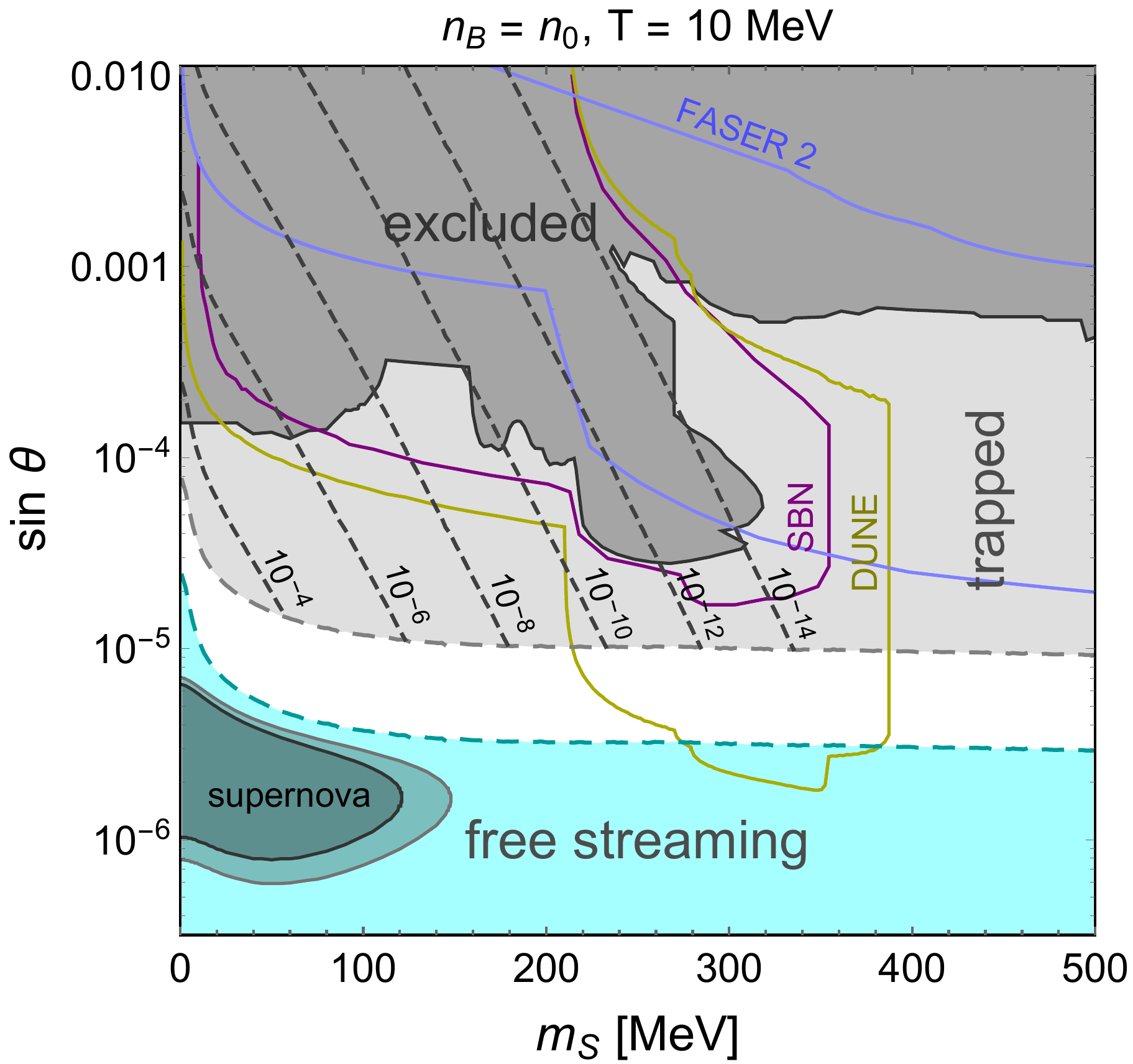}
\includegraphics[width=0.47\textwidth]{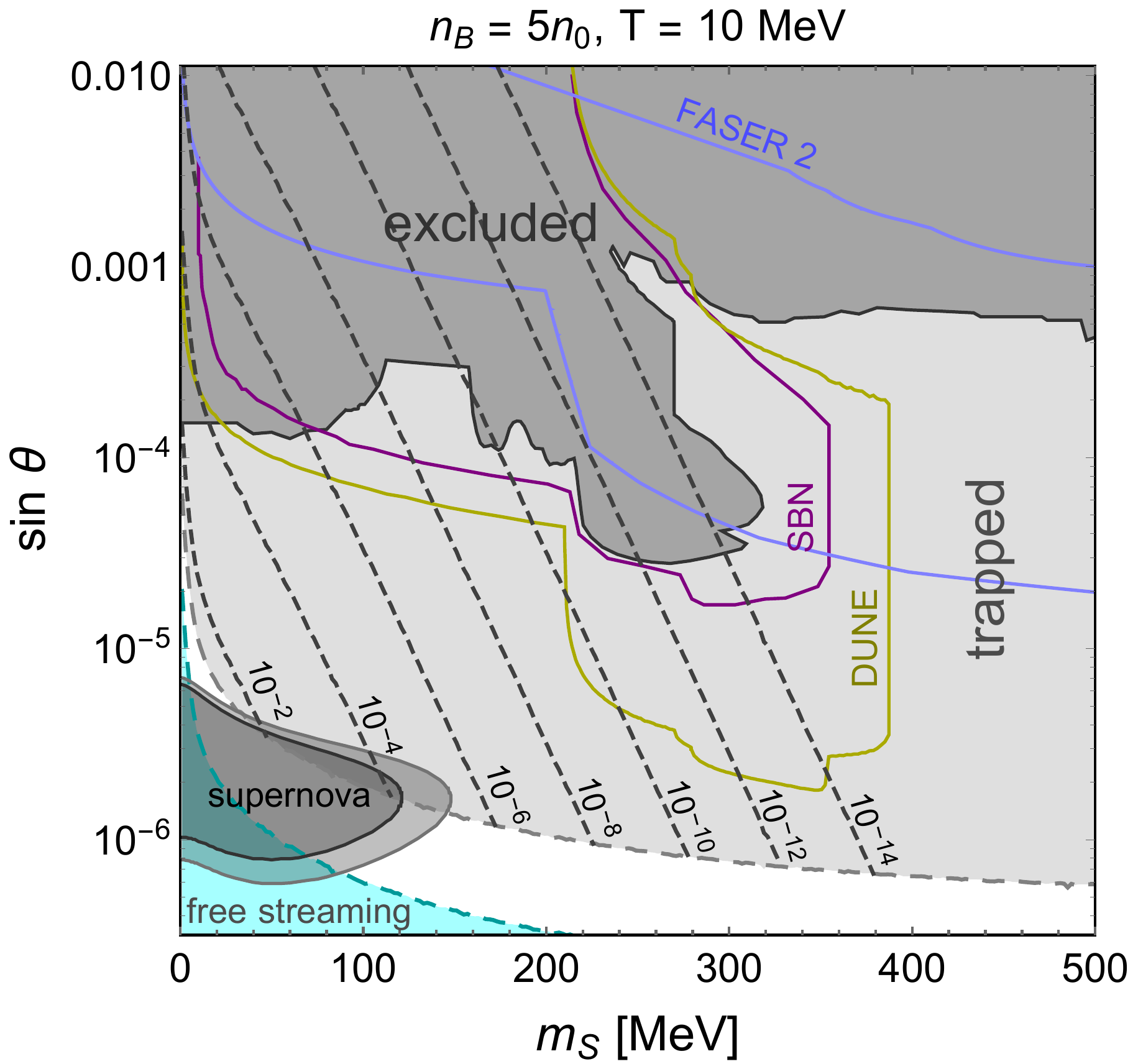}
\includegraphics[width=0.47\textwidth]{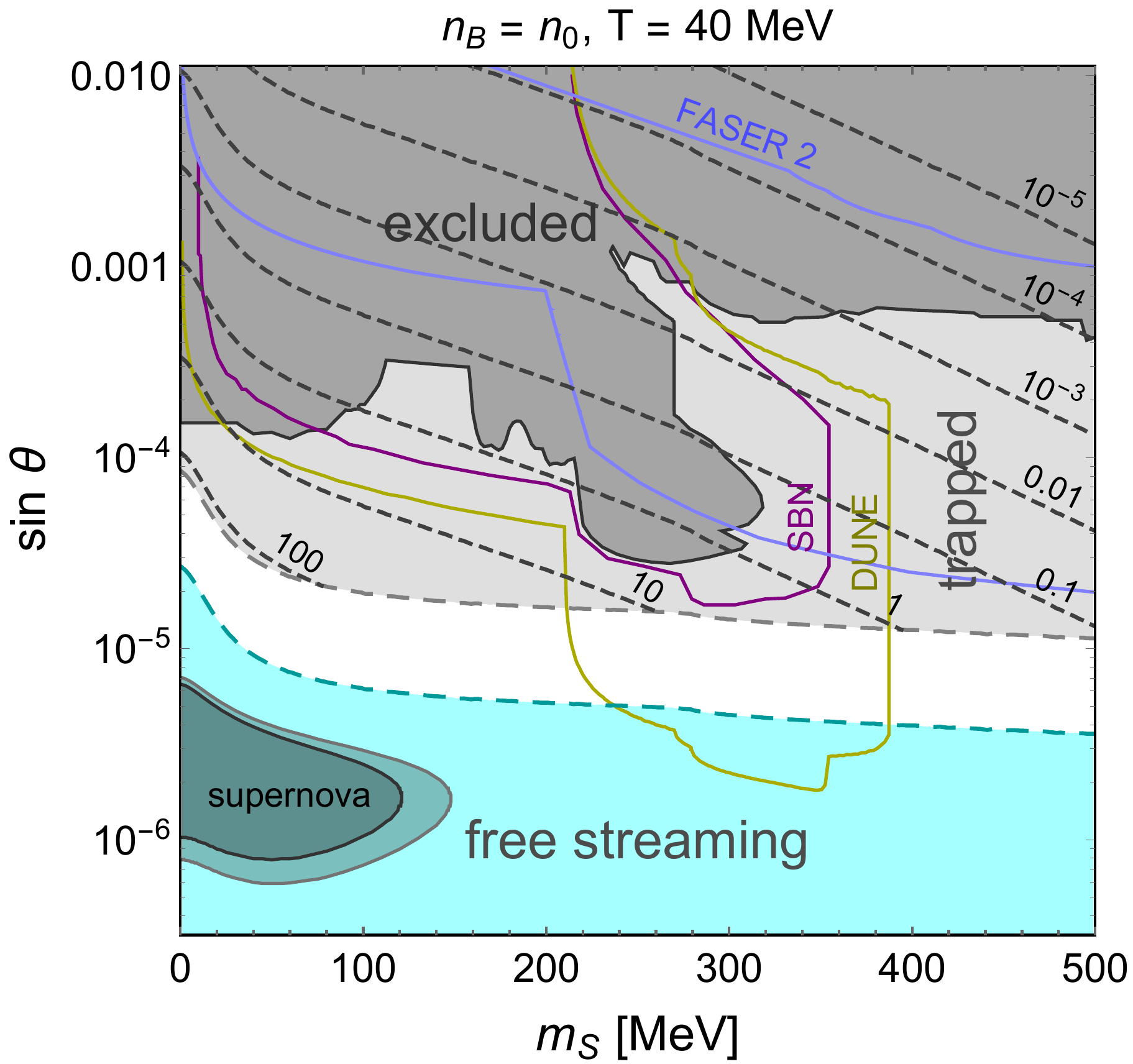}
\includegraphics[width=0.47\textwidth]{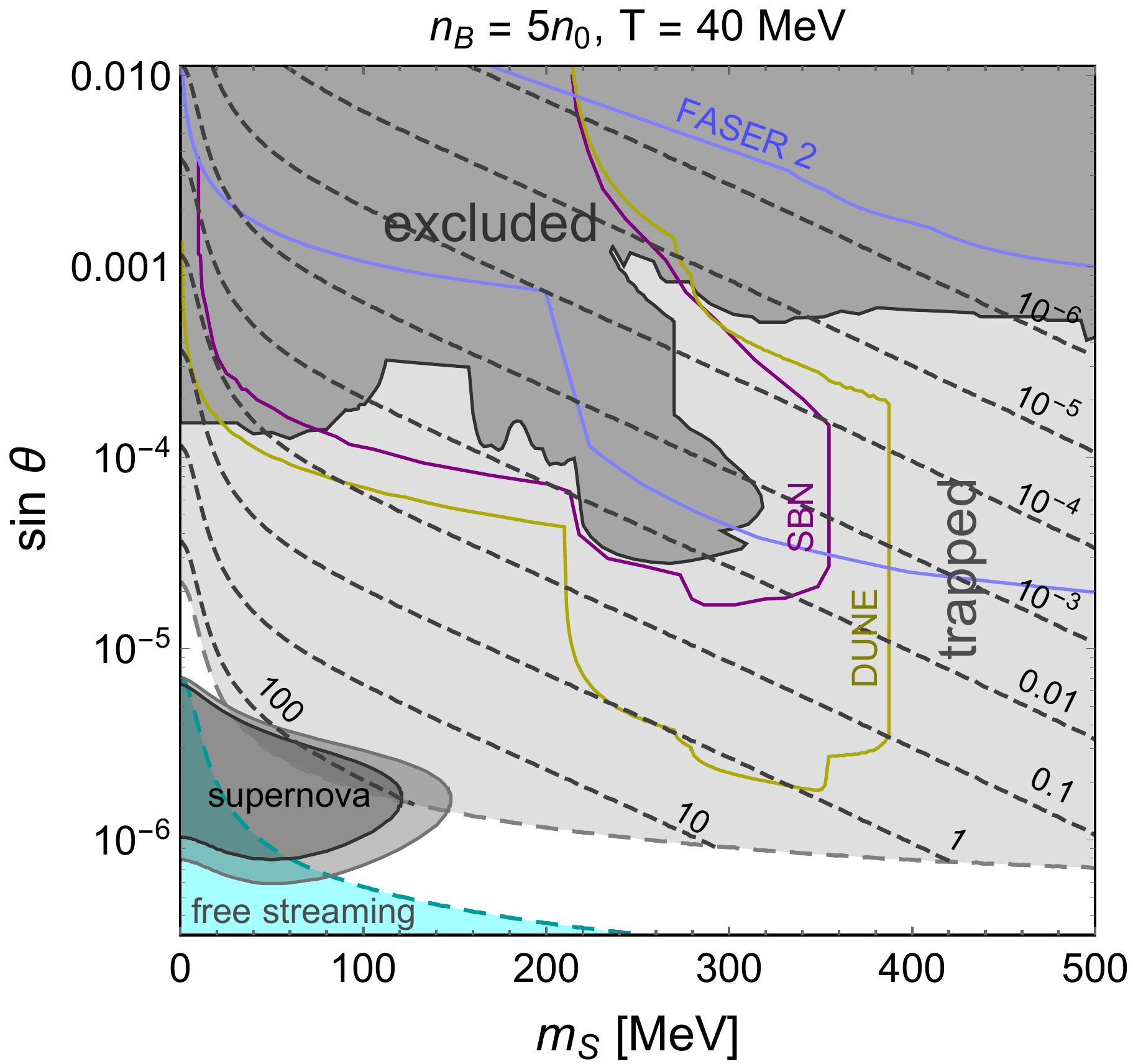}
\includegraphics[width=0.47\textwidth]{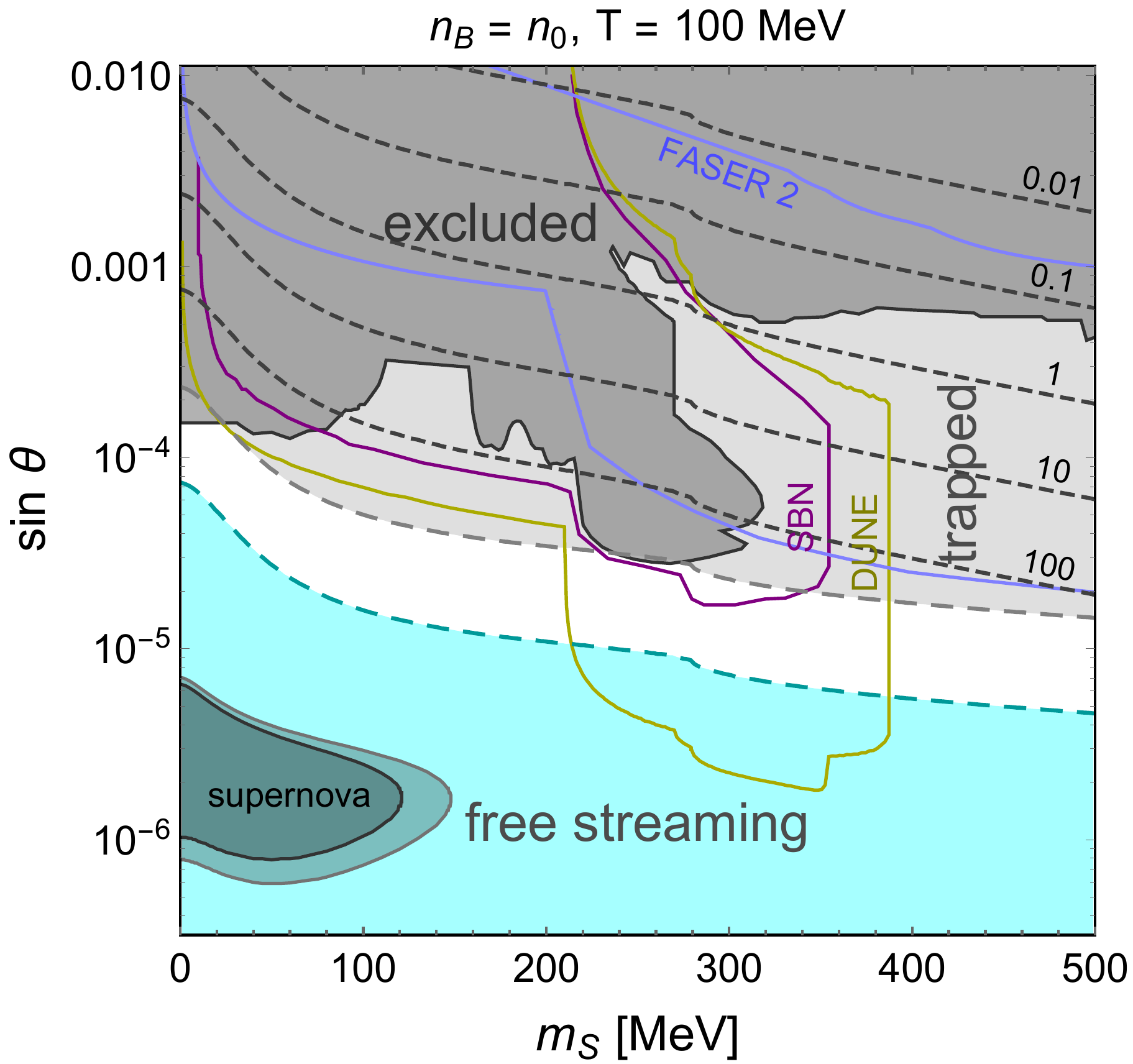}
\includegraphics[width=0.47\textwidth]{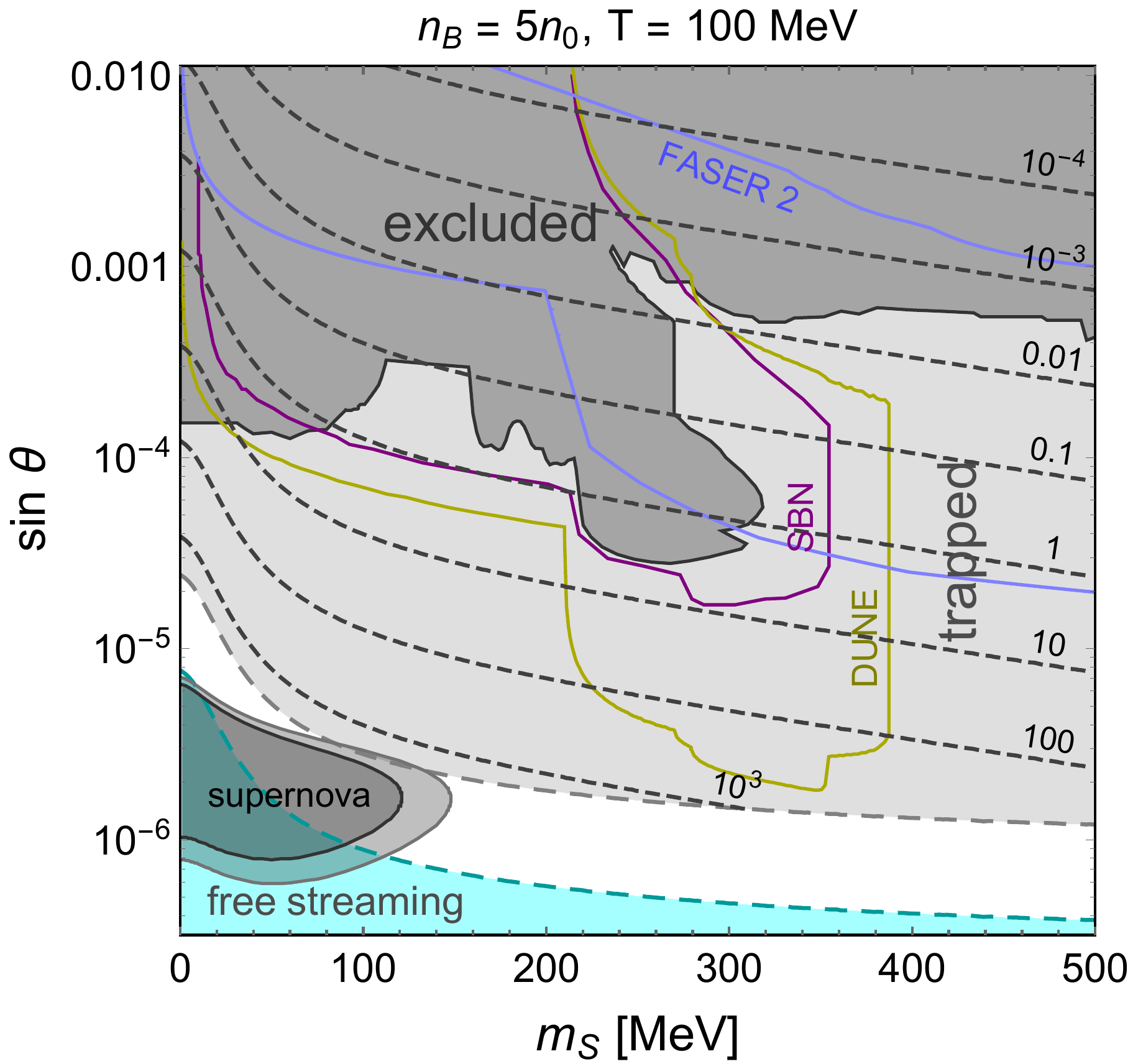}
\caption{Thermal conductivity ratio $\kappa_{S}/\kappa_{\nu}$ in the $m_S - \sin\theta$ plane, 
for the baryon density of $n_B =n_0$ (left) and $5n_0$ (right), and temperatures $T = 10$ MeV (upper), 40 MeV (middle) and 100 MeV (bottom). 
The labels are the same as in Figures~\ref{fig:S_mfp_ms_sintheta} and~\ref{fig:thermal_eq} with the current laboratory limits combined together as the black-shaded region.
}
\label{fig:kappa_ratio}
\end{figure}

While a trapped Bose gas of scalars can quickly even out temperature gradients in the merger remnant as shown in Figure~\ref{fig:thermal_eq}, a clear signature of BSM physics is only present if that rapid thermal conduction was unable to be attributed to SM particles, namely trapped neutrinos~\cite{Alford:2017rxf}.  In Figure~\ref{fig:kappa_ratio} we plot the ratio of the thermal conductivity $\kappa_S$ due to trapped scalars to that of trapped neutrinos $\kappa_\nu$, with baryon density $n_B = n_0$ and $5n_0$ respectively in the left and right panels.  For the sake of clarity, we have combined all the current laboratory limits, and show them in black in Figure~\ref{fig:kappa_ratio}. All other labels are the same as in Figures~\ref{fig:S_mfp_ms_sintheta} and~\ref{fig:thermal_eq}. 
In places where the ratio $\kappa_S/\kappa_\nu$ is larger than one, fast thermal equilibration is a signature of BSM physics, as no SM mechanism yields such a high thermal conductivity.  Figure~\ref{fig:kappa_ratio} shows that at moderate temperatures like $T=10\text{ MeV}$ (upper panels), the neutrino thermal conductivity is fairly high, since the neutrino MFP is long (neutrinos become free-streaming at temperatures just a few MeV lower) and so any contribution to the thermal conductivity from the scalars is hidden by the neutrino contribution.  One has to go to higher temperatures where the neutrino MFP shrinks to find situations where the scalars provide the dominant contribution to thermal equilibration inside the merger remnant.  In regions of the merger that reach $T=40 \text{ MeV}$ (middle panels), the scalars can quickly equilibrate temperature gradients, sometimes even one or two orders of magnitude faster than neutrinos.  This is only possible for scalars with masses below about 400 MeV, and couplings $10^{-6}\lesssim \sin{\theta} \lesssim 10^{-4}$.  If regions of the merger remnant were to reach temperatures of 100 MeV (lower panels), trapped scalars could quickly bring those regions to thermal equilibrium with nearby regions for a wide range of potential scalar masses. It is remarkable that some of the regions with $\kappa_S / \kappa_\nu >1$ can be directly probed at the future laboratory experiments such as SBN, DUNE and FASER 2.

\section{Discussion and conclusions} \label{sec:conclusion}

In this work, we have studied the effect of a CP-even scalar on the thermal transport properties of a neutron star merger remnant.  These scalars are produced in the hot, dense nuclear matter in the merger by nucleon bremsstrahlung processes.  The effect of the scalar depends on its MFP in the merger environment, which we plot in Figures~\ref{fig:S_mfp_nB_T} and~\ref{fig:S_mfp_ms_sintheta}. The parameter space of  $m_S-\sin{\theta}$ plane excluded by current laboratory limits and the supernova limits from SN1987A are also depicted in Figure~\ref{fig:S_mfp_ms_sintheta}. It is clear that taking into account the current constraints on the model, the MFP of the scalars could be as short as hundreds of micrometers. On the other hand, as the coupling between the scalar and nucleons becomes weaker, the MFP is free to grow arbitrarily large.

If the MFP is longer than the size of a particular macroscopic region of the merger remnant, the scalar will free-stream out of the region, cooling it.  If the scalar fails to encounter any region of the star where it has a short MFP throughout its trajectory, it will escape the remnant entirely, leading to a net loss of energy from the merger remnant as a whole.  In Figures~\ref{fig:S_emissivity_cooling_1d} and~\ref{fig:cooling}, we plot the time for a fluid element to cool to half of its original temperature by radiating scalar particles.  As long as the scalar is not too heavy, it can provide a mechanism to cool the hottest parts of the merger remnant on timescales shorter than the remnant lifetime. While it does not seem that this cooling affects the gravitational wave signal or the amount of ejecta significantly~\cite{Dietrich:2019shr}, the emission of scalars and their subsequent decay might be able to produce an additional electromagnetic signal, or the cooling to trigger a phase transition or affect the transport properties in the nuclear matter, which requires further investigation.

Perhaps more intriguingly, if the scalar is trapped in the merger remnant due to a short MFP, forming a Bose-Einstein distribution, it could contribute to transport processes in its own right. To this effect, we studied the contribution to the thermal conductivity coming from the Bose gas of CP-even scalar particles.  The timescale for the scalar to smooth temperature gradients in the merger remnant is plotted in Figures~\ref{fig:thermal_eq_1d},~\ref{fig:thermal_eq2}, and~\ref{fig:thermal_eq}.  We find that there is a significant amount of unconstrained space in the $m_S-\sin{\theta}$ plane where the scalars can cause fast thermal equilibration.  However, Figure~\ref{fig:kappa_ratio} shows that only some of that parameter space represents a case where the scalar thermal conductivity dominates over that coming from trapped neutrinos.  In particular, only at temperatures above 10 MeV can the scalar dominate the thermal equilibration of the merger remnant. This could potentially lead to some detectable signatures in future postmerger data. Furthermore, a part of this parameter space can also be directly probed at the future laboratory experiments such as SBN, DUNE and FASER 2 (cf.~the middle and lower panels of Figure~\ref{fig:kappa_ratio}), thus providing a complementary and independent probe of the model.

Before closing, we would like to comment on potential uncertainties and possible future improvements. First of all, in the nucleon bremsstrahlung process that produces $S$ particles, the nucleon scattering occurs via the strong interaction.  In our calculation of the matrix element for this process, following Ref.~\cite{Dev:2020eam} (and Refs.~\cite{1979ApJ...232..541F,Cullen:1999hc,Dent:2012mx,Mahoney:2017jqk,Shin:2021bvz} for bremsstrahlung processes producing other BSM particles), we used the OPE approximation.  It is well known that the OPE approximation underestimates the nucleon scattering cross section at low energies and overestimates it at high energies~\cite{Rrapaj:2015wgs}. In the future, this calculation of the $S$ production rate should be conducted in the soft-radiation approximation (SRA), which allows the $S$ production rate to be written in terms of the on-shell nucleon-nucleon scattering amplitude~\cite{Hanhart:2000ae,Rrapaj:2015wgs,Fortin:2021cog}.  For other bremsstrahlung processes, the SRA predicts a reduction of the production rate by a factor of a few compared to the rate calculated in the OPE approximation. 
    An alternative approach is to consider the exchange of heavier mesons or multiple pions, which also reduces the production rate relative to the OPE result~\cite{Carenza:2019pxu}.  If an improved treatment of the nuclear interaction were to decrease the emissivity of scalars compared to the OPE result we use here by, say, an order of magnitude (it would also increase the MFP by a similar factor due to the similarity of the phase space integrals for $Q$ and $\lambda$), our qualitative results would be unchanged, though the specific parameter space where, for example, trapped scalars would quickly smooth out temperature gradients in the system, would be different. 

Secondly, scalars with masses $m_S > 2m_{\pi^0}$ ($m_S > 2m_{\pi^{\pm}}$) can decay into neutral pions (charged pions).  For even higher mass scalars, decays into more than two pions or to heavier mesons are possible.  Nuclear matter in neutron star mergers likely also contains pions, either in a thermal population or in a condensate, either of which would alter the decay properties or even production mechanisms of a scalar particle.  In this work we allowed for a thermal population of pions at low density, which leads to Bose enhancement of the decay process $S\rightarrow \pi^+\pi^-$.  We treated the pions as free particles, although the inclusion of pion-nucleon interactions has been studied~\cite{Fore:2019wib}.  At high densities, the pions likely condense, but no calculation has yet been done of the decay rate of the scalar to pions in the presence of a pion condensate, nor is there a consensus on the density at which condensation occurs.  A unified, finite-temperature framework that can describe pions in and out of the condensed phase is needed.

Ultimately, to use neutron star mergers to find signatures of scalar particles, or to constrain them, will require numerical simulations which incorporate scalar particles.  In the free-streaming regime, this can be done by a leakage scheme where each fluid element has an energy loss term corresponding to the emissivity of scalars at the density and temperature of the fluid element.  This procedure was used to study cooling due to axion radiation~\cite{Dietrich:2019shr}.  In the regime where scalars are trapped, a thermal conduction term could be added to the relativistic hydrodynamic equations.  However, it is likely that for a scalar of a particular mass $m_S$ and coupling $\sin{\theta}$, there will be thermodynamic conditions in certain parts of the remnant where it will be trapped, but other parts of the remnant where it will free-stream.  Therefore, it would be necessary to develop a transport scheme for scalars similar to those developed for neutrinos in merger or supernova simulations~\cite{Foucart:2020qjb,Ardevol-Pulpillo:2018btx,Pan:2018vkx,Baiotti:2016qnr}. We hope the  results presented here will motivate the merger simulation community to implement the effects of light BSM particles like the one studied here into their simulations.

\section*{Acknowledgments}
We thank Mark Alford for discussions on thermal conductivity; Albino Perego for discussions on lepton number in neutron star merger simulations;  Katy Clough and Tim Dietrich for discussions on the possibility of including scalars in merger simulations; and Shyam Balaji, Bhaskar Dutta, Kevin Kelly, Rabi Mohapatra and Joe Silk for discussions on laboratory and astrophysical constraints on light scalars. The work of BD is supported in part by the U.S. Department of Energy under Grant No.~DE-SC0017987 and by a Fermilab Intensity Frontier Fellowship. This work was partly performed at the Aspen Center for Physics, which is supported by National Science Foundation grant PHY-1607611. The work of JFF is supported by NSERC.  The work of SPH is supported by the U.~S. Department of Energy grant DE-FG02-00ER41132 as well as the National Science Foundation grant No.~PHY-1430152 (JINA Center for the Evolution of the Elements).  The work of KS is supported in part by DOE Grant DE-SC0009956.  The work of YZ is supported by the National Natural Science Foundation of China under Grant No. 12175039, the 2021 Jiangsu Shuangchuang (Mass Innovation and Entrepreneurship) Talent Program No. JSSCBS20210144, and
the “Fundamental Research Funds for the Central Universities”.

\appendix 

\section{Mean free path for \texorpdfstring{$S$}{S} decay processes}
\label{appendix:decays}

In this appendix we list the MFPs for the $S$ particle through its decays with respect to the relevant SM particles.  For convenience, we define for particle species $i$
\begin{equation}
    \beta_i \equiv \left(1-\frac{m_S^2}{E_S^2}\right)\left(1-\frac{1}{\tau_i}\right),\label{eq:beta_i}
\end{equation}
where $\tau_i \equiv m_S^2/4m_i^2$. The matrix element for $S\rightarrow e^+e^-$ is
\begin{equation}
    \sum_{\text{spins}}\vert\mathcal{M}\vert_{e^+ e^-}^2=2\sqrt{2}G_F\sin^2{\theta}m_e^2(m_S^2-4m_e^2),
\end{equation}
and the MFP of the scalar from this decay is
\begin{align}
\lambda^{-1}_{e^+e^-} &= \int \frac{\mathop{d^3{\bf p}_+}}{(2\pi)^3}\frac{\mathop{d^3{\bf p}_-}}{(2\pi)^3}(2\pi)^4\delta^4(p_S-p_+-p_-)\frac{\sum_{\text{spins}}\vert\mathcal{M}\vert_{e^+ e^-}^2} {2^3E_sE_+E_-} \nonumber \\
& \quad \times (1-f(E_++\mu_e))(1-f(E_--\mu_e))\nonumber\\
& = \frac{\sin^2{\theta}}{8\pi v_{\rm EW}^2}\frac{m_e^2(m_S^2-4m_e^2)}{E_S\sqrt{E_S^2-m_S^2}}\frac{T}{1-e^{-E_S/T}} \Theta(m_S-2m_e) \nonumber \\
& \quad \times \ln{\left\{\frac{\cosh{(\mu_e/T)}+\cosh{[\frac{E_S}{2T}(1+\sqrt{\beta_{e}})]}}{\cosh{(\mu_e/T)}+\cosh{[\frac{E_S}{2T}(1-\sqrt{\beta_{e}})]}}\right\}},
\end{align}
where $p_{\pm}$ and $E_{\pm}$ are respectively the 4-momenta and energies of $e^\pm$, and $f$ denotes a Fermi-Dirac factor.  The MFP due to the decay $S\rightarrow\mu^+\mu^-$ is the same, but with the electron mass and chemical potential  replaced by those of the muon. The matrix element of the $S\rightarrow\gamma\gamma$ process is
\begin{equation}
\sum_{\text{spins}}\vert\mathcal{M}\vert_{\gamma\gamma}^2 = \frac{\sqrt{2}G_F\alpha^2m_S^4\sin^2{\theta}}{16\pi^2} \left| \sum_f N_C^f Q_f^2 A_{1/2} (\tau_f) + A_1 (\tau_W) \right|^2 \,,
\end{equation}
where the one-loop functions are 
\begin{eqnarray}
A_{1/2} (\tau) &  \equiv  & 2 \left[ \tau + (\tau-1) f(\tau) \right] \tau^{-2} \,, \\
A_1 (\tau) &\equiv& - \left[ 2 \tau^2 + 3\tau + 3 (2\tau-1) f(\tau) \right] \tau^{-2} \,, \\
{\rm with} \quad 
f(\tau) &  \equiv  & \left\{ \begin{array}{cl}
{\rm \arcsin}^2\sqrt{\tau} & ({\rm for}~\tau\leq 1) \\
-{\displaystyle \frac{1}{4}}\left[\log \left( \frac{1+\sqrt{1-1/\tau}}{1-\sqrt{1-1/\tau}}\right)-i\pi  \right]^2 & ({\rm for}~\tau>1)
\end{array}\right..
\label{fx}
\end{eqnarray}
The corresponding MFP is given by
\begin{align}
\lambda^{-1}_{\gamma\gamma} =& \int \frac{\mathop{d^3{\bf p}_1}}{(2\pi)^3}\frac{\mathop{d^3{\bf p}_2}}{(2\pi)^3}(2\pi)^4\delta^4(p_S-p_1-p_2)\frac{\sum_{\text{spins}}\vert\mathcal{M}\vert_{\gamma\gamma}^2}{2^3E_sE_1E_2}(1+g(E_1))(1+g(E_2)) \nonumber \\
=& \frac{\alpha^2\sin^2{\theta}}{128\pi^3 v_{\rm EW}^2}\frac{m_S^4}{E_S\sqrt{E_S^2-m_S^2}} \left| \sum_f N_C^f Q_f^2 A_{1/2} (\tau_f) + A_1 (\tau_W)  \right|^2 \nonumber\\ 
&\times  \frac{T}{1-e^{-E_S/T}}\ln{\left\{\frac{\sinh{\left[\frac{E_S}{4T}(1+\sqrt{\beta_{\gamma}})\right]}}{\sinh{\left[\frac{E_S}{4T}(1-\sqrt{\beta_{\gamma}})\right]}}\right\}},
\end{align}
where $p_{1,2}$ and $E_{1,2}$ are respectively the 4-momenta and energies of the two photons, and $g$ is the Bose distribution factor.

We also consider the possibility of $S$ decaying into pions.  We will treat the pions as a free Bose gas, neglecting for simplicity modifications to their dispersion relation due to the strong interaction (see Ref.~\cite{Fore:2019wib}).  Furthermore, the strong interaction would likely alter the pion mass~\cite{Balkin:2020dsr}, but we do not consider this effect, as pions have not yet been consistently included in a RMF theory.  The neutral pion has zero chemical potential, while the charged pions have chemical potential $\mu_{\pi^\mp}=\pm(\mu_n-\mu_p)$.
The matrix element for $S\rightarrow \pi^0\pi^0$ is
\begin{equation}
    \sum_{\text{spins}}\vert\mathcal{M}\vert_{\pi^0 \pi^0}^2= \frac{2}{81}\frac{\sin^2{\theta}}{v_{\rm EW}^2}\left(m_S^2+\frac{11}{2}m_{\pi^0}^2\right)^2,
\end{equation}
and the corresponding MFP is
\begin{align}
    \lambda_{\pi^0\pi^0}^{-1} =& \int \frac{\mathop{d^3{\bf p}_1}}{(2\pi)^3}\frac{\mathop{d^3{\bf p}_2}}{(2\pi)^3}(2\pi)^4\delta^4(p_S-p_1-p_2)\frac{\sum_{\text{spins}}\vert\mathcal{M}\vert_{\pi^0\pi^0}^2}{2^3E_SE_1E_2}(1+g(E_1))(1+g(E_2)) \nonumber \\
    =& \frac{\sin^2{\theta}}{324\pi v_{\rm EW}^2}\frac{\left(m_S^2+\frac{11}{2}m_{\pi^0}^2\right)^2}{E_S\sqrt{E_S^2-m_S^2}}\frac{T}{1-e^{-E_S/T}}\ln{\left\{\frac{\sinh{\left[\frac{E_S}{4T}(1+\sqrt{\beta_{\pi^0}})\right]}}{\sinh{\left[\frac{E_S}{4T}(1-\sqrt{\beta_{\pi^0}})\right]}}\right\}} \Theta(m_S-2m_{\pi^0}) \,,
\end{align}
where $p_{1,2}$ and $E_{1,2}$ respectively  the 4-momenta and energies for the two neutral pions. 
The matrix element for $S\rightarrow \pi^+\pi^-$ is
\begin{equation}
    \sum_{\text{spins}}\vert\mathcal{M}\vert_{\pi^+ \pi^-}^2= \frac{4}{81}\frac{\sin^2{\theta}}{v_{\rm EW}^2}\left( m_S^2+\frac{11}{2}m_{\pi^{\pm}}^2 \right)^2,
\end{equation}
and the corresponding MFP is
\begin{align}
    \lambda_{\pi^+\pi^-}^{-1} =& \int \frac{\mathop{d^3{\bf p}_+}}{(2\pi)^3}\frac{\mathop{d^3{\bf p}_-}}{(2\pi)^3}(2\pi)^4\delta^4(p_S-p_+-p_-)\frac{\sum_{\text{spins}}\vert\mathcal{M}\vert_{\pi^+ \pi^-}^2}{2^3E_sE_+E_-}\nonumber \\
    & \quad \times (1+g(E_++\mu_{\pi^-}))(1+g(E_--\mu_{\pi^-}))\nonumber\\
    =& \frac{\sin^2{\theta}}{324\pi v_{\rm EW}^2}\frac{\left(m_S^2+\frac{11}{2}m_{\pi^{\pm}}^2\right)^2}{E_S\sqrt{E_S^2-m_S^2}}\frac{T}{1-e^{-E_S/T}} \Theta(m_S-2m_{\pi^{\pm}}) \nonumber\\
    &\times\ln{\left\{\frac{\cosh{(\mu_{\pi^-}/T)}-\cosh{\left[\frac{E_S}{2T}(1+\sqrt{\beta_{\pi^{\pm}}})\right]}}{\cosh{(\mu_{\pi^-}/T)}-\cosh{\left[\frac{E_S}{2T}(1-\sqrt{\beta_{\pi^{\pm}}})\right]}}\right\}} \,,
\end{align}
where $p_{\pm}$ and $E_{\pm}$ respectively  the 4-momenta and energies for $\pi^\pm$. This formula for scalar decaying into charged pions is only valid when the pions have not formed a Bose-Einstein condensate.

\section{Dimensionless functions in nucleon bremsstrahlung matrix element}\label{appendix:nondimensional}

The dimensionless functions in the nucleon bremsstrahlung matrix elements were derived in Ref.~\cite{Dev:2020eam} and are reproduced here: 
\begin{align}
    I_A^{nn/pp} &= \frac{\mathbf{k}^4}{(\mathbf{k}^2+m_{\pi}^2)^2}+\frac{\mathbf{l}^4}{(\mathbf{l}^2+m_{\pi}^2)^2}-\frac{\mathbf{k}^2\mathbf{l}^2-2(\mathbf{k}\cdot \mathbf{l})^2}{(\mathbf{k}^2+m_{\pi}^2)(\mathbf{l}^2+m_{\pi}^2)},\\
    I_B^{nn/pp} &= \left(m_S^2+\frac{11}{2}m_{\pi}^2\right)^2\left[\frac{\mathbf{k}^4}{(\mathbf{k}^2+m_{\pi}^2)^4}+\frac{\mathbf{l}^4}{(\mathbf{l}^2+m_{\pi}^2)^4}-\frac{\mathbf{k}^2\mathbf{l}^2-2(\mathbf{k}\cdot \mathbf{l})^2}{(\mathbf{k}^2+m_{\pi}^2)^2(\mathbf{l}^2+m_{\pi}^2)^2}\right],\\
    I_C^{nn/pp} &= \left(m_S^2+\frac{11}{2}m_{\pi}^2\right)\bigg\{ \frac{\mathbf{k}^4}{(\mathbf{k}^2+m_{\pi}^2)^3}+\frac{\mathbf{l}^4}{(\mathbf{l}^2+m_{\pi}^2)^3}\nonumber\\
    &\hspace{5cm}-\frac{1}{2}\frac{(\mathbf{k}^2+\mathbf{l}^2+2m_{\pi}^2)[\mathbf{k}^2\mathbf{l}^2-2(\mathbf{k}\cdot \mathbf{l})^2]}{(\mathbf{k}^2+m_{\pi}^2)^2(\mathbf{l}^2+m_{\pi}^2)^2} \bigg\},\\
    I_A^{np} &= \frac{\mathbf{k}^4}{(\mathbf{k}^2+m_{\pi}^2)^2}+\frac{4\mathbf{l}^4}{(\mathbf{l}^2+m_{\pi}^2)^2}+2\frac{\mathbf{k}^2\mathbf{l}^2-2(\mathbf{k}\cdot \mathbf{l})^2}{(\mathbf{k}^2+m_{\pi}^2)(\mathbf{l}^2+m_{\pi}^2)},\\
    I_B^{np} &= \left(m_S^2+\frac{11}{2}m_{\pi}^2\right)^2\left[\frac{\mathbf{k}^4}{(\mathbf{k}^2+m_{\pi}^2)^4}+\frac{4\mathbf{l}^4}{(\mathbf{l}^2+m_{\pi}^2)^4}+2\frac{\mathbf{k}^2\mathbf{l}^2-2(\mathbf{k}\cdot \mathbf{l})^2}{(\mathbf{k}^2+m_{\pi}^2)^2(\mathbf{l}^2+m_{\pi}^2)^2}\right],\\
    I_C^{np} &= \left(m_S^2+\frac{11}{2}m_{\pi}^2\right)\bigg\{ \frac{\mathbf{k}^4}{(\mathbf{k}^2+m_{\pi}^2)^3}+\frac{4\mathbf{l}^4}{(\mathbf{l}^2+m_{\pi}^2)^3}\nonumber\\
    &\hspace{5cm}+\frac{(\mathbf{k}^2+\mathbf{l}^2+2m_{\pi}^2)[\mathbf{k}^2\mathbf{l}^2-2(\mathbf{k}\cdot \mathbf{l})^2]}{(\mathbf{k}^2+m_{\pi}^2)^2(\mathbf{l}^2+m_{\pi}^2)^2} \bigg\},
\end{align}
where $\mathbf{k}=\mathbf{p}_2-\mathbf{p}_4$ and $\mathbf{l}=\mathbf{p}_2-\mathbf{p}_3$ measure the nucleon 3-momentum transfers. 
We do not choose to average over the direction of the momentum of the emitted BSM particle, as was done for axion emission, for example, in Ref.~\cite{Brinkmann:1988vi}.

\section{MFP phase space integral}\label{appendix:MFP}
In this appendix, we detail the calculation of the MFP of the $S$ particle due to absorption via $n+n+S\rightarrow n +n$, and then comment on the other two inverse bremsstrahlung absorption processes.  The phase space integral in Eq.~(\ref{eq:MFP_integral}) can be reduced to a five-dimensional integral via the following steps.  First, we neglect the 3-momentum of $S$ in the momentum-conserving $\delta$-function; however, we keep its energy in the energy $\delta$-function.  We perform the 3-dimensional integral over $\mathbf{p}_1$ using the 3-momentum $\delta$-function and then we convert each of the remaining 3-momentum integrals to spherical coordinates.  We choose our $z$-axis to align with $\mathbf{p}_2$, and then choose the $xz$-plane to be the plane in which $\mathbf{p}_2$ and $\mathbf{p}_3$ lie, with angle $r\equiv\cos{\theta}$ between them.  This choice of axes allows us to write $\mathbf{p}_2$, $\mathbf{p}_3$, and $\mathbf{p}_4$ as
\begin{align}
    \mathbf{p}_2 &= p_2(0,0,1),\nonumber\\
    \mathbf{p}_3 &= p_3(\sqrt{1-r^2},0,r),\nonumber\\
    \mathbf{p}_4 &= p_4(\sqrt{1-s^2}\cos{\phi},\sqrt{1-s^2}\sin{\phi},s),
\end{align}
where $s$ is the cosine of the angle between $\mathbf{p}_2$ and $\mathbf{p}_4$. 
This choice of coordinate system renders three of the angular integrals trivial, yielding a factor of $8\pi^2$.  The MFP is now given by
\begin{align}
    \lambda_{nn}^{-1} =& \frac{\alpha_{\pi}^2f^4\sin^2{\theta}}{16\pi^4m_*^2E_S}\int_0^{\infty}\mathop{dp_2}\mathop{dp_3}\mathop{dp_4}\int_{-1}^1\mathop{dr}\mathop{ds}\int_0^{2\pi}\mathop{d\phi} p_2^2p_3^2p_4^2 \frac{f_1f_2(1-f_3)(1-f_4)}{E_1^*E_2^*E_3^*E_4^*} \nonumber  \\
    &\times\delta(E_1^*+E_2^*-E_3^*-E_4^*+E_S)\left(\frac{y_{hnn}^2m_S^4}{4E_S^4}I_A^{nn} + \frac{m_*^2}{192v_{\rm EW}^2}I_B^{nn} +\frac{2y_{hnn}m_S^2m_*}{9E_S^2v_{\rm EW}}I_C^{nn}      \right) \,,
\end{align}
where $E_1^*=\sqrt{(\mathbf{p}_3+\mathbf{p}_4-\mathbf{p}_2)^2+m_*^2}$.  Now, we integrate over $\phi$, using the energy $\delta$-function.  The argument of the $\delta$-function can only go through zero if
\begin{equation}
    E_3^*+E_4^*>E_2^*+E_S. \label{eq:mfp_condition_1}
\end{equation}
If
\begin{align}
    \vert m_*^2+E_S^2/2+E_3^*E_4^*-E_2^*E_3^*-&E_2^*E_4^*+E_S(E_2^*-E_3^*-E_4^*)+p_2p_3r+p_2p_4s-p_3p_4rs\vert\nonumber\\ &\leq p_3p_4\sqrt{1-r^2}\sqrt{1-s^2},\label{eq:mfp_condition_2}
\end{align}
then there are two values of $\phi$ between $0$ and $2\pi$ which are picked out by the $\delta$-function.  Both values $\phi_0$ have the same cosine.  If Eq.~(\ref{eq:mfp_condition_2}) does not hold, then the integral is zero.  Condition (\ref{eq:mfp_condition_1}) can be turned into an upper bound on $p_2$ along with the further constraint
\begin{equation}
    E_3^*+E_4^*-E_S\geq m_*. \label{eq:mfp_condition_3}
\end{equation}
Therefore the MFP of the scalar $S$ from the neutron-neutron inverse bremsstrahlung process can be written as 
\begin{align}
    \lambda_{nn}^{-1} =& \frac{\alpha_{\pi}^2f^4\sin^2{\theta}}{8\pi^4m_*^2E_S}\int_0^{\infty}\mathop{dp_3}\mathop{dp_4}\int_0^{\sqrt{(E_3^*+E_4^*-E_S)^2-m_*^2}}\mathop{dp_2}\int_{-1}^1\mathop{dr}\mathop{ds} \frac{p_2^2p_3^2p_4^2}{E_2^*E_3^*E_4^*}f_1f_2(1-f_3)(1-f_4)\nonumber\\
    &\times \left(\frac{y_{hnn}^2m_S^4}{4E_S^4}I_A^{nn} + \frac{m_*^2}{192v_{\rm EW}^2}I_B^{nn} +\frac{2y_{hnn}m_S^2m_*}{9E_S^2v_{\rm EW}}I_C^{nn}      \right) \nonumber \\
    &\times \bigg(p_3^2p_4^2(1-r^2)(1-s^2)-(m_*^2+E_S^2/2+E_3^*E_4^*-E_2^*E_3^*-E_2^*E_4^* \nonumber \\
    &+E_S(E_2^*-E_3^*-E_4^*)+p_2p_3r+p_2p_4s-p_3p_4rs)^2\bigg)^{-1/2} \,,
\end{align}
with the restrictions (\ref{eq:mfp_condition_2}) and (\ref{eq:mfp_condition_3}).  The Fermi-Dirac factors are 
\begin{align}
    f_1 &= \frac{1}{1+e^{(E_3^*+E_4^*-E_2^*-E_S-\mu_n^*)/T}} \, , \nonumber \\ 
    f_2 & = \frac{1}{1+e^{(E_2^*-\mu_n^*)/T}} \, ,\nonumber\\
    1-f_3 &= \frac{1}{1+e^{-(E_3^*-\mu_n^*)/T}} \, , \nonumber\\
    1-f_4 & = \frac{1}{1+e^{-(E_4^*-\mu_n^*)/T}} \, .
\end{align}
The integration variables $p_2, p_3, p_4$ can be nondimensionalized by $m_*$, if desired.

The MFP due to absorption via $p+p+S\rightarrow p+p$ is the same, except that the neutron chemical potential is replaced by that of the proton.  The MFP due to the absorption  $n+p+S\rightarrow n+p$ has the neutron chemical potentials of particles 2 and 4 replaced by the proton chemical potential, and the result is divided by four (or, $\lambda_{np}^{-1}$ is multiplied by a factor of four) due to the difference in matrix elements (cf.~Eqs.~(\ref{eq:nn_matrix}) and (\ref{eq:np_matrix})).

\section{Emissivity phase space integral}

To evaluate the phase space integral in Eq.~(\ref{eq:emissivity_integral}) for, say, $n+n\rightarrow n+n+S$, we neglect the 3-momentum of the scalar particle in the 3-momentum conserving $\delta$-function and then, considering the integral over $\mathbf{p}_S$ in spherical coordinates, integrate over the angular part, giving a factor of $4\pi$.  To simplify the 3-momentum conserving $\delta$-function, we change the coordinate system from $\{\mathbf{p}_1,\mathbf{p}_2, \mathbf{p}_3, \mathbf{p}_4\}$ to $\{\mathbf{p}_+,\mathbf{p}_-,\mathbf{a},\mathbf{b}\}$ where 
\begin{align}
    \mathbf{p}_+ & = \frac12 (\mathbf{p}_1+\mathbf{p}_2) \,, & 
    \mathbf{a} & = \mathbf{p}_3-\mathbf{p}_+ \,, \nonumber \\
    \mathbf{p}_-&= \frac12 (\mathbf{p}_1-\mathbf{p}_2) \,, &
    \mathbf{b}&=\mathbf{p}_4-\mathbf{p}_+ \,.
\end{align}
The Jacobian gives a factor of 8, and the emissivity becomes
\begin{align}
    Q_{nn} =& \frac{\alpha_{\pi}^2f^4\sin^2{\theta}}{32\pi^8m_*^2}\int\mathop{d^3 {\bf p}_+}\mathop{d^3{\bf p}_-}\mathop{d^3{\bf a}}\mathop{d^3{\bf b}}\int_0^{\infty}\mathop{dp_S} p_S^2\delta(E_1^*+E_2^*-E_3^*-E_4^*-E_S) \delta^{(3)}(\mathbf{a}+\mathbf{b}) \nonumber \\
    &\times \frac{f_1f_2(1-f_3)(1-f_4)}{E_1^*E_2^*E_3^*E_4^*}\left(\frac{y_{hnn}^2m_S^4}{E_S^4}I_A^{nn} + \frac{m_*^2}{81v_{\rm EW}^2}I_B^{nn} +\frac{2y_{hnn}m_S^2m_*}{9E_S^2v_{\rm EW}}I_C^{nn}\right).
\end{align}
Now we do the integral over $\mathbf{b}$ using the 3-momentum $\delta$-function and then the integral over $p_S$, making use of the energy $\delta$-function.  This yields
\begin{align}
    Q_{nn} =& \frac{\alpha_{\pi}^2f^4\sin^2{\theta}}{32\pi^8m_*^2}\int\mathop{d^3 {\bf p}_+}\mathop{d^3{\bf p}_-}\mathop{d^3{\bf a}}(E_1^*+E_2^*-E_3^*-E_4^*)\sqrt{(E_1^*+E_2^*-E_3^*-E_4^*)^2-m_S^2}\nonumber\\
    &\times \frac{f_1f_2(1-f_3)(1-f_4)}{E_1^*E_2^*E_3^*E_4^*}\left(\frac{y_{hnn}^2m_S^4}{4E_S^4}I_A^{nn} + \frac{m_*^2}{192v_{\rm EW}^2}I_B^{nn} +\frac{2y_{hnn}m_S^2m_*}{9E_S^2v_{\rm EW}}I_C^{nn}\right),
    \end{align}
    with the constraint 
\begin{equation}
    E_1^*+E_2^*-E_3^*-E_4^*\geq m_S.\label{eq:Q_restriction}
\end{equation}
We choose a coordinate system such that
\begin{align}
    \mathbf{a}&=a(0,0,1),\nonumber\\
    \mathbf{p_-}&=p_-(\sqrt{1-r^2},0,r),\nonumber\\
    \mathbf{p_+}&=p_+(\sqrt{1-s^2}\cos{\phi},\sqrt{1-s^2}\sin{\phi},s),
\end{align}
with $r$ and $s$ here respectively the cosines of the angles between ${\bf p}_{\mp}$ and ${\bf a}$, and then do the integrals over three trivial angles, yielding
\begin{align}
    Q_{nn} =& \frac{\alpha_{\pi}^2f^4\sin^2{\theta}}{4\pi^6m_*^2}\int_0^{\infty}\mathop{dp_+}\mathop{dp_-}\mathop{da}\int_{-1}^{1}\mathop{dr}\mathop{ds}\int_0^{2\pi}\mathop{d\phi}p_+^2p_-^2a^2(E_1^*+E_2^*-E_3^*-E_4^*)\nonumber \\
    &\times \sqrt{(E_1^*+E_2^*-E_3^*-E_4^*)^2-m_S^2} \frac{f_1f_2(1-f_3)(1-f_4)}{E_1^*E_2^*E_3^*E_4^*}\nonumber\\
    &\times\left(\frac{y_{hnn}^2m_S^4}{4E_S^4}I_A^{nn} + \frac{m_*^2}{192v_{\rm EW}^2}I_B^{nn} +\frac{2y_{hnn}m_S^2m_*}{9E_S^2v_{\rm EW}}I_C^{nn}\right),
\end{align}
where the condition (\ref{eq:Q_restriction}) holds, and $E_S=E_1^*+E_2^*-E_3^*-E_4^*$.  In the functions $I_{A,B,C}$ (cf.~Appendix~\ref{appendix:nondimensional}), the momentum transfer variables are \begin{align}
\mathbf{k}^2=a^2+p_-^2-2p_-ar, \quad \mathbf{l}^2=a^2+p_-^2+2p_-ar, \quad \mathbf{k}\cdot\mathbf{l}=p_-^2-a^2.    
\end{align}
In addition, the former integration variables are now expressed as functions of the new integration variables
\begin{align}
  {\bf p}_1^2 & = p_-^2+p_-^2+2p_+p_-rs+2p_+p_-\sqrt{1-r^2}\sqrt{1-s^2}\cos{\phi} \,, \nonumber \\
  {\bf p}_2^2 & = p_+^2p_-^2-2p_+p_-rs-2p_+p_-\sqrt{1-r^2}\sqrt{1-s^2}\cos{\phi} \,, \nonumber \\ 
  {\bf p}_3^2 & = p_+^2+a^2+2p_+as \,, \nonumber \\ 
  {\bf p}_4^2 & = p_+^2+a^2-2p_+as \,.
\end{align}

\section{Higher-order corrections for scalar production amplitude}
\label{appendix:relativistic_corrections}

We perform in this appendix a more complete calculation for the nucleon bremsstrahlung process Eq.~(\ref{eqn:brem}), following the procedure in Ref.~\cite{Dev:2020eam}. In this section, we use $m_N$ to label the nucleon mass, which in a RMF theory is replaced with the Dirac effective mass $m_*$ (see Section~\ref{sec:nuclear_matter_in_mergers}).  For the diagrams (a) through (d) in Figure~\ref{fig:diagram}, the amplitudes are
\begin{eqnarray}
{\cal M}_{a,b,c,d} \propto \frac{1}{(p_i \mp k_s)^2 -m_N^2} \,,
\end{eqnarray}
where the ``$-$'' and ``$+$'' signs are respectively for $i=1,\,2$ and  $i=3,\,4$, and 
\begin{eqnarray}
(p_i + k_s)^2 -m_N^2 &=& \mp 2 (p_i \cdot k_s) + m_S^2 \nonumber \\
&=& \mp 2 m_N E_S \left[ 1 + \frac{E_i - m_N}{m_N} 
- \frac{{\bf p}_i \cdot {\bf k}_S}{m_N E_S} \mp \frac{m_S^2}{2m_N E_S} \right] \,.
\end{eqnarray}
For convenience, we define the small dimensionless parameters
\begin{eqnarray}
	a_i \equiv \frac{E_i - m_N}{m_N} \,, \quad
	b_i \equiv -\frac{{\bf p}_i \cdot {\bf k}_S}{m_N E_S} \,, \quad
	c_i \equiv \mp c \,,
\end{eqnarray}
where
\begin{equation}
    c \equiv \frac{m_S^2}{2m_N E_S}\,.
\end{equation}
Then the propagators become
\begin{eqnarray}
	\frac{1}{(p_i \mp k_s)^2 -m_N^2} &=& \mp \frac{1}{2m_N E_S}
	\frac{1}{1 + a_i + b_i + c_i} \,,\\
{\rm with}\quad 
	\frac{1}{1 + a_i + b_i + c_i} &=&
	1 + \sum_{k=1}^{\infty} (-1)^{k} \left( a_i^{k} + b_i^{k} + c_i^{k} \right) \nonumber \\
	&& + \sum_{k=2}^{\infty} \sum_{j=1}^{k-1} (-1)^{k} \frac{k!}{j!(k-j)!}  \left( a_i^{j} b_i^{k-j} + a_i^{j} c_i^{k-j} + b_i^{j} c_i^{k-j} \right) \nonumber \\
	&& + \sum_{k=3}^{\infty} \sum_{j,m=1}^{k-1} (-1)^{k} \frac{k!}{j!m!(k-j-m)!}   a_i^{j} b_i^{m} c_i^{k-j-m} \,.
		\label{eqn:series}
\end{eqnarray}

When we sum up the (a) through (d) diagrams, the ``1'' term in Eq.~(\ref{eqn:series}) cancels out, which is the cancellation mentioned in Ref.~\cite{Dev:2020eam}. For the $a_i$ terms, let us first consider the leading order terms with $k=1$, which produce
\begin{eqnarray}
\label{eqn:ai1}
	- \sum_i \mp a_i &=& 
	\frac{E_1 - m_N}{m_N} + \frac{E_2 - m_N}{m_N} - \frac{E_3 - m_N}{m_N} - \frac{E_4 - m_N}{m_N}  = a \,, \\
	\label{eqn:bi1}
	- \sum_i \mp b_i &=& 
	-\frac{{\bf p}_1 \cdot {\bf k}_S}{m_N E_S} - \frac{{\bf p}_2 \cdot {\bf k}_S}{m_N E_S}
	+ \frac{{\bf p}_3 \cdot {\bf k}_S}{m_N E_S} + \frac{{\bf p}_4 \cdot {\bf k}_S}{m_N E_S} 
	= - b \,, \\
	\label{eqn:ci1}
	(-1)^k \sum_i \mp c_i^k &=& 
	-2 \left( 1 + (-1)^{k-1} \right) c^k \,,
\end{eqnarray}
where we introduced
\begin{eqnarray}
a \equiv \frac{E_S}{m_N} \,, \quad
b \equiv \frac{E_S^2 -m_S^2}{m_N E_S} \,.
\end{eqnarray}
For the high-order terms $a_i^{k}$ with $k=2,3,4...$ in Eq.~(\ref{eqn:series}), in the center-of-mass frame and in the limit of $m_S$, $E_S \to 0$, $\Delta E_i = E_i - m_N$ are all the same; as a result the $(E_i-m_N)^k$ terms with $k\geq2$ cancel out. In other words, these higher-order terms are highly suppressed by powers of $(E_S/m_N)^k$. 
For the higher-order terms $b_i^k$ (with $k = 2,3,4...$) in the center-of-mass frame and in the limit of $m_S,\,E_S \to 0$, the scalar products
\begin{eqnarray}
	{\bf p}_i \cdot {\bf k}_S = | {\bf p}_i | |{\bf k}_S| \cos\theta_i \,,
\end{eqnarray}
where $\theta_i$ are the angles between ${\bf p}_i$ and ${\bf k}_S$, then
\begin{eqnarray}
	\sum_i \mp b_i^k &\propto& 
	\left( \frac{| {\bf p}_i | |{\bf k}_S|}{m_N E_S} \right)^k
	\left( \cos^k\theta_1 + \cos^k\theta_2 - \cos^k\theta_3 - \cos^k\theta_4 \right) \,.
\end{eqnarray}
Integrating over all the parameter space, the angles $\theta_{1,2,3,4}$ are averaged, thus the factor 
\begin{eqnarray}
	\label{eqn:angle}
	\left( \cos^n\theta_1 + \cos^n\theta_2 - \cos^n\theta_3 - \cos^n\theta_4 \right) \to 0 \,.
\end{eqnarray}
In other words, the higher-order terms of $b_i^k$ (with $k\geq 2$) are also highly suppressed by $(E_S/m_N)^k$. 

For the cross terms $a_i^{j} b_i^{k-j}$ in Eq.~(\ref{eqn:series}), let us first consider the first term with $j=1$ and $k=2$, which turns out to be
\begin{eqnarray}
\label{eqn:aibi1}
	\sum_i \mp 2a_i b_i &=& 
	\frac{2}{m_N^2 E_S} \left[ E_1 {\bf p}_1 \cdot {\bf k}_S + E_2 {\bf p}_2 \cdot {\bf k}_S - E_3 {\bf p}_3 \cdot {\bf k}_S - E_4 {\bf p}_4 \cdot {\bf k}_S \right] 
	- \frac{2 {\bf k}_S^2}{m_N E_S} \,.
\end{eqnarray}
To evaluate the $E_i {\bf p}_i\cdot {\bf k}_S$ terms, we take the limit of $E_i - m_N \to \Delta E \simeq {\bf p}_{N}^{\,2}/2m_N^2$ in the center-of-mass frame with the limit of $m_S,\, E_S \to 0$. In this limit, Eq.~(\ref{eqn:aibi1}) can be greatly simplified to 
\begin{eqnarray}
\sum_i \mp 2a_i b_i &=& 2bd \,,
\end{eqnarray}
where we have defined $d \equiv \Delta E/m_N$. Generalizing to higher-order terms of the form $a_i^{k-1} b_i$ (with $k\geq2$) in Eq.~(\ref{eqn:series}), we have
\begin{eqnarray}
(-1)^k \sum_i \mp k a_i^{k-1} b_i &=& 
(-1)^k k bd^{k-1} \,.
\end{eqnarray}
The $a_i^{k-j} b_i^j$ terms (with $j \geq2$) are vanishing as
\begin{eqnarray}
	a_i^m b_i^n \propto 
	\left( \frac{\Delta E}{m_N} \right)^{m}
	\left( \cos^n\theta_1 + \cos^n\theta_2 - \cos^n\theta_3 - \cos^n\theta_4 \right) \,.
\end{eqnarray}
Similarly it is straightforward to obtain that 
\begin{eqnarray}
	\sum_i \mp a_i^j c_i^{k-j} &=& \frac12 c^{k-j} 
	\left[ 4 d^{j} \left( 1 + (-1)^{k-j-1} \right) - \delta_{j1} a \left( 1 + (-1)^{k-j} \right) \right] \,, \\
	\sum_i \mp b_i^j c_i^{k-j} &=& \frac12b c^{k-j}  \delta_{j1} \left( 1 + (-1)^{k-j} \right)\,, \\
	\sum_i \mp a_i^j b_i^m c_i^{k-j-m} &=& \frac12 b d^j  c^{k-j-m} \delta_{m1} \left( 1 + (-1)^{k-j-m} \right) \,.
\end{eqnarray}

Summing up all the contributions, we get
\begin{eqnarray}
	&& a  -b 
	-4 \sum_{n=1}^\infty c^{2n-1} 
	+\sum_{n=1}^\infty (-1)^{n+1} (n+1) d^n b \nonumber \\
	&& + 4 \sum_{m = 1}^\infty \sum_{n = 1}^\infty (-1)^{m+2n-1} C_{m+2n-1}^m d^m c^{2n-1} 
	+ \sum_{n=1}^\infty (1+2n) a c^{2n} \nonumber \\
	&& - \sum_{n=1}^\infty (1+2n) b c^{2n} 
	+ \sum_{m = 1}^\infty \sum_{n = 1}^\infty (-1)^{m+2n+1} \frac{(m+2n+1)!}{m!(2n)!} d^m b c^{2n} \nonumber \\
	&=& \frac{(1+c^2)a}{(1-c^2)^2}
	- \frac{[(1+d)^2 + c^2]b}{[(1+d)^2-c^2]^2} - \frac{4c}{(1+d)^2-c^2} \,.
\end{eqnarray}
As all the parameters $a$, $b$, $c$ and $d$ are very small, the summation can be approximated to be
\begin{eqnarray}
a - b - 4c = \frac{E_S}{m_N} - \frac{E_S^2 - m_S^2}{m_N E_S} - \frac{2 m_S^2}{m_N E_S}
= - \frac{m_S^2}{m_N E_S}\,,
\end{eqnarray}
which corresponds to the lowest order terms $a_i$, $b_i$ and $c_i$ in Eq.~(\ref{eqn:series}). This means that we have to scale the amplitudes for the (a) through (d) diagrams in Ref.~\cite{Dev:2020eam} by (roughly) a factor of 1/2.

\section{Neutrino thermal conductivity}
\label{app:neutrino_kappa}
The thermal conductivity of a neutrino gas has been calculated for one neutrino species in the degenerate limit~\cite{Goodwin:1982hy} and at arbitrary neutrino degeneracy (though assuming nondegenerate nuclear matter)~\cite{vandenHorn:1984zz}.  In mergers, there exist neutrinos and antineutrinos of electron and muon flavors, so we perform an estimate of the neutrino thermal conductivity using the simple expression given by Eq.~(\ref{eq:kappa_i}) that includes all four particle types and does not make assumptions about the degeneracy of neutrinos.  However, Eq.~(\ref{eq:kappa_i}) assumes that the MFP is independent of the neutrino energy, which is certainly not true for neutrinos in dense matter (see, for example, Figure~1 in Ref.~\cite{Fischer:2018kdt} or Figures~5 and 6 in Ref.~\cite{Guo:2020tgx}), so the following calculation is just a rough estimate.

The neutrino contribution to the thermal conductivity is the sum of the individual thermal conductivities of electron and muon neutrinos and their antineutrino counterparts.  The specific heat of arbitrary degeneracy gases of $\nu_e$ and $\bar{\nu}_e$ are 
\begin{align}
    c_{V,{\nu_e}}&=\frac{3T^3}{\pi^2}\left[\frac{\mu_{\nu_e}}{T}\phi_3(-e^{\mu_{\nu_e}/T})-4\phi_4(-e^{\mu_{\nu_e}/T})\right]\label{eq:cv_neutrino},\\
    c_{V,{\bar{\nu}_e}}&=-\frac{3T^3}{\pi^2}\left[\frac{\mu_{\nu_e}}{T}\phi_3(-e^{-\mu_{\nu_e}/T})+4\phi_4(-e^{-\mu_{\nu_e}/T})\right],\label{eq:cv_antineutrino}
\end{align}
where $\phi_n(x)$ is the polylogarithm function of order $n$.  The sum of the specific heat of electron neutrinos and antineutrinos is 
\begin{align}
c_{V,{\nu_e}}+c_{V,\bar{\nu}_e} = \frac{7}{30} \pi^2T^3 + \frac12 \mu_{\nu_e}^2T \,, 
\end{align}
which is consistent with the expression given for the energy density of a fermion-antifermion gas in Ref.~\cite{greiner1995thermodynamics}.  But, we will treat neutrinos and antineutrinos as separate species for this calculation because they have different MFPs.  The specific heat expressions for muon neutrinos and antineutrinos are analogous to Eqs.~(\ref{eq:cv_neutrino}) and (\ref{eq:cv_antineutrino}).

The neutrino thermal conductivity is then given by
\begin{equation}
    \kappa_{\text{all neutrinos}} = \frac{1}{3}\left( c_{V,{\nu_e}}\lambda_{\nu_e}+c_{V,\bar{\nu}_e}\lambda_{\bar{\nu}_e} +c_{V,{\nu_\mu}}\lambda_{\nu_\mu}+c_{V,\bar{\nu}_\mu}\lambda_{\bar{\nu}_\mu}   \right),
\end{equation}
where the velocity term in the thermal conductivity expression (see Eq.~(\ref{eq:kappa_i})) has been set to one because the neutrinos are essentially massless.  The neutrino/antineutrino MFPs were calculated in the same manner as in Refs.~\cite{Fischer:2018kdt,Guo:2020tgx} (that is, using nucleon dispersion relations obtained from a RMF theory and conducting the full integration over phase space), but neglecting for simplicity the weak magnetism and pseudoscalar parts of the hadronic current.  The following processes were included in the calculation:
\begin{align}
    \nu_e+n\rightarrow p + e^-,\nonumber\\
    \nu_e + n \rightarrow \nu_e + n,\nonumber\\
    \bar{\nu}_e+p\rightarrow e^++n,\nonumber\\
    \bar{\nu}_e+p+e^-\rightarrow n,\nonumber\\
    \bar{\nu}_e+n\rightarrow \bar{\nu}_e+n,\nonumber\\
    \nu_{\mu}+n\rightarrow p + \mu^-,\nonumber\\
    \nu_{\mu} + n \rightarrow \nu_{\mu} + n,\nonumber\\
    \bar{\nu}_{\mu}+p\rightarrow \mu^++n,\nonumber\\
    \bar{\nu}_{\mu}+p+\mu^-\rightarrow n,\nonumber\\
    \bar{\nu}_{\mu}+n\rightarrow \bar{\nu}_{\mu}+n. 
\end{align}
We evaluate the neutrino MFP at $E_{\nu}\approx 3T$, which is its average energy when the chemical potential is much less than the temperature \cite{Bergstrom:1999kd}.  A proper calculation of the neutrino thermal conductivity would involve solving the Boltzmann equation, properly accounting for the full distribution of neutrinos and the energy-dependence of the neutrino MFP.  

\bibliographystyle{JHEP}
\bibliography{merger}
\end{document}